
\input phyzzx\input graphicx\input epsf
\overfullrule=0pt
%
%
%
%
\newcount\lemnumber   \lemnumber=0
\newcount\thnumber   \thnumber=0
\newcount\conumber   \conumber=0
\def\mysec#1{\conumber=0\thnumber=0\lemnumber=0
             \equanumber=0\chapter{#1}}
\def\myeq{{\rm \chapterlabel.\the\equanumber}}

\def\Lemma{\par\noindent\global\advance\lemnumber by 1
           {\bf Lemma\ (\chapterlabel\the\lemnumber)}}
\def\Corollary{\par\noindent\global\advance\conumber by 1
           {\bf Corollary\ (\chapterlabel\the\conumber)}}
\def\Theorem{\par\noindent\global\advance\thnumber by 1
           {\bf Theorem\ (\chapterlabel\the\thnumber)}}

%
%
\def\e{\adveq\eqno{\rm (\chapterlabel.\the\equanumber)}}
\def\adveq{\global\advance\equanumber by 1}
\def\twoline#1#2{\displaylines{\qquad#1\hfill(\adveq\myeq)\cr\hfill#2
\qquad\cr}}

\def\manyeq#1{\eqalign{#1}\e}

%
%
\font\tensl=cmsl10
\font\tenss=cmssq8 scaled\magstep1
\outer\def\quote{
   \begingroup\bigskip\vfill
   \def\endquote{\endgroup\eject}
    \def\par{\ifhmode\/\endgraf\fi}\obeylines
    \tenrm \let\tt=\twelvett
    \baselineskip=10pt \interlinepenalty=1000
    \leftskip=0pt plus 60pc minus \parindent \parfillskip=0pt
     \let\rm=\tenss \let\sl=\tensl \everypar{\sl}}
\def\from#1(#2){\smallskip\noindent\rm--- #1\unskip\enspace(#2)\bigskip}
\def\fro#1{\smallskip\noindent\rm--- #1\bigskip}

\def\WIS{\address{Department of Particle Physics\break
      Weizmann Institute of Science\break
      Rehovot 76100, Israel}}

\def\r#1{$\lb \rm#1 \rb$}

%
%
\def\rarrow{\rightarrow}
\def\d#1{{\rm d}#1\,}

\def\calH{{\cal H}}

\def\ch{{\rm ch}}

\def\semidirect{\mathrel{\raise0.04cm\hbox{${\scriptscriptstyle |\!}$
\hskip-0.175cm}\times}}

\def\mod{\mathop{\rm mod}\nolimits}

\def\n{\mathopen{:}}
\def\nn{\mathclose{:}}
\def\ref#1{$^{#1}$}

\def\pr#1{{#1^\prime}}

\def\wiggle{\tilde}

\def\Im{\mathop{\rm Im}\nolimits}
\def\half{{1\over2}}
\def\lb{\lbrack}
\def\rb{\rbrack}

\def\pr{\prime}
\def\diam{{\hbox{\hskip-0.02in
\raise-0.126in\hbox{$\displaystyle\bigvee$}\hskip-0.241in
\raise0.099in\hbox{ $\displaystyle{\bigwedge}$}}}}
\def\dop{\mathop{{\diam}}\limits}
\def\dw#1#2#3#4{
{\scriptstyle{#4}}\,
{\dop_{#3}^{#1}}
{\scriptstyle{#2}}  }
\def\bw#1#2#3#4#5{{w\left(\matrix{#1&#2\cr#3&#4\cr}\bigg\vert #5\right)}}
\def\sw#1#2#3#4#5{{S\left(\matrix{#1&#2\cr#3&#4\cr}\bigg\vert #5\right)}}


\def\sqr#1#2{{\vcenter{\hrule height.#2pt
      \hbox{\vrule width.#2pt height#1pt \kern#1pt
        \vrule width.#2pt}
      \hrule height.#2pt}}}

\def\underwig#1{	
	\setbox0=\hbox{\rm \strut}
	\hbox to 0pt{$#1$\hss} \lower \ht0 \hbox{\rm \char'176}}

\def\bunderwig#1{	
	\setbox0=\hbox{\rm \strut}
	\hbox to 1.5pt{$#1$\hss} \lower 12.8pt
	 \hbox{\seventeenrm \char'176}\hbox to 2pt{\hfil}}

\def\gcd{\mathop{\rm gcd}\nolimits}
\def\pr#1{{#1^\prime}}
\def\wig{\tilde}
\def\SU{\mathop{\widehat{\rm SU}}\nolimits}
\overfullrule=0pt
\Pubnum={}
\date{September, 2018}
\titlepage
\title{Foundations of Rational Quantum Field Theory}
\author{Doron Gepner}
\WIS
\abstract
We study the connection between Rational Conformal Field Theory (RCFT),
$N=2$ massive supersymmetric field theory, and solvable Interaction Round
the Face (IRF)
lattice models. Specifically, one identifies the
fusion rings with the chiral rings.
The theories so obtained are conjectured, and largely shown, to
be integrable. A variety of examples and the structure of
the metric in moduli space are given. The kink scattering theory is given by
the Boltzmann weights of an IRF model, which is built entirely in terms
of the conformal data of the original RCFT. This procedure produces all
solvable IRF models in terms of projection operators of the RCFT.
The soliton structure and their scattering amplitudes are
described.  A host of new rational conformal field theories
is constructed generalizing most,
if not all, of the known ones.
\endpage

\mysec{Introduction}
Two dimensional field theories play an important role in the understanding
of phenomena as diverse as string unified theories, exact quantum field
theories and a variety of solid state systems. Importantly, such theories
offer the possibility of exact solvability. There are four related, though
different
approaches to solvable two dimensional physics. These are integrable field
theories, solvable lattice models, factorizable soliton systems and rational
conformal field theories. Our purpose in this paper is to establish that
all four approaches are essentially the same, and that solving or classifying
one of them, leads to  a solution in all the others.
More precisely, in the language of category theory, we establish an explicit
isomorphism between the following four categories: 1) integrable $N=2$
supersymmetric field theory; 2) rational conformal field theories;
3) solvable fusion interaction round the face lattice models; 4)
factorizable soliton systems. The isomorphisms are general to all
such systems, and in particular cover all the examples known in
the literature.

The initial idea is the following conjecture that we set out to establish:
all integrable $N=2$ supersymmetric field theories are in one to one relation
with rational conformal field theories. The mapping of a given RCFT to
an $N=2$ supersymmetric field theory is done by the identification of the
fusion ring of the rational conformal field theory (RCFT) with the
chiral ring of the $N=2$ supersymmetric theory. This is the first isomorphism
of categories
discussed above, and the key to the others. Our purpose in this paper, is to
establish this basic theorem.

The second isomorphism  maps any given RCFT onto a solvable
interaction round the face lattice model (IRF). It  is done is by
setting the vertex variables of the lattice model to be the primary fields
of the RCFT. The admissibility condition for the allowed lattice configuration
is given by the fusion rules of the RCFT, $a\sim b$ if, and only if,
$N_{a,\alpha}^b>0$,
where $a$ and $b$ are vertex primary fields, $\alpha$ stands for some fixed primary
field, and $N_{a,\alpha}^b$ is the fusion coefficient (actually, two
different,
horizontal and vertical, admissibility conditions are eventually used, for
the general case). The Boltzmann weights need to satisfy the star triangle
equation (STE) for the lattice model to be solvable, and for them to lead to
commuting transfer matrices (see, e.g., ref. \REF\Book{R.J. Baxter,
``Exactly solved models in statistical mechanics'', Academic Press, London,
England, 1982}
\r\Book).
We solve the STE relation by letting the Boltzmann
weights be the extension of the braiding matrices of the rational anomalous
field theories. These
automatically satisfy the braid group relations, and thus the STE at
the extreme UV limit, as well as giving the correct admissibility
relation. We then describe a universal way to introduce
spectral parameters, using properties of the braiding matrices of RCFT
and, in particular, the fact that they obey the $n$'th order
braid algebra introduced here, which generalizes the $A$ type Hecke algebra,
(which corresponds to $n=2$). The Boltzmann weights are given
by projectors onto each eigenvalue times a universal function which depends
only on the conformal dimensions. All the known
solvable IRF models (in the trigonometric
limit) are rederived by this mapping, along with a considerable
number of new
ones. This gives a unified formulation for IRF models, and a
way to treat them using RCFT methods.

From a solvable IRF model we build an integrable soliton system by letting
the vacua stand for the vertex primary fields, and letting the solitons
stand for the primary fields appearing in the admissibility conditions. The
Boltzmann weights then become essentially the soliton scattering amplitudes
(up to factors ensuring unitarity and crossing symmetry, which are described
in general), and the spectral parameter becomes the relative rapidity of
the scattered solitons. This type of map was established before on a
case--by--case basis (for a review see, e.g.,
\REF\rev{M. Wadati, T. Deguchi and Y. Akutsu,
Physics Reports 180 (4\&5) (1989) 247, and ref. therein}
ref. \r\rev) and is described here, in general.

The STE relation
then implies the crucial factorization equation \REF\ZamZam{A.B. Zamolodchikov
and Al. B Zamolodchikov, Annals of Physics 120 (1979) 253} ref. \r\ZamZam,
which ensures the integrability of
the soliton scattering theory. Thus we find a unique way (up to, so called,
Castillejo-Dalitz--Dyson
(CDD)  ambiguities) to map any
solvable IRF model onto any integrable soliton system, recovering,
in particular, all the known soliton systems.

Finally, the soliton systems described here are in fact the solutions to the
massive
$N=2$ supersymmetric field theory based on any given RCFT
(up to a trivial duplication of solitons required for supersymmetry),
thus giving the last isomorphism, along with the solution for the particle
content of the theories, and their scattering amplitudes.

This completes the circle of categorical isomorphisms, establishing that
all the four categories are equivalent. Further, various operations,
such as fusion (IRF), fusion (RCFT) and Bootstrap (RCFT), are category
morphisms, i.e., they commute with the isomorphisms, or in physical language,
are equivalent to each other, under the aforementioned maps.
Another example is the quotient procedure (orbifoldization) in RCFT,
versus that in IRF models \REF\FenGin{P. Fendley and P. Ginsparg,
Nucl.Phys. B324 (1989) 549} \r\FenGin. Yet another example, is the
transition to an extended algebra in RCFT, versus the folding procedure
in IRF models (e.g., \REF\Zuber{J.B. Zuber and P. Di Francesco, Nucl.Phys.
B338 (1990) 602} refs. \r{\Zuber,\rev}).  

The original conjecture about the integrability of $N=2$ massive theories
is then verified by a thermodynamic Bethe ansatz calculation of the algebraic
central charge, along with a calculation of the metric in moduli space of
the $N=2$ theory, which will be reported in future work, thus proving
it.

Our  work connects four central physical problems
and shows their equivalence. Further, it allows the use of methods from
one category to another. For example, from RCFT data one builds an IRF model,
which is
then solved to give a multiple critical point structure that gives back the
{\rm full} RCFT, solving the so called reconstruction `fantasy' problem of
RCFT.

This paper is organized as follows. In section (2), an introduction to this
framework is described using as a concrete set of examples the known
integrable $N=2$ models and showing that they all stem from RCFT's.
In section (3) the connection with metric and kinks is introduced.
In sections (4), (5) and (6) new rational conformal field theories are
introduced, including new types of cosets. The conformal data is introduced
in sections (4) and (5) and the realizations in section (6). The systems
generalize all known RCFT systems (with the exclusion, perhaps, of some
rational orbifolds) along with providing a host of new ones.
In sections (7) (the two block case) and (9), the general case, fusion IRF
models are constructed from their corresponding RCFT theories and the
Boltzmann weights are explicitly described. In section (8)
the graph state IRF case is explored, along with a number of examples.
In section (10) fusion soliton theories are treated, giving their $S$
matrices.
Finally, in section (11) we discuss this work, indicating also
future directions.
\REF\Jimbo{M. Jimbo, T. Miwa and M. Okado, Lett. Math. Phys. 14 (1987) 123}
Our convention about equations is that if the equation is in the same section, 
the section number is deleted from its reference.
\mysec{Fusion Scalar Field Theories}
Let us consider the potential,
$$V(q_1,q_2,\ldots,q_m)=\sum_{n=1}^m {q_n^k\over k}-\lambda q_n\e$$
expressed, in terms of the symmetric variables
$$x_r=\sum_{i_1<i_2<\ldots<i_r} q_{i_1} q_{i_2}\ldots q_{i_r}.\e$$
Here $\lambda$ is the coupling constant. The potential $V$ may be used
as a superpotential of an $N=2$ scalar field theory, or a Landau--Ginzburg (LG)
theory. As was discussed
in \REF\DGCMP{D. Gepner, Comm. Math. Phys. 141 (1991) 381}\REF\Vafa{C.
Vafa and S. Cecotti, Nucl. Phys. B367 (1991) 359} refs. \r{\DGCMP},
for $\lambda=0$, this potential describes the hermitian symmetric model (h.s.s.)
\REF\Suzuki{Y. Kazama and H. Suzuki, Nucl. Phys. B321 (1989) 232}\r\Suzuki\
$SU(m+1)/SU(m)\times U(1)$. The perturbation at $\lambda\neq0$ is known
to be integrable, and the solitonic spectrum and $S$ matrices have been found
or have been convincingly conjectured
\REF\Lerche{W. Lerche and N.P. Warner, Nucl. Phys. B358 (1990) 571}
\REF\Fendley{P. Fendley, W. Lerche, S.D. Mathur and N.P. Warner,
Nucl. Phys. B348 (1991) 66}
\REF\Fendtwo{P. Fendley and K. Intriligator, Nucl. Phys. B380 (1992) 265;
Nucl. Phys. B 372 (1991) 533}
\r{\Lerche,\Fendley,\Fendtwo}.

A closely related set of ideas \r{\DGCMP,\Vafa}\
is the tight connection between
fusion potentials in rational conformal field theory and integrable
$N=2$ supersymmetric models. We would like to examine this class of theories
from this particular angle. Specifically, we will construct a rational
conformal
field theory whose fusion ring is given by the potential $V$.
We shall then further establish that the rational conformal field
theory essentially solves the $N=2$ one, and allows one to derive the
solitonic scattering amplitudes (section (10)).

Consider then the potential $V$, eq. (1), with the coupling set to
$\lambda=(-1)^m$. As was discussed in ref. \r\DGCMP\ the points of the variety
$${\partial V\over \partial x_i}=0\e$$
form a diagonal basis for the fusion ring. We shall denote the
solutions of eq. (3) by $x_i$, $i=1,2,\ldots,q$, and refer to this
set of points as the fusion variety.
The primary fields $\lb p \rb$ are also numbered as $p=1,2,\ldots q$, and
form a basis for the fusion ring $P\lb x_i\rb/\partial_i V$ (the algebra of polynomials
modulo the derivatives of $V$).
At the points $x_i$, the primary fields assume the values \r\DGCMP,
$$\lb p\rb(x_i)=S_{p,i}^\dagger/S_{p,1},\e$$
where $S_{p,q}$ is the matrix of torus modular transformations,
and ``$1$'' denotes the unit primary field.

For the potential $V$, eq. (1), the fusion variety is readily found
via eq. (3),
$$q_i=e^{2\pi i (r_i-(m-1)/2)/k},\e$$
where $0\leq r_1\leq r_2\leq\ldots\leq r_m<k$ is a monotonic sequence
which labels the solutions. The $x_i$ are computed by substituting the
$q_i$, eq. (5), into eq. (2).

We face the problem of how to define the primary fields. Clearly,
this is a very special basis for the fusion ring, where the structure
constants are non--negative integers, and, in which, eq. (4) holds.
There is no systematic way of
doing this, and in fact the same potential may in principle correspond to
different rational conformal field theories, albeit,
no such examples are known\foot{It is conjectured below that the primary
fields basis is the unique one which diagonalizes the metric, as will be explained below. 
This is true
for all examples in which the metric have been calculated. This implies,
in particular, that for each potential there is a unique basis of primary
fields which is consistent, and which can be unequivocally found by calculating
the metric (c.f., section (3)).}.
Thus, we essentially need to guess the expression
for the primary fields, and then check that it is a consistent one.
The natural form is a Weyl--type formula,
$$\lb r_1,r_2,\ldots,r_m\rb={\sum_{w\in S_m} (-1)^w
\prod_i q_{w(i)}^{r_i} \over \prod_{i<j} (q_i-q_j)}.\e$$
The denominator, being the Vandermonde determinant, can be written
also by a summation formula
$$\prod_{i<j}(q_i-q_j)=\sum_{w\in S_m} \prod_i(-1)^w q_{w(i)}^{i-1},\e$$
which shows that the field $\lb 0,1,\ldots,m-1\rb$ indeed corresponds
to the unit polynomials, using eq. (6).
Here $S_m$ stands for the permutation group
of $m$ elements, and $(-1)^w$ is $1$ ($-1$) when $w$ is an even (odd)
permutation. The polynomials, eq. (6), are symmetric functions of
the $q_i$, and thus can be expressed in terms of the $x_i$ which are
the generators of the symmetric polynomials.
The matrix $S$ can now be computed from eq. (4),
$$S_{r_i;s_i}={1\over [m(k+m)]^\half}
\sum (-1)^w e^{2\pi i (r_i-(m-1)/2)(s_{w(i)}-
(m-1)/2)/k}.\e$$
It can be verified that this $S$ matrix is unitary, $S^\dagger=S$, which
is required for the consistency of the theory.
Since $S$ is symmetric in $r$ and $s$, and further
since this $S$ matrix obeys eq. (4) above, it follows that the
fusion rules defined by $S$ are identical to the product of polynomials,
eq. (6), along with the relations derived from the potential
$V$, eq. (3). Moreover, it can be checked that $S^2=(ST)^3=C$, where $C$ is the
charge conjugation matrix
$$C_{r_i;s_i}=\sum_w \prod_i \delta(r_i+s_{w(i)}-m+1),\e$$
where $\delta(q)=0$ unless $q$ is divisible by $k$, and then
$\delta(q)=1$. The sum in eq. (9), is over all permutations $w$.
For each primary field $p$, there is a unique
conjugate one $\bar p$, such that $C_{p,\bar p}=1$, and all other
matrix elements of $C$ vanish. Clearly, also, $C^2=1$.
$T$ is the generator of the modular transformation $\tau\rarrow \tau+1$,
which is the diagonal matrix
$$T_{p,q}=\exp\lb 2\pi i (\Delta_p-c/24)\rb \delta_{p,q},\e$$
where $\Delta_p$ is the dimension of the primary field $p$.
The dimension formula
can be seen to be
$$\Delta_{r_i}={1\over 2k} \sum_i \lb r_i-(m-1)/2\rb^2-{1\over24k}
(m^2-1)(m+2).\e$$
The constant $c$ which is the central charge of the Virasoro algebra
is
$$c={(k-m)(m^2-1)\over k}+1.\e$$
The above formulae for $S$, $T$, $C$, $\Delta$ and $c$ follow directly
from consistency with respect to the proposed fusion rules.
It is shown in the sequel that the fusion coefficients are all, indeed,
non--negative integers.

It is convenient to introduce Young tableau notation for the fields in the
theory by letting the `signature' series $r_i - i + 1$ stand for
the width of the $i$'th row
in the tableau.
See an example in fig. (2.1).
It is also convenient to introduce a dual notation, where
$\lb r_i\rb \equiv (s_1, s_2, \ldots, s_k)$, $r_m \leq k + m - 1$, and $0
\leq s_i \leq s_{i + 1} \ldots, \leq m$, where $s_i$ stands for the height
of $i$'th column in the Young tableau.  The relation between $s_i$ and $r_i$
is, that the
series $\{r_i - i + 1\} \cup \{s_i - 1\}$ is identical to the series
$\{0,1,2,\ldots,
\max (k,m)\}$.  Alternatively $[s_i+i-1]$ describes the transpose of the
young tableau of $\lb r_i\rb$.
Now, one can write down a
Schubert--like calculus for the fusion rules of the theory.
The generators are the `fundamental fields',
$y_r=(0,0,\ldots,0,r)= \lb0,1,\ldots,m-r, m-r + 2, m-r + 3\ldots,m + 1\rb$.
The products of the field $y_r$
with any primary field in the theory are expressed by the
Pieri formula (see, for example, ref.
\REF\Hiller{H. Hiller, ``Geometry of Coxeter groups'', Pitman, London, 1982}
\r\Hiller),
$$y_r  \cdot (s_1,s_2,\ldots,s_k)=\sum_{s_i\leq p_i<s_{i+1}\atop
\sum p_i=\sum s_i+r } (p_1,p_2,\ldots,p_m).\e$$
{}From the Pieri formula one may derive also a Giambelli
formula expressing the general primary field as a polynomial in the
generators. In fact, the primary fields are the usual Schure functions
for $S_m$, and the formula expressing them in terms of the generators
is the Jacoby--Trudi formula (see, e.g.,
\REF\And{G.E. Andrews,
Theory of partitions,
Encyclopedia of mathematics and its applications, Vol. 2,
(1976).}\r\And),
$$(s_i)=\det_{i,j} y_{s_i+i-j},\e$$
where $y_0 = 1$, and $y_n = 0$ for $n>m$, by convention.
The two formulas, eq. (13--14) form the so called Schubert calculus.
As is well known, there is a duality upon the exchange of
$k$ and m, which is essentially the transposition of Young
tableau \r\And.  This is equivalent to the exchange of  $r_i - i + 1$ with
$s_i$.  Eq. (14), the Jacobi-Trudi, still holds in terms of the
generators which are the
totally symmetric Young tableaux, $Z_r = (0,0,1,\ldots,1)$, where there are
$r$ $1$'s, i.e.,
$$\lb r_i\rb = \det_{i,j} Z_{r_i +1-j},\e$$
where $Z_0=1$ and by convention, $Z_i=0$ for $i<0$.

Rather than dwell any further on the aforementioned relations, we will directly
establish a realization of these fusion rules as conformal field
theories, from which they will follow. To do so, consider the conformal field
theory,
$$U(1)_{m(k+m)}\times SU(m)_k,\e$$
which can be realized as a free boson propagating on a circle of radius
$\sqrt{m(k+m)}$ (for the U(1) part) and a level $k$ $SU(m)$ WZNW
theory
\REF\Witten{E. Witten, Comm. Math. Phys. 92 (1984) 455}
\REF\KZ{V.G. Knizhnik and A.B. Zamolodchikov, Nucl. Phys. B247 (1984) 83}
\REF\GW{D. Gepner and E. Witten, Nucl. Phys. B278 (1986) 493}
\r{\Witten,\KZ,\GW}.
The general primary field in the theory is represented by the product
$$\Phi^\lambda_q=g^\lambda e^{iq\phi/\sqrt{m(m+k)}},\e$$
where $q$ is an integer, $\lambda$ is an integrable highest weight
at level $k$ of $\SU(N)$, and $g^\lambda$ is a primary fields
with this weight and highest weight.
The dimension of the primary fields is
$$\Delta={m\lambda(\lambda+2\rho)+q^2\over 2 m(k+m)}.\e$$
Here $\rho$ is half the sum of positive roots.
Actually, the theory eq. (16) admits
an extended chiral algebra, and we can organize the primary fields
into representations of this larger algebra. We assume that $k$ is an
odd integer. Then, we can use as an extended current algebra the
holomorphic field of integral dimension
$H=\Phi^0_{m(k+m)}$,
Further,
we can define the extended algebra currents,
$$J_r=\Phi^{k\lambda_r}_{r(k+m)},\e$$
where $\lambda_r$, $i=1,2,\ldots,m$,
are any of the fundamental weights of $SU(m)$. It can be checked that
the $J_r$ are holomorphic fields of the integral dimension,
$$\Delta_r={r(k+r)\over2}.\e$$
The extension of the algebra to include the field $H$
implies that the charge $q$ is defined modulo $m(k+m)$, as the fields
$\Phi^\lambda_q$ and $\Phi^{\lambda}_{q+m(k+m)}$,
are related by the
appropriate operator product expansion with respect to $H$.
Next, there is a requirement that the fields will be local with
respect to the new currents $J_r$. This implies that in the operator product
$J_r(z) \Phi^\lambda_q(w)=\Phi^{\sigma_r(\lambda)}_{q+r(k+m)}(w) (z-w)^
\delta+{\rm h.o.t}$
only integral values of $\delta$ are allowed to
appear. Here $\sigma^r(\lambda)$ denotes the external automorphism of the
extended Dynkin diagrams.
Using eq. (18), it follows that
$$c(\lambda)=m\lambda\lambda_m=q\  \mod m,\e$$
where $c(\lambda) \ \mod\, m$ is the $m$-ality of the weight $\lambda$.
In addition, the action, in the operator product sense, of the new
currents identifies the fields
$$\Phi^{\sigma^r(\lambda)}_{q+r(k+m)},\e$$
for any integer $r$ (defined modulo $m$).
In terms
of the Dynkin labels $l_i$, defined below, the
automorphism is
$$\sigma(l_1,l_2,\ldots,l_m)=(k-\sum l_i,l_1,l_2,\ldots,l_{m-1}).\e$$

Let us demonstrate that indeed the conformal field theory $SU(m)
\times U(1)$ is a realization of the fusion ring described above.
First, note that the central charges are indeed identical, since the
$U(1)$ factor contributes the central charge $c=1$, while $SU(m)_k$ has the
central charge $c=k(m^2-1)/(k+m)$. Shifting $k$ to $k-m$ we find
that the central charge so obtained is
identical to the one found earlier, eq. (12).
To describe the correspondence between
the fields, it is convenient to introduce the signature of the Young
tableau $t_i=\sum_{j=m+1-i}^{m-1} l_j+i-1$, where $i=2,3,\ldots,m$,
$t_1=0$, and
where the $l_j$, $j=1,2,\ldots,m-1$,  are the Dynkin
labels of the representation; $\lambda=\sum_i l_i \lambda_i$, where
$\lambda_i$ are the fundamental weights.
The $t_i$ form a strictly monotonically
increasing sequence, $0=t_1<t_2<\ldots<t_m<k+m-1$. Conversely,
any such sequence describes an integrable highest weight at the level $k$.
Now, the primary field $\lb r_i\rb$, defined by eq. (6), uniquely
maps to the primary field
$\Phi^t_q$, where the signature $t$ is defined as $t_i=r_{i+1}-r_1$,
and the charge $q$ is given by $q=\sum r_i-i$, or,
$$\lb r_1,r_2,\ldots,r_m \rb\longrightarrow \Phi^{t_2,t_3,\ldots,t_m}
_q.\e$$
To prove this equation, we make use of the Weyl character formula to
show that the $S$ matrix, eq. (8), decomposes into a product of an
$SU(m)$ modular matrix, at the level $k$, with a $U(1)$ factor,
at the level $m(k+m)$. Denote the weight by $\lambda$, as before.
The simple roots of $SU(m)$ are given by $e_i-e_{i+1}$, where
$i=1,2,\ldots,m-1$, and the $e_i$ form an orthonormal set of
unit vectors. Up to an overall addition of a term proportional to
$e_1+e_2+\ldots+e_m$, the fundamental weights are given by
$\lambda^r=-\sum_{r+1}^m e_i$. Denoting $\rho=\sum
\lambda_i$, we find that the weight $\lambda+\rho$ assumes the form
$$\lambda+\rho=-\sum_{i=1}^m t_i e_i\,\mod\, e_1+e_2+\ldots+e_m,\e$$
where $t_i$ is the signature of the Young tableau, introduced
above. The modular matrix for the group $G$ current algebra
at the level
$k$ can be written explicitly as \r\GW,
$$S_{\lambda,\mu}={m^\half\over (k+m)^{m-1\over2}}\,\sum_{w\in W} (-1)^w
e^{2\pi i w(\lambda+\rho)(\mu+\rho)/(k+g)},\e$$
Where $W$ is the Weyl group of $G$, which for the Lie algebra $SU(m)$ acts
as permutations on
the $e_i$, and $g=m$. $(-1)^w$ is $\pm1$ if the permutation is even or odd.
Substituting the expressions for $\lambda$ and $\rho$, eq. (25),
we find
$$S_{\lambda,\mu}=C\sum_{p\in S_m} (-1)^p
e^{2\pi i \sum_r r_i s_{p(i)}/(k+g)},\e$$
where $r_i$ and $s_i$ are
the signature
sequences for the representations $\lambda$ and $\mu$.
This $S$ matrix is identical to eq. (8), where the only
difference is that $\sum r_i=\sum{s_i}$ is required to vanish, in order to
describe $SU(m)$ weights.
It follows that we can factor out this term to get a product of a
$U(1)$ modular matrix times an $SU(m)$ one,
$$S_{r_i;s_i}=\sum_{p\in P} e^{2\pi i \sum_r r_i s_{p(i)}/(k+m)}
=e^{\pi i M_r M_s \over m(k+m)} S_{\lambda,\mu},\e$$
where $\lambda$ and $\mu$ are the $SU(m)$ weights with the signatures
$r_i-r_1$ and $s_i-s_1$ respectively, and the $U(1)$ charge is
given by $M_r=\sum (r_i-i+1)$.  This proves that
we get also the correct fusion rules, making use of eq. (4)
(or, alternatively,
the relation found in ref. \REF\Verlinde{E. Verlinde, Nucl. Phys. B300 (1988)
360}\r\Verlinde).
It can be checked that the fusion rules so obtained are precisely
those of the $SU(m)_k\times U(1)_{m(k+m)}$ with the extended algebra
described above. The Schubert--like calculus, developed for
$SU(m)$ in ref. \r\DGCMP, can then be used to derive the full Schubert
calculus, eqs. (13,14).

Let us turn now to the potential $V$.  Denote by $V_m$ the function,
$$V_m = q_1^k + q_2^k + \ldots + q_m^k,\e$$
expressed in terms of the symmetric variables $y_r=x_r$, eq. (2).  A
recursion relation for $V$ follows from the polynomial equation obeyed by
$q_i$ (see the appendix to ref. \r\DGCMP),
$$0=\prod_{i = 1}^m (X - q_i)=X^m - y_1 X^{m-1}+y_2 X^{m-2}+
\ldots+(-1)^m y_m.\e$$
Multiplying by $X^s$, and summing over the solutions, $q_i$, implies
the recursion relation,
$$V_{m+s}-y_1V_{m+s-1}+y_2V_{m+s-2}+\ldots+(-1)^m y_mV_m=
0.\e$$
clearly, $V_0=m$, $V_1=y_1$, $V_2=y_1^2-2y_2$, etc.  Thus we can use
the relation eq. (31) to compute the potentials.  Another relation that
holds for $V_m$ is,
$${1\over m} ~ {\partial V_m\over \partial y_s} = (-1)^{s - 1} Z_{m-s},\e$$
where $Z_n$ denotes, as before, the polynomial expressing the totally symmetric representation,
$$Z_n = \lb 0,1,2,\ldots, m-1, m+n\rb.\e$$

Now, the potential of the theory, eq. (1), can be written as
$$V = V_{k+m+1}-\lambda y_1,\e$$
where $\lambda = (-1)^{m-1}$ is the coupling constant of the
perturbation $y_1$.
The fusion ring is, as usual \r\DGCMP\ the ring
generated by the $y_r$, $r=1,2,\ldots,m$, along with the relations,
${\partial V\over\partial y_r} = 0$, or it is the Jacobian variety,
$$R \approx {P\lb y_i\rb\over (\partial_i V)} .\e$$
Using eq. (32), the relations can be written as,
$$Z_{k+m-r} = \lambda \delta_{r,1} ~{\rm for}~ r = 1,2,\ldots, m-1,\e$$
where the $Z_m$'s are expressed as polynomials in the $y_i$ variables,
which can  be computed, for instance,
using the Jacobi--Trudi, eq.
(14).  The number of solutions to this set of quasi--homogeneous
equations, is the product
of the degrees of the equations, divided by the product of the degrees of
the generators, where $\deg (y_r) = r$, (see, e.g.,
\REF\Harris{P. Griffiths and J. Harris, ``Principles of algebraic geometry'',
New--York, Wiley, 1978}
\r\Harris)
$$N = \prod_{r=1}^m {k+m-r\over r}.\e$$
Indeed, this is exactly the correct number of the primary
fields as defined by eq. (6).

Let us turn, now, to some examples.  Consider the case of $m = 3$, in the
notation of eq. (1).
The potentials $V_m$, eq. (29), are readily calculated using eq. (31),
$V_0 = 3$, $V_1 = y_1$, $V_2 = y_1^2 - y_2$,
$V_3 = y_1^3 - 3y_1 y_2 + 3y_3$, $V_4 = y_1^4 - 4y_1^2 y_2 + 4y_1 y_3 +
2y_2^2$,
$V_5 = y_1^5 - 5y_1^3 y_2 + 5y_1y_2^2 + 5y_1^2 y_3 - 5y_2 y_3$, etc.
In the case of $k = 1$ we obtain, using the Jacobi--Trudi, eq. (14),
the following relations,
$$Z_2 = \left|\matrix{y_1 & 1\cr y_2 & y_1\cr}\right| = y_1^2 - y_2 =
0,\e$$
$$Z_3 = \left|\matrix{y_1 & 1 & 0\cr y_2 & y_1 & 1\cr y_3 & y_2 &
y_1\cr}\right| = y_1^3 - 2 y_1 y_2 + y_3 = 0, \e$$
$$Z_4 = \left|\matrix{y_1 & 1 & 0 & 0\cr y_2 & y_1 & 1 & 0\cr y_3 & y_2
& y_1 & 1\cr 0 & y_3 & y_2 & y_1\cr}\right| = y_1^4 - 3y_1^2 y_2 + 2y_1
y_3 + y_2^2 = 1. \e$$
It follows that $y_2 = y_1^2$ (eq. (38)),  $y_3 = y_1^3$ (eq. (39)),
and $y_1^4 = 1$ (eq. (40)).  Thus, we recover the fusion ring\foot{
Note that the superpotentials $V_k^m$ and $V_m^k$ lead to identical
superconformal field theories, that is there is the rank--level duality
of exchanging
$m$ and $k$. This follows from the usual duality of the h.s.s. theory,
$SU(m+1)_k/SU(m)\times U(1)$ of exchanging $m$ with $k$
\REF\Geplong{D. Gepner, Comm. Math. Phys. 142 (1991) 433}
\r{\Suzuki,\Geplong}. This h.s.s. theory 
is the $\lambda=0$ limit. Further, since the duality entails the
transposition of the Young tableaux, the perturbation $y_1$ is duality
invariant (as well as any other perturbation which is transposition
invariant). Thus, on physical grounds $V_m^k\approx V_k^m$. Alternatively,
one may use the duality of the Schubert calculus \r\And, to prove this
from the mathematical viewpoint. (Or, alternatively, give a physical
proof for this duality.) In particular, $V_{m,k}$ with $k=1$ is dual
to $V_{1,m}=q^{m+2}/(m+2)-q$. Since this is the fusion potential of $SU(m+1)$
it follows that all $k=1$ examples give the fusion ring of
$SU(m+1)_1$. In terms of the corresponding
RCFT's, the duality translates to the equivalence
relation of the theories $SU(n)_k\approx SU(k)_{-n}$,
up to a free boson theory.
This is a consequence of the known RCFT relationship $SU(mk)_1\approx
U(1)\times SU(m)_k\times SU(k)_m$, with a particular modular invariant,
which follows from the decomposition of $nk$ fermions according to flavor
and color current algebras
\REF\GR{D. Gonzales and A.N. Redlich, Nucl. Phys. B256 (1985) 621}
\REF\Affleck{I. Affleck, Nucl. Phys. B265 [FS15] 409; 448}
\REF\Mlawer{E.J. Mlawer, S.G. Naulich, H.A. Riggs, H.J. Schnitzer,
Nucl. Phys. B352 (1991) 863}
\r{\GR,\Affleck,\DGCMP,\Mlawer}.
Inverting $SU(k)_m$ to the other side of this equation,
implies the duality of exchanging $m$ and $-k$.
In principal, we could
have taken $SU(m)_{k/p}$ where $p$ is any integer strange to $m(k+m)$,
leading to the same fusion rules, see section (5).
However, only the values of $p=\pm1$ are allowed due to time reversal
symmetry of the soliton scattering amplitudes (c.f. section (10)),
and these are equivalent under this duality.
It follows that the $N=2$
LG leads to a unique RCFT, as it should, since the scattering amplitudes are
built in terms of this particular RCFT braiding matrices (section (7)).
}\ 
of $SU(4)_1$, $R\equiv{P\lb y_1\rb\over (y_1^4 - 1)}$.
Using eqs. (6,24) we can
write the primary fields as polynomials in $y_1$.  We find,
$y_1 = \Phi_1^{\lambda_1}$, $y_2 = y_1^2=\Phi_2^{\lambda_2}$, $y_3=
y_1^3=\Phi_3^0$, and indeed the relation $y_1 y_3 =
\Phi_1^{\lambda_{1}} \cdot \Phi_3^0
= \Phi_4^{\lambda_{1}} = \Phi_0^0 = 1$ holds (where we used the external
automorphism action, eq. (23)).  The central charge is, eq. (12),
$c = {3^2-1\over 3+1} +
1 = 3$, i.e., the same value as for $SU(4)_1$. We conclude that this theory is
equivalent to $SU(4)_1$.  For $k = 2$, the calculations are slightly
more involved, and are mostly left as an exercise to  the reader.  The
relations in the fusion ring can be written as
$$ \eqalign{2y_1 y_2 - y_1^3 &= y_3,\cr
y_1^4 - y_2^2 - y_1^2 y_2 &= 0,\cr
-3 y_1 y_2^2 + 2y_1^3 y_2 + 1 &= 0.\cr}\e$$
Again, $y_1 = \Phi_1^{\lambda_{1}}, y_2 = \Phi_2^{\lambda_{2}}, y_3 =
\Phi_3^0$ and there are indeed ten fields, which can be expressed as
polynomials using eq.
(14).  For example, $y_1^2=\Phi_1^{\lambda_1}\cdot \Phi^{\lambda_1}_0=
\Phi_2^{2\lambda_1} + \Phi_2^{\lambda_2}$,
showing that indeed
$\Phi_2^{2\lambda_1}=
\left|\matrix{y_1 & 1\cr y_2 & y_1\cr}\right|=y_1^2 - y_2$.
\mysec{Metric and kinks}

The potential $V$ can be taken to be a superpotential of an $N=2$
scalar field theory (or a Landau-Ginzburg Wess--Zumino model in two dimensions)
\REF\ABZM{A.B. Zamolodchikov, Sov. J. Nucl. Phys. 441 (1987) 529}
\REF\Martinec{E. Martinec, Phys. Lett. 217B (1989) 431}
\REF\Wartwo{N.P. Warner and C. Vafa, Nucl. Phys. B324 (1989) 371}
\REF\Scalar{D. Gepner, Nucl. Phys. B322 (1989) 65}
refs. \r{\ABZM,\Martinec,\Wartwo,\Scalar}.  The
lagrangian  of the theory is,
$${\cal L}=\int K(x,\bar x) {\rm d}^2\theta \, {\rm d}^2\bar\theta+
\int V (x){\rm d}^2\theta+{\rm c.c.},\e$$
where the $x_i$ are now $N=2$ superfields, $V$ is the superpotential,
and $K$ is a suitable kinetic term.
The resulting theories are the perturbation
by the most relevant operator of the h.s.s. conformal field theories
which are the superconformal cosets $SU(m+1)/SU(m)\times U(1)$
\r{\DGCMP,\Vafa}.
These massive field theories are known to be integrable
\r{\Lerche,\Fendley,\Fendtwo}.
Thus we see nicely the connection between the integrability of the
potential, and the fact that it is the fusion potential of a rational
conformal field theory, a conjecture that was already addressed in ref.
\REF\GS{D. Gepner and A. Schwimmer, Nucl. Phys. B380 (1992) 147}
\r\GS.  In a precise form, this statement is

\noindent{\bf Conjecture}:
Every rational conformal field theory gives rise to an integrable
$N=2$ superconformal field theory, where the chiral algebra of the
latter is isomorphic to the fusion ring of the former. Further,
any $N=2$ field theory is integrable if, and only if, it originates
from some rational conformal field theory.

Our aim in this paper
is to prove, and establish this statement.

The merit of this conjecture is the linking of two hard, yet very
different problems, i.e.,
the classification of rational conformal field theories and the study of two
dimensional integrable systems.

The class of examples, discussed in section (2), fits nicely into this
conjecture. Indeed, the $N=2$ theories with the potential eq. (2.1) were
the first family of integrable $N=2$ LG theories to be described.
The others were, subsequently, studied by directly substituting the
known fusion potentials of $SU(m)_k$ \r{\Lerche,\Fendtwo}
and $C_m$ \r\GS.
Both families of
theories have been argued to be integrable. Thus, the $`U(m)$' family,
described in section (2), essentially completes the verification of
this conjecture for all the known integrable $N=2$ theories, and
almost all the rational conformal field theories for which a fusion
potential is known.

There are more detailed aspects of this correspondence, in part
noted in ref. \r\GS. In particular, it was found there, in some examples,
that the metric
in the moduli space of the $N=2$ theory is diagonalized by the primary
fields, and so this preferred, `typically rational conformal field
theory' basis can be defined entirely in the $N=2$ context, as the
unique basis that diagonalizes the metric. If true, in general, this
would allow one to construct a rational conformal field theory by
calculating the metric of the corresponding $N=2$ one, a relatively
manageable
task. When the rational conformal field theory is known, it can,
on the other hand, be used to diagonalize the metric and to make its
calculation easier.

Now, the ground states of the $N=2$ scalar theory, eq. (1) are given by the
minima
of the scalar potential, $W = \sum_i |\partial_i V|^2$, implying, in
turn,
that $\partial_iV=0$. Recall that, these are
precisely the points of the fusion variety, $x_p$, which are labeled by
the primary fields $p$, as explained in section (2).
The value of the primary field $q$ at the
point $x_p$ is, eq. (2.4), $\lb q\rb(x_p)=S_{p,q}^\dagger/S_{1,q}$,
where $S$ is the matrix of modular transformations. Between some
pairs of admissible `neighboring' vacua $p$ and $q$, there are some
interpolating kinks. The following kink structure has been found
\r\GS\ in some examples, using a metric calculation, and as is shown in
this paper holds in general.  The kinks are labeled by some `fundamental'
representations, $f_i$.  The vacua $p$ and $q$ are connected by the
kink $f_i$ if, and only if, the fusion coefficient
$N_{p,f_i}^q$ does not vanish. We are thus led to a complete
scattering picture for the kink theory, where the $S$ matrices are
essentially given by the Boltzmann weights of the interaction round
the face theory $w\left(\matrix{p&q\cr r&s\cr}\bigg|u\right)$, where
$u$ is the relative rapidity of the incoming kinks. This $S$ matrix represents
the scattering amplitude of $K_{s,p}K_{p,q}\rarrow K_{s,r}K_{r,q}$, as
will be described in detail in the proceeding sections.
The Young--Baxter integrability condition for the process is, in turn,
the star triangle equation (STE) for this Boltzmann weight, which ensures
the integrability of the lattice model, whose partition function is
$$Z=\sum_{\{p_i\}}\prod_{\rm plaq.}
w\left(\matrix{p_i&p_j\cr p_k&p_l\cr}
\bigg\vert u\right).\e$$
Importantly,  all the known integrable lattice models, which
have a second order phase transition point, can be written as fusion IRF
theories. Thus, each such model is associated to some unique integrable $N=2$
supersymmetric
theory, along with some rational conformal field theory.  Explicit examples
are elaborated in section (9).
In fact, many constructions of integrable  lattice models have been, in
part, based on fusion rules of known rational conformal field
theories, refs. \r{\Jimbo, \Zuber}. Not surprisingly,
there is a close relationship between
the lattice models and their conformal field theory counterparts.
For example, the critical theory itself is a rational conformal
field theory, having more or less the same fusion rules. In the
case of $G=SU(m)_k$, it was shown \r\Jimbo\ to be the coset model
$G_k\times G_1/G_{k+1}$. Thus, the conformal field theory figure both in the
definition of the theory and in its solution\foot{
The exact relation between the generic fusion IRF model and its
multicritical RCFT's
is not yet resolved in its entirety, as it behooves us to disentangle the
full-fledged, i.e. thermalized, lattice model.
The following `empirical' rule seems to hold. The multi--critical points of
the model IRF$({\cal O},x,x)$ are described by the RCFT $\cal O$ as perturbed
by the field $x$. Note, however, that there are a number of choices for the
RCFT $\cal O$ and $x$ which lead to the same IRF model, and one must choose
the correct RCFT which gives the particular multicritical point. For example,
the models IRF$(SU(n)_k,[n],[n])$ (see section (8) for more)
are described in the ferroelectric regime by the
RCFT $G_{k-1}\times G_1/G_{k}$, where $G=SU(n)$, perturbed by an operator $x=
\Phi^{0,0}_{\rm ad}$
\REF\Kuniba{A. Kuniba, T. Nakanishi and J. Suzuki, Nucl. Phys. B 356
(1991) 750}
ref. \r\Kuniba.
Since the operator $x$ is a singlet in the $G_{k-1}
\times G_1$ theory, it follows that it, essentially,
gives back the fusion rules of $G_k$.}.
Further, certain
expectation values on the lattice, the so called local height
probabilities, expressing the probability of a given lattice site to
have a certain value, are given directly in terms of the characters
of the conformal field theory, and obey the same modular transformation
properties, an observation made in ref. \REF\Laguna{D. Gepner, in the
proceedings of, ``Nonperturbative methods in field theory'', Irvine (1987),
World Scientific, p. 105, and Princeton University preprint PUPT 1049 (1987).}
\r\Laguna. Thus, we see here a triad between rational conformal field
theories, integrable superconformal models and integrable lattice models.
Our object is to show that these  are equivalent physical problems.

Let us examine the family of rational conformal field theories
described in section (2) for the purpose of verifying the picture drawn
above. This is a relatively straightforward task, as these conformal field
theories describe the simplest family of known integrable $N=2$
supersymmetric theories. In fact, in essentially all other known
integrable supersymmetric field theories, this picture was already
verified \r\GS. The first question then, is regarding the metric:
is it diagonal in the basis of the primary fields? To see that this
is so, let us recall the calculation of the metric, refs. \r{\Vafa,\GS}.
Using a change of variables, we can write the superpotential, eq. (2.1),
as
$$V=\lambda(\sum_{i=1}^m {q_i^{k+1}\over k+1}-q_i),\e$$
where $\lambda$ is a coupling constant. For $\lambda \rightarrow 0$, the
superpotential $V$ describes the h.s.s. theory $SU(m+1)/SU(m)\times
U(1)$. For $\lambda=(-1)^m$ it is the fusion potential of the
rational conformal field theory $\lb SU(m)_k\times U(1)_{k(k+m)}\rb/\sigma$,
described in section (2).
For any $\lambda$, it is an integrable theory.
The metric $g_{\lambda,\mu}=\langle \lambda|\mu\rangle$ is defined
as the overlap of the Ramond vacua $\mu$ with the one obtained on the
left by `spectral flow'. It is the normalization of the chiral fields,
as well as the metric in the moduli space. The metric obeys the equation
\REF\Ferrara{S. Cecotti, S. Ferrara and L. Girardello, Phys. Lett.
B213 (1988) 443; Int. J. mod. Phys. 4 (1989) 2475;
Phys. Lett. 213B (1988) 443}
\REF\Stroming{A. Strominger, Comm. Math. Phys. 133 (1990) 163}
\r{\Ferrara,\Stroming,\Vafa}
$$\partial(g\partial g^{-1})+\lb V,gV^\dagger g^{-1}\rb=0\e$$
where $\partial=\partial/\partial\lambda$, and the equation is
written in a matrix notation. In ref. \r{\Vafa} it was shown that for this
family of potentials, the matrix elements $g_{\lambda,\mu}$ obey
an affine Toda equation. We will show that the off diagonal elements
vanish in the primary fields basis. One computes the solutions
using the pullback to the decoupling variables $q_i$. The Jacobian,
for this change of variables from $x_i$  to  the $q_i$
is the Vandermonde determinant $D=\prod_{i<j} (q_i-q_j)$.
The metric for the potential $V$, in terms of the $q_i$, is
of the form
$$g_{t_i,n_i}=g(t_i-n_i)=\prod_i e^{L(r_i)},\e$$
where $t_i$ and $n_i$ label the  primary fields, and are
the same labels that were used for the $U(m)$ primary fields in eq. (2.6).
The functions $L(r_i)$ are the solutions of an $\hat A_m$ Toda equation, where
the $r_i$ label the specific boundary condition, determined by the
behavior of $L$ at the origin, i.e., the metric at the conformal point.
The metric in the fusion variety basis is given by the Fourier components of
$g(n_i - m_i)$,
$$g_{r,s}=e^{2\pi i(t_1r_1+t_2r_2+\ldots+t_mr_m)/k} e^{-2\pi i
(s_1 n_1+s_2 n_2+\ldots+s_m n_m)/k} g(n_i-m_i).\e$$
Now, pulling back the metric to the symmetric variables, involves
two things: 1) Multiplying by $J(r_i)^* J(s_i)$,
where $J$ is the Jacobian for this change of variables, which is
identical to the
Vandermonde determinant.
2) Anti--symmetrizing with respect to the permutations of the
different $i$'s.  Thus, we find that the metric in the fusion variety
basis of the $U(m)$ theory is
$$g_{r_i,s_i}=J(r_i)^* J(s_i)\sum_{t_i,n_i} \left(
\sum_{p\in S_m} (-1)^p  e^{2\pi ir_it_{p(i)}}\right)
\left(\sum_{q\in S_m} (-1)^q  e^{2\pi is_in_{q(i)}}\right) g(t_i-n_i).\e$$
Note, that the Vandermonde is essentially the denominator of the
Weyl character formula, while the permutations correspond to the
Weyl group.
Hence, the transition to the primary fields basis is implemented by the
matrix $S_{r;m}^\dagger/S_{r;0}$ which is the `Weyl' character
of $r_i$ specialized to the points $m_i$, eqs. (2.6, 2.27). It follows
that we find
exactly the orthonormality condition for the Weyl characters, using
as a measure
the denominator, $D = J$,
$$g_{t_i,n_i}=\sum_{r_i,s_i} {S_{r;t}\over S_{r;0}} {S_{s;n}^\dagger
\over S_{s;0}} g(r_i;s_i)=\delta_{t_i;n_i} \prod_i e^{L(r_i)},\e$$
and we proved that the metric is diagonal in the primary fields
basis.

In the primary fields basis, $p=\lb r_1,r_2,\ldots,r_m\rb$, the equation
for the metric assumes the form,
$$\partial(g_p\partial g^{-1}_p)+\sum_i g_{p+e_i}g^{-1}_p-g_pg^{-1}_
{p-e_i}=0,\e$$
where $e_i=\lb 0,0,\ldots,1,0\ldots,0\rb$. Note that the combination
that appears in the equation is the multiplication by $V=x\mod\partial_i
V$, according to the Schubert calculus, eq. (2.13). This equation is solved
by the product of $\hat A_m$ Toda solutions in each of the variables.
One still needs to determine the boundary conditions.  Unfortunately,
the connections between the boundary conditions and the asymptotic behavior
of the
metric at $x \rightarrow \infty$ is not known in general\foot{
The problem can be analyzed, as usual, via an isomonodromy problem for a
linearized Lax pair representation of the equation, which is mathematically
straightforward, but was not done yet. Alternatively, the equation can be
solved numerically.}.
However, the
result of the calculation when done, would show that for $\lambda
\rightarrow \infty$ the metric assumes the form (in the fusion variety
basis),
$$g_{r,s} = N_{r,f_i}^s e^{-2|\lambda|M_{f_i}},\e$$
where the $f_i$ are the solitons of the theory,
$M_{f_i}$ are their masses, and $N^s_{r,f_i}$ are
the fusion coefficients.
Since the asymptotic behavior of the metric is dominated by the single soliton
transitions, refs. \r{\Vafa,\GS},
this shows that, indeed, the
$f_i$--soliton connects
the $r$ and $s$ vacua if and only if $N_{r,f_i}^s>0$, in accordance with
the precedingly described IRF picture.
\par
Alternatively, we can study the allowed
solitonic
transitions of the integrable field theories eq. (2.1) directly.
The allowed transitions for
this class of potentials were
conjectured in refs. \r{\Lerche,\Fendley,\Fendtwo}.
It was shown there that the Bogomolni lower bound
for the mass of the solitons interpolating between the $x_i$ and $x_j$
vacua is given by $|V(x_i)-V(x_j)|$, where $V$ is the superpotential.
Assuming that the solitons are chiral, implies that this bound is
saturated. In this case, we can calculate the
masses of the possible solitons in the theory. Specializing to
the case of $m=1$, $V={q^{m+1}\over m+1}-q$, we find that the classical
ground states are given by $q_r=e^{2\pi ir /m}$, and that the mass
of the soliton which interpolates the $r$ and $s$ vacua is
$$M_{rs}=|V(q_r)-V(q_s)|,\e$$
and we find for the potential $V$, in particular,
$M_{rs}={2k\over k+1}\sin\lb{\pi (r-s)\over k}\rb$.
This is precisely the mass spectrum of the $\hat A_m$ S-matrices
\REF\KS{R. Koberle and J.A. Swieca, Phys. Lett. 86B (1979) 209}
\r\KS,
and it follows that all such transitions are allowed, and that the $s$
soliton mediates between the $r$ and $r+s$ vacua. In other
words, the allowed transitions indeed correspond precisely to the fusion
rules of the
bosonic rational conformal field theory which has $V$ as a fusion
potential.

For the potential, eq. (1), with $m>1$, one can deduce the allowed
transitions by making the ansatz   that the solitonic masses in the
theory are identical to those that appear for $m=1$.
This implies that the vacua $\lb r_1,r_2,\ldots,r_m\rb$ and
$\lb n_1,n_2,\ldots,n_m\rb$ are connected, if and only if,
$r_i=n_i$ for all $i$, except for a unique $s$, for which $r_s \not=
n_s$, and then they
are connected by the $t=r_s-n_s$ soliton, whose mass is
$M_t=\sin(\pi t/k)$. When one draws the connectivity graph of the allowed vacua
and their connecting solitons, the conjectured polytopes of ref.
\r\Lerche\ are found, now, reinterpreted as the fusion rules of the
corresponding RCFT
(for more detail, see section (8), which treats such graphs, in general).

We wish to examine whether this  solitonic structure
is consistent with the fact that these are fusion theories.
Straightforwardly, we find that the lowest soliton, $t=1$, implies
the transition,
$$p=\lb r_1,r_2,\ldots,r_m\rb\rarrow \sum_s \lb r_1,r_2,\ldots,r_s+1,
\ldots,r_m\rb.\e$$
which is identical to a fusion by the fundamental field of the
rational conformal field theory, $x=q_1+q_2+\ldots+q_m$ using the Pieri formula,
eq. (2.13). In other
words, the vacua $p$ and $q$ are connected by the fundamental
soliton if and only if $q$ appears in the operator product of $x$
and $p$. This is, precisely, the afore discussed fusion structure of the IRF
theories, and is consistent with eq. (10).
Indeed the same fusion rules, again, describe
the theory, and its solitonic structure. The higher solitons are mediated
by the fields in the theory, which are
of the form $x_r=q_1^r+q_2^r+\ldots+q_m^r$.
In other words, the $p$ and $q$ vacua are connected by the $t$ soliton
if and only if $q$ appears in the product of $p$ and $x_t$. Again this
is a fusion structure, albeit, for $r>1$ the field $x_r$ is not a
primary field, but a combination of several ones.  We conclude that the
general framework drawn in this paper is, indeed, verified for
the class of potentials, eq. (2.1). This completes our voyage through
the integrable $N=2$ supersymmetric theories which have been constructed
explicitly to date\foot{
There are a number of sporadic examples
not covered here.  These are the $D-E$ perturbations found
in \REF\Ltwo{W. Lerche and N.P. Warner, Caltech preprint CALT--68--1747
(1991)}
\r\Ltwo, where the $A$ perturbation is the usual Chebishev $SU(2)_k$ one.
The RCFT correspondence, drawn here,
works perfectly for the $D$ and $E$ cases, as well.
It can be shown that all the A-D-E
LG potentials are the fusion potentials of some rational conformal theories
The $S$ matrix of each of the RCFT is,
essentially, the eigenvalue matrix
of the corresponding Cartan matrix. This is consistent with eq. (2.4) and
the observation made
in ref. \r\Ltwo\ regarding the locations of the vacua of the LG theories.
Further, the $S$ matrices of the soliton scattering
amplitudes are given by the extension of the braiding matrices, as
is discussed, in general, in sections (7-10).}.
\mysec{New Rational Conformal Field Theories}
Our purpose in this section is to introduce a novel class of rational
conformal field theories, closely related to current algebras, and
in particular new type of coset conformal field theories.
These theories will serve us as a framework
in which one can further address the questions raised previously.

Consider the following system of fusion rules:
$$A_1 A_r=A_{r-1}+A_{r+1}\qquad {\rm for\ } 1\leq r\leq N,\e$$
with the convention that $A_0=1$ and $A_{N+1}=A_N$.
The rest of the products
are computed by associativity. The case of $N=1$ corresponds to the fusion
rules of $(G_2)_1$. It can be checked that this is a consistent set of fusion
rules, where every field has a unique conjugate, which is itself, (i.e.,
the theory is real),
 and the fusion rules are fully symmetric
when one lowers the index with the conjugation bilinear form.  Thus, eq.
(1) represents a consistent conformal fusion ring.

Next, we wish to find the table of representations of the fusion algebra
(which are all one dimensional), as these correspond to the $S$ matrix.
Note the similarity of these fusion rules to those of the $SU(2)$
current algebra; the only modification is in the
last condition $A_{N+1}=A_N$ instead of $A_N=0$. Thus, again, we
represent the primary fields as polynomials in $A_1=2\cos\phi$ where $\phi$
is some angle. The polynomials are the same Chebishev polynomials of the
second kind, which appear for $SU(2)$, since the recursion relation, eq.
(1) is identical to the SU$(2)$ one \r\DGCMP,
$$A_r(2\cos\phi)={\sin(r+1)\phi\over\sin\phi}.\e$$
The only additional relation is $A_{N+1}=A_N$, which implies
$$\sin\lb(N+2)\phi\rb=\sin\lb(N+1)\phi\rb,\e$$
and is solved to give the points of the fusion variety
$$\phi={\pi(2r-1)\over 2N+3},\qquad {\rm where\ } r=1,2,\ldots,N+1,\e$$
where we excluded the solution $\phi=0$, since it leads to the
vanishing of the denominator, $\sin\phi$.
Upon a reordering of the primary fields,
$$m\rarrow 1+(-1)^m m/2 \,\mod \,(2N+3),\e$$
we find that the $S$
matrix of the theory, which, up to a constant, is a table
of the representations, is
$$S_{i,j}={2\over\sqrt{2N+3}} \sin\left[{\pi(2i-1)(2j-1)\over
2N+3}\right],\e$$
where $1\leq i,j\leq N+1$.
Evidently, $S$ is a real symmetric matrix, obeying $S^2=SS^\dagger=1$,
and thus is physically acceptable.
We find the matrix $T$ from the consistency relation $(ST)^3=1$,
$$T_{i,j}=e^{2\pi i(\Delta_i-c/24)}\delta_{i,j},\e$$
where
$$\Delta_i={i(i-2)\over 2N+3},\e$$
is the dimension of the $i$th primary field. The central charge
of the theory is
$$c={6N+3\over 2N+3}-(-1)^N.\e$$
The dimensions are found modulo an integer, while the central charge,
$c$, is found modulo $8$.
For $N=1$, we find $c=14/5$ and $\Delta=2/5$, agreeing with the well
known values for the $(G_2)_1$ current algebra. $N=0$ implies
that $c$ and $\Delta$ vanish,
the well known values for a theory with only one primary field
\r\Verlinde.

Since each of the fields is a Chebishev polynomial in the generating
primary field $A_1$, it follows that the fusion ring is described by
$P\lb x\rb/(p(x))$, where $p(x)$ generates the ideal
of relations in the ring, and where $x=A_1$.
The potential of the theory, $V$, is related to $p$ by $p=\d V/\d x$,
and is
$$V(x)=\sum_{i=1}^2 {q_i^{N+2}\over N+2}-{q_i^{N+1}\over N+1},\e$$
where $q_1q_2=1$ and $x=q_1+q_2$.

The potential $V(x)$, eq. (10), is, like all fusion potentials generated by one
conformal field, a massive perturbation of the $N=2$ minimal conformal
models. More precisely, if one takes the superpotential of an
$N=2$ LG theory to be $W(x)=\lambda V(x\lambda^{-1\over N+2})$, where
$\lambda$ is the coupling constant, then for $\lambda=1$, the
superpotential is identical to the fusion potential, whereas, for
$\lambda=0$, it reduces to the homogeneous part, $W\propto x^{N+2}$,
which is the superpotential of the $N$th minimal model.

The fusion potential can be written as a
difference of Chebishev polynomials of the first kind,
$$V(x)={T_{N+2}(x)\over N+2}-{T_{N+1}(x)\over N+1},\e$$
where $T_n(2\cos\phi)=2\cos(n\phi)$. The Chebishev polynomials
are $T_0=2$, $T_1=x$ and obey the recursion relation $x T_n=T_{n+1}+
T_{n-1}$, from which they can be readily calculated.
For $N=1$ the fusion potential assumes the form $V(x)={x^3\over3}-
{x^2\over2}-x+1$. This potential describes a massive perturbation of
the $k=1$ minimal model. By a shift of $x$, $x\rarrow x+c$ the potential
is seen to describe the most relevant perturbation $V=x^3-\lambda x$,
which is equivalent to a particular sine-Gordon theory,
a well known integrable model.
For $N=2$, we find the potential $V_2={x^4\over4}-
{x^3\over3}-x^2+x+{1\over2}$, which is a perturbation of the second
minimal model, and is a theory of one scalar field, along with one fermion
field. Similarly, one can calculate the potentials for higher $N$.
The fusion relation is ${\d V\over\d x}=0$. For
$N=1$, it is $x^2=x+1$, the well known fusion rule of $(G_2)_1$.

The class of theories described above affords a generalization to all
$SU(N)$, for any $N\geq2$. To define the theories, it is convenient to
start directly with the current algebra $SU(N)_k$, where $k$ and $N$ are
mutually strange numbers, $(k,N)=1$.
Recall the external automorphisms of SU$(N)$, used in section (2).
It is well known
\REF\GepFuchs{D. Gepner and J. Fuchs, Nucl. Phys. B294 (1987) 30}
\REF\Gep{D. Gepner, Nucl. Phys. B290 (1987) 10}
\r{\GepFuchs,\Gep,\Geplong} that these are external
automorphisms of the fusion ring of $G_k$, for any $G$ and $k$,
$$\lambda\sigma(\mu)=\sigma(\lambda)\mu=\sigma(\lambda,\mu),\e$$
for any integrable highest weights $\lambda$ and $\mu$ and for any external
automorphism $\sigma$. For $SU(N)_k$ where $(k,N)=1$ all the external
automorphisms act without fixed points, i.e., $\sigma(\lambda)=\lambda$
implies $\sigma=1$. $SU(N)_k$ is the only group which admits
no fixed points for a generic $k$.
The level one simply laced groups are the only other such examples.
Since $\sigma$ is an automorphism of the fusion rules,
eq. (12), we may consistently identify  fields
which are related
by the action of $\sigma$, and define the new fusion ring $R/\sigma$.
Namely,
we define $\lb\lambda\rb$ to be
the equivalence class of the weight $\lambda$ modulo the action of
$\sigma$. We further define the fusion coefficient of such classes by
$$\lb \lambda\rb \times \lb \rho\rb=\sum_{\mu \mod \sigma}
N_{\lambda,\rho}^\mu \lb \mu\rb,\e$$
where $N_{\lambda,\rho}^\mu$ is the fusion coefficient of $SU(N)_k$.
This definition is consistent, since $\sigma$ acts as an external
automorphism, eq. (12). Further, the resulting fusion rules are
acceptable for a conformal field theory. To establish this, note
that for $(k,N)=1$, the $N$-ality of $\sigma(\lambda)$ and $\lambda$
differ by one. Thus, every orbit under $\sigma$ contains exactly
one  element of $N$-ality zero (color singlet) which may be chosen
as a representative, denoted by $\hat\lambda$.
For example, for $SU(2)_2$
there are two orbits represented by $1$ (singlet) and $8$ (adjoint).
Thus, the new fusion rules may be thought of as a restriction of the
usual fusion rules to the $N$-ality singlet sector. It follows,
that they represent a fully consistent operator
product algebra, where the conjugate of each field is the usual one,
$C_{\lambda,\rho}=\delta_{\lambda,\bar\rho}$. Using $C$ to lower
the index $\rho$, $N_{\rho\lambda,\mu} = C_{\rho\theta}
N_{\lambda\mu}^\theta$, we indeed get a fully
symmetric three point function $N_{\rho\lambda\mu}$.
We conclude that these
are acceptable fusion rules.

The $S$ matrix of the theory is given (up to a factor) by the usual $S$
matrix of $SU(N)$, eq. (2.26), restricted to the singlet sector,
$$S_{\lambda,\mu}=\sqrt{N} S_{\lambda,\mu}^{SU(N)}.\e$$
This $S$ matrix is unitary, $SS^\dagger=1$, obeys $S^2=C$, and
diagonalizes the fusion rules eq. (13). To see this, we use the relation
$$S_{\lambda,\sigma(\rho)}=e^{-2\pi i\Lambda_\sigma \lambda}
  S_{\lambda,\rho}\e,$$
which is shown by a direct calculation on the $S$ matrix ref.
\REF\GepCosets{D. Gepner, Phys. Lett. 222B (1989) 207}
\r{\GepCosets}. To prove, for example, the relation with the fusion
rules we may compute \r\Verlinde
$$ \sum_t {S_{t,a} S_{t,b} S_{t,c}^\dagger\over S_{t,1}}=
  \sum_{t,\sigma} {S_{\sigma(t),a}S_{\sigma(t),b} S_{\sigma(t),c}^
\dagger\over S_{\sigma(t),1}}=N_{a,b}^c,\e$$
proving that the restricted $S$ matrix eq. (14) diagonalizes the
restricted fusion rules. Similarly, one shows that $S^2=C$ and $SS^
\dagger=1$.

The central charge of the theories is
$$c={k(N^2-1)\over k+N}-\Theta(k,N) \mod 8,\e$$
where $\Theta(k,N)$ is an integer defined by the Gauss-type sum,
$$e^{2\pi i\Theta(k,N)/8}= {1\over\sqrt N}
\sum_{n=0}^{N-1} e^{\pi i k n(N-n)/N}.\e$$
The sum, eq. (18), can be computed through the modular properties of the
Dedekind's eta function\REF\Imbimbo{U. Imbimbo, INFN Genoa preprint,
GEF-TH-5-1992 (1992)}, and is
\r\Imbimbo
$$\Theta(k,N)=8S(k,2N)-4S(k,N),\e$$
where $S(k,N)$ denotes the Dedekind's symbol,
$$S(k,N)=\sum_{n=1}^{N-1} \lb n/N-1/2\rb \lb nk/N-1/2\rb,\e$$
and where $\lb x\rb$ denotes the fractional part of $x$,
$$\lb x\rb=x \mod\,1, {\qquad \rm and \ } -1/2< \lb x\rb\leq 1/2.\e$$
The dimensions of the fields are identical to that of $SU(N)_k$,
restricted to the singlet sector,
$$\Delta_\lambda={\lambda(\lambda+2\rho)\over 2(k+g)},\e$$
where $\lambda$ is the highest weight, $\rho$ is half the sum of
positive roots and $g=N$ is the dual Coxeter number. The matrix $T$ is
$$T_{\lambda,\mu}=e^{2\pi i(\Delta_\lambda-c/24)} \delta_{\lambda,\mu},\e$$
and obeys the required relation, $(ST)^3=C$.

To prove the above relations, and also for instructive purposes,
it is convenient to describe this class of theories as quotient
theories. For $(k,N)=1$ the action of the external automorphisms is
faithful (without fixed points), and thus every weight $\lambda$ can
be written as $\lambda=\sigma^i(\hat\lambda)$ where $\sigma$ is the
generating external automorphism (associated to the first fundamental
weight, $\lambda_1$), and where $\hat\lambda$ is a singlet under the
center ($0$ N-ality).
Using eq. (15), the $S$ matrix for $SU(N)_k$, can be written as
$$S_{\sigma^n(\hat\lambda),\sigma^m(\hat\mu)}=S_{\hat\lambda,\hat\mu}
S^k_{n,m},\e$$
where $S^k_{n,m}$ is the matrix
$$ S^k_{n,m}=e^{2\pi i knm/N}.\e$$
The matrix $S^k$, eq. (25), corresponds to the $S$ matrix of a conformal
field theory with the fusion rules of $SU(N)_1$, but which
is a $k$th power of the usual $S$ matrix. Similarly, the dimensions of the
fields are
$$\Delta^k_n={kn(N-n)\over 2N} \,\mod 1.\e$$
The central charge can be computed from the consistency
relation $(ST)^3=C$, where $C=\delta_{n+m}$, and we find,
$$c=\Theta(k,N) \mod 8,\e$$
where $\Theta(k,N)$ is given by eq. (19). Clearly, for $k=1$, we find
$c=N-1$ which is the central charge of $SU(N)_1$. The fusion relations
of the theory are independent of $k$, $\lb n \rb \lb m\rb=\lb n+m\rb$.
The matrix $T$ is, as usual,
$$T^k_n=e^{2\pi i \lb kn(N-n)/N-c/24\rb}.\e$$
It can be checked that all the usual consistency relations hold.

For the matrix $T$ of $SU(N)_k$, we find the same tensor decomposition
as for $S$,
$$T_{\sigma^n(\hat\lambda)}=T_{\hat\lambda} T^k_{n}.\e$$
Thus, owing to this tensor product structure, it follows immediately
that $(ST)^3=C$, for the restricted theory, as well as giving an
alternative proof for all the other relations. In view of it, we may
think of the restricted theory as the quotient (`new coset') model
$SU(N)_k/SU(N)_{1/k}$, as the center theory is formally $SU(N)$ at the
rational level $1/k$.  (See section (5) for more details.)

Let us explore a few examples of the resulting fusion rings of the
quotient theories $SU(N)_k/SU(N)_{1/k}$. For $k=1$ the theories are
trivial, and indeed eqs. (17,22) imply, $c=\Delta=0$.
Consider $SU(3)_2$. The only nontrivial
product involving the singlet representations is $\lb 8\rb\times
\lb 8\rb=\lb 1\rb+\lb 8\rb$.
Denoting $x=\lb 8\rb $, this becomes the fusion relation $x^2=1+x$,
which is the fusion relation of $(G_2)_1$.
We conclude that
$${SU(3)_2\over SU(3)_{1/2}}\approx (G_2)_{-1},\qquad
{SU(2)_3\over SU(2)_{1/3}}\approx
(G_2)_1.\e$$
(For the explanation of the notation $(G_2)_{-1}$ see the next section.)

The theories described in this section represent new rational
conformal field theories, and the problem of finding realizations for
them is addressed in section (6).

\mysec{General new cosets}
\par
The class of theories described here affords even a further generalization.
Consider the general fusion rules of the current algebra of a group $G$
at the level $k$. From these fusion rules we get 
an $S$ matrix via Verlinde's formula \r\Verlinde.
However, the relation is highly ambiguous and a large number
of different theories have the exact same fusion rules, but different $S$
and $T$ matrices. Let $p$ be any integer which is strange to $d(k+g)$,
$(p,d(k+g))=1$, where
$g$ is the dual Coxeter number, and $d$ is the index of $M^*$ in $M$,
that is
$d$ is the least integer such that $d M^*\subset M$.  (For $SU(N)$ $d=g=N$.)
The general $S$ matrix corresponding
to the $G_k$ fusion rules is
$$S_{\lambda,\mu}^{(p)}=\left\vert {M^*\over (k+g) M}\right\vert^{1/2}
\sum_{w\in W} (-1)^w e^{-2\pi ipw(\lambda+\rho)(\mu+\rho)/
(k+g)}.\e$$
Note that $S_{\lambda,\mu}^{(1)}$ is the usual $S$ matrix
of $G_k$, eq. (2.26), while other
values of $p$ correspond to new $S$ matrices. The following relation
ensures that the resulting fusion rules are independent of $p$
$$S^{(p)}_{\lambda,\mu}=s(\mu) S^{(1)}_{\lambda,h(\mu)},\e$$
where $h(\mu)$ is some permutation of the integrable highest weights at
level $k$, which depends on $p$, and $s(\mu)=\pm1$ is some sign factor.
By examining the expression for $S$,
eq. (1), we find that $h$ is defined (uniquely) by
$$h(\mu)+\rho=pw(\mu+\rho)\, \mod \,(k+g)M,\e$$
where $w$ is some Weyl element, $s(\mu)=(-1)^w$, and the fact that this
relation holds is guaranteed by $(p,d(k+g))=1$. It follows that the
fusion rules are
$$N_{a,b}^{c (p)}=\sum_t {S_{a,h(t)} S_{b,h(t)} S_{c,h(t)}^\dagger\over
S_{0,h(t)}}=
\sum_t {S_{a,t}S_{b,t}S_{c,t}^\dagger\over S_{0,t}}=N_{a,b}^{c (1)},\e$$
and thus do not depend on $p$.
It can be shown that the dimension formula for the primary fields assumes
the form,
$$\Delta_\lambda={p[(\lambda+\rho)^2-\rho^2]\over2(k+g)}\ \mod1.\e$$
and the central charge of the theory is
$$c={pkD\over k+g}+\Theta(p,k,G)\ \mod 8,\e$$
where $D$ is the dimension of the algebra $G$, and
where $\Theta(p,k,G)$ is an integer defined by
$$e^{\pi i \Theta(p,k,G)/4}=\left\vert {M^*\over kM}\right\vert^{-1/2}
\sum_{\lambda\in M^*\mod kM} e^{-\pi i p\lambda^2/k}
.\e$$
Note that this relation holds for any even lattice,
not necessarily one which is a root lattice of some group,
and generalizes to any lattice the gauss
sums encountered earlier, eq. (4.18), which correspond to $SU(N)_1$.

A straight forward calculation shows that the usual relations for the
modular group hold, $(ST)^3=S^2=C$, where $C=\delta_{\lambda,\bar\mu}$
is the charge conjugation matrix.

The calculation of the generalized gauss sum, eq. (7),
(and the proof of this formula) proceeds
by observing that this sum is given by specializing the theta function
$$\Theta_M(\tau)=\sum_{\lambda\in M^*} e^{\pi i\lambda^2\tau},\e$$
to the value of $\tau=-p/q+i\epsilon$, where $\epsilon$ is an infinitesimally small parameter.
Up to an irrelevant real factor,
$$e^{\pi i\Theta(p,k,G)/4}=\Theta_M(-{p\over q}+i\epsilon).\e$$
The latter expression may be evaluated using various theta functions
identities, relating it to the multiplier system of the $\eta$
function, which is given, in turn, by the Dedekind symbols, eq. (4.20).
For example, in the case of $G=SU(3)$ we find, $\Theta(p,k,G)=
\Theta(p,1,G)=8S(p,6)-4S(p,3)$. The RCFT SU$(2)_{p/q}$ was previously
discussed in ref. \r\Imbimbo.

Let us denote the conformal field theories defined above as $G_{k/p}$,
as formally, at least, they correspond to taking the rational level
$k/p$ in the definition of the $S$ and $T$ matrices\foot{We are slightly
cavalier, in this notation, about the shift of $k \rightarrow k + g$,
between the bosonic and affine cases.  To be precise, $G_{k/p}$ stands
for $G_{(k+g)/p}$, in an obvious abuse of notation.}.

Although for different values of $p$ the theories have the same
fusion rules,
they are, in fact, different conformal field theories. Such examples
are found even within standard conformal field theory, e.g., $(E_7)_1\approx
SU(2)_{-1}$, $(E_6)_1\approx SU(3)_{-1}$. However,
it is a difficult problem to find realizations for the generic conformal
field theory $G_{k/p}$.\foot{A possible solution for this stems from the
general
treatment of the reconstruction `fantasy' problem, is described in the
sequel. The braiding matrix of $G_{k/p}$ can be shown to be equal to
those of $G_k$ with the simple replacement of $k+g$ by $(k+g)/p$.  Thus,
as described in sections (7,8,9), a corresponding IRF lattice model can be
constructed.  A multi-critical point of this lattice model would then
realize the full RCFT $G_{k/p}$.}

The current algebra--type theories described above, admit a simplification,
as bosonic (or $U(1)^n$) theories at fractional levels. A  realization
for the latter theories can be easily contemplated. Consider a theory
of $l$ free bosons propagating on the $l$-torus specified by the lattice
$\sqrt{k} M$, where $k$ is the level, and $M$ is any even lattice of rank
$l$. The vertex operators, which describe the primary fields in the theory,
are, $V_p=\exp(i\vec p\vec \phi/\sqrt k)$, where $\vec\phi$ is a canonical
free boson, and $\vec p$ is the momenta, which takes its values in
$p\in M^*\mod kM$. The partition function for the block with a given
value of $p$ is expressed as a theta function in the
lattice,
$$\Theta_p(\tau)=\sum_{\lambda=p \mod kM}  e^{\pi i\lambda^2\tau/k},\e$$
from which one derives the modular transformations,
$$S_{p,q}=e^{-2\pi i pq/k},\e$$
and
$$T_{p,q}=e^{-\pi i l/12}e^{\pi i p^2/k}\delta_{p,q}.\e$$
These transformations are compatible with the relations $S^2=(ST)^3=C$
where $C_{p,q}=\delta_{p,-q}$, as well as with the fusion rule
$$V_p\times V_q=V_{p+q}.\e$$

Now, the point is, that all of the above results remain valid
if we define $k$ to be any rational number, which is not necessarily
an integer,
i.e., $k=a/b$ where $a$ and $b$ are two strange integers.
Substituting this value in the equations for $S$ and $T$ we find
$$S_{p,q}=e^{-2\pi i a pq/b},\e$$
and
$$T_{p,q}=e^{-\pi i c/12} e^{\pi i a p^2/b}\delta_{p,q}.\e$$
where the central charge is defined by the same Gauss sum encountered
earlier, eq. (9),
$$c=\Theta(a,b,M)\mod 8.\e$$
The realization of these bosonic theories, denoted by $M_{a/b}$,
is treated in the next section.
\mysec{Realizations}
\par
In this section we would like to address the problem of the realization of
the conformal data systems described
earlier, i.e., the new `bosonic' systems, the affine theories at fractional
levels, and the new quotient type theories, as full fledged rational
conformal field theories.
As we shall show, these systems are interrelated, and it will be
established  that given a realization for the `bosonic' (thereafter
referred to as
group algebras) systems, the other theories may be realized as well.

Recall the construction of the group algebra systems. One starts from an
arbitrary even lattice, $M$. $M$ is a sublattice of the dual lattice, $M^*$.
The primary fields are labeled by the momenta $\vec p$, which are elements of
$M^* \mod kM$, where $k$ is some integral level.
The fusion rules assume the usual form, $[\vec p]\times [\pr{{\vec p}}]=
[\vec p+\pr{{\vec p}}]$.
Note that the fusion rules form the group algebra based on the
abelian group ${M^*\over kM}$.

The $S$ matrix assumes the form,
$$S^{(q)}_{p,\pr p}={1\over\sqrt N} e^{-2\pi i q p\pr p/k},\e$$
where $q$ is an arbitrary integer strange to $|M^*/kM|$, which
labels the different conformal systems, denoted by $(q,M)$.
For $q=1$ we recover the usual bosonic systems based
on the lattice $M$. The conformal dimensions and central charges of the
group algebra systems are given by eqs. (4.26, 4.27). For notational
simplicity, we
assume that $k=1$, without any loss of generality,
by redefining the lattice $M$,
$M\rarrow \sqrt{k}M$. Thus, we shall, henceforth implicitly assume that $k=1$,
unless otherwise specified.

The key to realizing the group algebras systems is, that while for
$q=1$ they correspond to free bosons propagating on the lattice $M$,
for $q\neq1$ they corresponds to free bosons propagating on some other
lattice $M_q$, whose rank is given by the integral central charge of the
theory. In other words, we have the equivalence of conformal systems
$$(q,M)\approx (1,M_q).\e$$
For example, $(-1,SU(2))\approx (1,E_7)$ and $(-1,SU(3))\approx
(1,E_6)$, where we refer to the root lattice by the name of the group.
To realize the theory, we construct the appropriate lattice
$M_q$, for each of the pairs $(q,M)$. The lattice $M_q$ needs to have
the following properties,
$${M_q\over M_q^*}\approx {M\over M^*},\e$$
by a group isomorphism denoted by $\phi$, $\phi(\lambda)\in M_q$ for any
$\lambda$ in $M$ (modulo the actions of $M_q$ and $M$).
This insures the correct set of primary fields and their fusion rules.
Further, it remains only to demand the dimension formula,
$$\phi(\lambda)^2=q\lambda^2\mod 2Z,\qquad {\rm for\ all\ } \lambda\in M_q.\e$$
Provided that the lattice $M_q$ obeys the properties eqs. (3,4), it realizes
the conformal system $(q,M)$ described above, as a usual bosonic theory
on this new lattice. So, for example, $SU(2)_{-1}$ is realized by seven
free bosons propagating on the maximal torus of the group $E_7$.

Note, that we have the equivalence of realizations $(q,M)\approx
(qm^2,M)$ for any integer $m$ (strange to $|M^*\mod M|$). This follows
from the map of primary fields $\phi(x)=mx\mod\, M$, where
the strangeness of
$m$ ensures that $\phi$ is an isomorphism. Consequently, we simply rename the
primary fields on the same lattice $M$.
In addition,
if $r$ is the index of $M$, i.e., $r$ is the least positive integer such that
$rM^*\i M$, then $q$ is defined modulo $r$, since the shift $q\rarrow q+r$
does not change the $S$ matrix, eq. (1). It follows that $q$ is defined modulo
the index $r$,  up to a multiplication by a square of an arbitrary integer. In
particular,
for $r$ which is a prime number, this implies that $q=\pm1$ are the
only distinct possibilities.

Now, it is not easy to find, for each such pair $(q,M)$, a lattice $M_q$
obeying eqs. (3,4). However, the abelian nature of the theory, i.e., the
fact that all the conformal blocks are one dimensional,
very strongly suggests that the theory is a free bosonic one, and
so the lattice $M_q$ exists.
The construction of each such lattice resembles a well
paved mathematical path, not unlike, for example, the construction
of the Leech lattice. In
particular, using the modular properties of the characters on this lattice,
and using the properties of the principal congruence subgroups of the
modular group, the characters of the lattice $M_q$ may be fully computed.
These, in turn, encode the number of vectors on the lattice of any given
length.
Together with the fusion rules, this provides enough information
to determine the lattice $M_q$.
In what follows, we shall assume that the lattice $M_q$ indeed exists for
any pair $(q,M)$, and leave systematic determination of such lattices
to further work.

Let us describe now the realizations of the new quotient conformal
field theories. Consider the theory
$$U_k\approx SU(N)_k\times SU(N)_{1/q},\e$$
where $SU(N)_{1/q}$ is realized as explained previously. The primary fields
in the theory are labeled by the pairs $(\lambda_1,\lambda_2)$, where
$\lambda_1$ is an integrable  highest weight of $SU(N)_k$, and
$\lambda_2$ is a highest weight of $SU(N)_1$. Alternatively, we may describe
the fields of the theory $\Phi^{\lambda_1}_{\lambda_2}$ as the products
$$\Phi^{\lambda_1}_{\lambda_2}=G^{\lambda_1} H^{\lambda_2},\e$$
where $G$ and $H$ are the corresponding primary fields of $SU(N)_k$ and
$SU(N)_{1/q}$ (which are labeled in the same way as those of $SU(N)_1$).

In this theory, consider the diagonal external automorphisms
$\wig\sigma=(\sigma_1,\sigma_1^\prime)$, where $\sigma_1$ and $\sigma_1^\prime$ are  the
generator of the external automorphisms of the two $SU(N)_k$ and
$SU(N)_1$. The automorphism $\wig\sigma$ and its powers generate a
$Z_N$ group of external automorphisms, which may be described as
the operator product with respect to the primary field \r\GepCosets,
$$J_r=G^{k\Lambda^r} H^{\Lambda^r},\e$$
where $\Lambda^r$ is the $r$th fundamental weight, and $G^\lambda$ stands
for the highest weight field $\lambda$, along with the weight $\lambda$.
Similarly $H^\lambda$ stands for the highest weight field of $H$.
More precisely, we have the operator product relation,
$$J_1(z_1) \Phi^{\lambda_1}_{\lambda_2}(z_2)=\Phi^{\sigma_1(\lambda_1)}_
{\sigma_1^\prime (\lambda_2)}(z_1-z_2)^d +{\rm h.o.t.},\e$$
where the h.o.t. are in the same conformal block, and
$d=\Delta(\sigma_1(\lambda_1))+\Delta(\sigma_1^\prime (\lambda_2)-\Delta(\lambda_1)
-\Delta(\lambda_2)$.
The dimensions of the field $J_1$ (up to an integer) is
$$\Delta_1={k(N-1)\over2N}+{q(N-1)\over2N}={(k+q)(N-1)\over2N}\e$$
(using eqs. (4.22)). Now, note that we can select the value of $q=-k$
(where $q$ is defined modulo $N$, for odd $N$, and modulo $2N$, for
even $N$), in which case the dimensions $\Delta_r$ all become integers.
Thus, for $k=-q$ the fields $J_r$ are candidates for an extended current
algebra for the theory, which would include, in addition to $J_r$,
the affine current
algebras of $SU(N)_k$ and $SU(N)_1$. We shall denote by $\cal O$ the
infinite dimensional current algebra
generated by the moments of $J_r$, along with the affine Lie algebras.
To ensure that $\cal O$ is a symmetry of the theory, it
is enough to require
the locality of $J_r(z)$ with
respect to all the fields in the theory. From eq. (8), the field
$\Phi^{\lambda_1}_{\lambda_2}$ will be local with respect to all the
currents $J_r$ if, and only if,
$$c(\lambda_1)+c(\lambda_2)=0\mod N,\e$$
where $c(\lambda)=N\vec\lambda\vec\Lambda^1$ is the $N$-ality of $\lambda$.
Thus, we impose this condition on the fields of the theory,
and eliminate from the spectrum all the fields which do not obey eq.
(10). This results in an $\cal O$ symmetric theory.
The fields of the theory, $\Phi^{\lambda_1}
_{\lambda_2}$, where $c(\lambda_1)+c(\lambda_2)=0$, are organized into
irreducible highest weight representations of the algebra $\cal O$, i.e.,
the extended conformal blocks. Since $J_r$ acts on the field $\Phi^{\lambda_1}
_{\lambda_2}$ via the external automorphism $\wig\sigma_r$ eq. (8),
the primary fields correspond to the
pairs $(\lambda_1,\lambda_2)$, for which $c(\lambda_1)+c(\lambda_2)=0$,
modulo the action of $\wig\sigma$.
As $\gcd(k,N)=1$, the automorphism $\wig\sigma$ acts faithfully
on the pairs $(\lambda_1,\lambda_2)$, and from each conformal bock, we
may choose a unique representative by demanding that
$c(\lambda_1)=c(\lambda_2)=0$.
In other words, the conformal blocks of the theory are labeled by the
$N$-ality singlets of $SU(N)_k$, which is precisely what we found for the
new cosets defined in section (4). The fusion rules of theory $U_k$ are
given by
$$\Phi^{\lambda_1}_0 \times \Phi^{\lambda_2}_0 =\sum_{\lambda
} N^\lambda_{\lambda_1,\lambda_2} \Phi^\lambda_0,\e$$
where we used the conservation of $N$-ality, $c(\lambda)=c(\lambda_1)
+c(\lambda_2)$.
$N^\lambda_{\lambda_1,\lambda_2}$ are the fusion coefficients of
$SU(N)_k$, i.e., the fusion rules are simply those of $SU(N)$
restricted to the $N$-ality singlet sector.
Again, these are identical to the fusion rules of the new cosets
described earlier, eq. (4.13). Finally the dimensions and central charges are
computed
using those of each of the component theories, $SU(N)_k$ and $SU(N)_1$, and are
found to be
$$\Delta^{\lambda_1}_{\lambda_2}={\lambda_1(\lambda_1+\rho)\over k+N}+{
q\lambda_2 (\lambda_2+\rho)\over N+1}\ \mod 1,\e$$
$$c={k(N^2-1)\over k+N}-\Theta(k,N)\ \mod 8.\e$$
which, again, are the dimensions and central charges of the new cosets.
We conclude that the theories $U_k$ are full fledged conformal field
theories which realize the new quotient systems $SU(N)_k/SU(N)_{1/k}$.

It is straightforward, now, to deduce the character formula, or the one
loop partition functions, of the theory $U_k$, by adding up products of
characters from each of the ingredient conformal field theories,
$$\chi^{\lambda_1}_{\lambda_2}(\tau)=\ch(\Phi^{\lambda_1}_{\lambda_2})(\tau)=
\sum_{n\mod N} \chi_{(k)}^{\sigma^n(\lambda_1)}(\tau) \chi_{(-1/k)}^{\sigma^n(
\lambda_2)}(\tau),\e$$
where $\chi^\lambda_{(m)}(\tau)$ stands for the
\def\SU{\mathop{\widehat{\rm SU}}\nolimits}
$\SU(N)_m$ character of
the representation with the highest weight $\lambda$, where $\tau$ is
the toroidal modulo, and the characters
of $SU(N)_{(-1/k)}$ are taken as the characters of the lattice $M$ which
realizes them, as explained earlier.
The $0$-point function on the torus is then seen to be
$$Z(\tau)=\sum_{\lambda\atop c(\lambda)=0} \chi^\lambda_0(\tau)^*
\chi^\lambda_0
(\tau),\e$$
and is modular invariant, as is seen from the unitarity of the $S$ matrix,
eq. (4.14).

Other, non--diagonal invariants are obtained from any invariant of
$SU(N)_k$.
The partition function then
becomes
$$Z(\tau)=\sum_{\lambda,\bar\lambda\atop c(\lambda)=c(\bar\lambda)=0}
N_{\lambda,\bar\lambda} \chi_\lambda(\tau)
\chi_{\bar\lambda}(\tau)^*,\e$$
where $N_{\lambda,\bar\lambda}$ is any invariant of $SU(N)$ which
obeys $c(\lambda)=c(\bar\lambda)$. (For a prime $N$, all the $SU(N)$
invariants are of this form, up to conjugation of $\lambda$.)
This completes the classification of modular invariants for the theories,
as they are all of the form eq. (16).

Before proceeding, let us discuss in some detail the case where $N$ is a
prime number. As was mentioned earlier, the realizations of $SU(N)_{1/p}$ and
$SU(N)_{1/q}$ are the same conformal field theory, provided that $p=m^2 q
\mod sN$ where $s=2$, for an even $N$, $s=1$, for an odd $N$,
where  $m$ is an
arbitrary integer. For $N$ which is prime (such that $N>2$) it is well known, that there are
exactly two possibilities for each integer $p$ modulo $N$, $p=m^2 \mod N$
or $p=-m^2\mod N$ for some $m$ (since $Z_N$ is a finite field this follows
from the basic theorem of algebra on solutions of polynomial equations
over a finite  field, see, for example, \REF\Jacobson{N. Jacobson, ``Basic
Algebra'', vol. 4, W.H. Freeman, New--York (1985)}ref. \r\Jacobson).
More succinctly put, we may assume, that
$$q\rarrow \left({q\over N}\right)=\pm1,\e$$
where $\left({q\over N}\right)$ is the Legendre symbol defined as
$$\left({q\over N}\right)=\cases{+1& $q=m^2 \mod N$,\cr -1 &$q=-m^2\mod
N$.\cr}\e$$
The theory with $q=\pm m^2 \mod N$ is realized by making the substitution
of the fundmental weight $\lambda_{mr}$ instead of the fundamental weight
$\lambda_r$, in the theory $SU(N)_{\pm1}$. In this case the extended current
algebra is generated by the current $J_1=G^{\Lambda^1} H^{\Lambda^m}$ where
$H$ is the corresponding primary field of $SU(N)_{\pm1}$, which has the
dimension ($\mod Z$),
$$\Delta_1={k(N-1)\over2N}\pm {m(N-m)\over 2N},\e$$
which is an integer.

The theory with $q=1$ can be realized by the usual $SU(N)_1$ current
algebra, while the theory with $q=-1$ is realized by its complex conjugate
theory, which is a theory of $1-N\mod 8$ free bosons propagating on some
lattice
$M$, such that $M^*\times SU(N)_1^*$ contains an even self dual
lattice of the same rank\foot{
For $SU(N)_{-k}$ it is not hard to find realizations for any $N$ and $k$, using
the decomposition of $Nk$ fermions into color and flavor (see the footnote
after eq. (2.40)). It follows that $SU(N)_k\approx SU(k)_{-N} \times SU(kN)_1
$, with a particular modular invariant.}.
For example, for $SU(2)$ we may choose the root lattice of
$E_7$ and for $SU(3)$ we may choose the root lattice of $E_6$.
This utilizes the well known relations of conformal field theory,
$$E_8\approx SU(2)_1\times (E_7)_1\approx SU(3)_1\times (E_6)_1\e$$
(for a particular choice of modular invariants),
which shows that the modular transformations of $E_6$ ($E_7$) are the
complex conjugate of those of $SU(3)$ ($SU(2)$), since $E_8$ is modular
invariant (as its root lattice is even and self dual),
implying that indeed $SU(2)_{-1}\approx (E_7)_1$ and $SU(3)_{-1}\approx
(E_6)_1$.

Now, since $\Theta(\pm1,N)=\pm(N-1)\mod8$, it follows that the central charge
of $U_k$ for a prime $N$ is
$$c={k(N^2-1)\over k+N}- \left({k\over N}\right)(N-1),\e$$
where $\left({k\over N}\right)=\pm1$ is the Legendre symbol, eq. (18).
In the case of $SU(2)$ we find the two subfamilies of theories,
$SU(2)_k\times SU(2)_1$ for $k=-1\mod4$ and $c={3k\over k+2}+1$,
and $SU(2)_k\times (E_7)_1$ for $k=1\mod 4$ and $c={3k\over k+2}+7$.
In the case of $SU(3)$ the two sequences are,
$SU(3)_k\times SU(3)_1$ for $k=2\mod3$ with the central charge $c={8k\over k+3}
+2$, and the second sequence $SU(3)_k\times (E_6)_1$, for $k=1\mod3$,
with the central charge $c={8k\over k+3}+6$. All theories are defined
with the aforementioned extended algebras and the restriction on the fields.
For $k=1$, any $N$, the theories are always
trivial, $c=0$ mod 8, while the $k=2$ case of $SU(2)$ is
equivalent to $(G_2)_1$ and $SU(3)$ is equivalent to the conjugate
theory $(G_2)_{-1}$. Evidently, these are full fledged conformal field
theories,  related to current algebras.

Let us return now to the general case. Owing to the structure of the theory
(up to the restriction on the fields) as a product of current algebras,
$$\Phi^{\lambda_1}_{\lambda_2}(z)=G^{\lambda_1}(z) H^{\lambda_2}(z),\e$$
where $G^{\lambda_1}(z)$ is the corresponding field in $SU(N)_k$, and
similarly $H$ is the appropriate bosonic operator, the correlation functions
on the sphere are immediately calculated to be,
$$\langle\prod_i \Phi^{\lambda_1^i,\nu^i}_{\lambda_2^i,\mu^i}(z_i)\rangle=
\langle \prod_i G^{\lambda^i_1}_{\nu^i}(z_i)\rangle \langle
\prod_i H^{\lambda_2^i}_{\mu^{i}}(z_i)\rangle,\e$$
where we have reinserted the `magnetic' quantum numbers $\mu$ and $\nu$
which are weights in the representations of the corresponding {\bf affine}
Lie algebras $SU(N)_k$ and $SU(N)_1$, and which label the different fields
in the corresponding conformal blocks\foot{Actually, we need to put in the
right moving part, which depend on $\bar z$, for the full description of
the theory. According to eq. (15), the torus partition function,
the right moving $\lambda$'s are the same as the left moving ones, whereas the
right moving $\mu$'s and $\nu$'s are arbitrary. Alternatively, eq.
(23)
may be interpreted in the language of conformal blocks (see sect. (7)).
Since $H$ has only one conformal block per channel, this equation holds
for each conformal block separately.  See eq. (36) for the description
of the general situation.}.
Since the $N$-ality is conserved in each
of the component theories, there is an exact correspondence of the conformal
blocks in the intermediate channels, and no `wrong' cross products, which
ensures this relation on the sphere. Note, that for higher genus this relation
is no longer valid, as we have separate fields flowing in the loops, and
in particular, the character formula eq. (14), which corresponds
to the toroidal 0-point function, is not of this form.

As the foregoing discussion above was slightly abstract, it is useful to
illustrate
one example in full detail. Consider, then, the theory $(G_2)_1\approx {SU(2)_2
\times SU(2)_1}$ (modulo the extended algebra). This example is closely related
to the quotient relation $SU(2)_1={(G_2)_1\over SU(2)_3}$,
which is a usual coset
theory (where $SU(2)_3$ is the algebra generated by the short simple root
of $G_2$). According to the previous discussion, the extended algebra $\cal O$
is generated by the dimension one currents: 1) $J_a(z)$ which are the
three currents of $SU(2)_3$. 2) $K_a(z)$ which are the currents of $SU(2)_1$.
(Here $a=3,+,-$, are the usual magnetic numbers.) 3) The fields $G_i^{3/2} H_j^
{1/2}$, where $3/2$ and $1/2$ are the isospin, and $i=\pm 3/2$, $\pm
1/2$,  and $j=\pm1/2$
are the magnetic numbers. Counting, we find $14$ currents of dimension
one, which is precisely the dimension of the Lie algebra $G_2$. To see that
these currents indeed obey the correct operator product relations for the
affine algebra $(\hat G_2)_1$, we may use a bosonization technique.
We express the $SU(2)_1$ fields as
$$\manyeq{
K_3 (z)&={i\over\sqrt2} \partial\rho (z), \qquad K_\pm (z) =
\n e^{\pm i\sqrt2 \rho(z)}\nn,\cr
H_{\pm 1/2}^{1/2}(z)&=\n e^{\pm {i\over\sqrt{2}} \rho(z)}\nn ,\cr}$$
and the $SU(2)_3$ fields as,
$$\manyeq{
J_3(z)&={i\sqrt3\over\sqrt2} \partial\phi (z) \qquad J_\pm
(z)=\n e^{\pm i{\sqrt2\over\sqrt3} \phi(z)}\nn \psi_{\pm2}^0(z)\cr
G_r^{3/2} (z)&=\n e^{ {ir\phi(z)\over \sqrt6}}\nn\psi_{2r}^3,\cr}$$
and so
$$G_r^{3/2}(z) H_{t}^{1/2} (z) = \psi_{2r}^3(z) \n
e^{{2ir\over\sqrt6}\phi(z)+{2it\over \sqrt2} \rho (z)}\nn,$$
where $\psi_{\pm 2}^0 (z)$, $\psi_{\pm 1}^3$ and $\psi_{\pm3}^3=1$ denote
the corresponding parafermionic fields (for explanation and notation see
refs. \REF\ZF{A.B. Zamolodchikov and V.A. Fateev, Sov Phys. Jetp 62 (1985) 215;
(Zh. Eksp. Theor. Fiz. 89 (1985) 380)}
\REF\GQ{D. Gepner and Z. Qiu, Nucl.
Phys. B285 (1987) 423}
\r{\ZF,\GQ}),  $\phi (z)$ and $\rho(z)$ are canonical free
bosons, and where `$\n$' stands for the usual normal ordering.  The fields
$\psi_2^0 = \psi_{-1}^3$ and $\psi_{-2}^0 = \psi_1^3$ correspond to the
critical three states Potts model, and have the dimension $\Delta =
{2\over 3}$, \r{\ZF,\GQ}.
The entire content of eqs.
(24,25) may be summarized in the relation
$$O(z)=S(z) e^{i\vec m\vec\theta(z)},\e$$
where $O(z)$ is any of the twelve currents not in the Cartan subalgebra
(which is generated by $J_3$ and $K_3$),
$\vec\theta(z)=(\phi(z),\rho(z))$ is a vector boson,
$\vec m$ is a root lattice of the Lie algebra $G_2$, and $S(z)$ is unit for a
long root, and is
a parafermion of the dimension ${2\over3}$,
for a short root. Recall from ref. \r\Gep\ that this is
precisely
the vertex operator construction of the current algebra $(G_2)_1$, and thus
we have established that ${\cal O}\approx (G_2)_1$. Finally, there are
exactly seven fields of dimension ${2\over5}$, the 4-plet
$\Phi_{1/2}^{3/2} = G^{3/2} H^{1/2}$ and the triplet $\Phi_0^1 = G^1$,
in the notation of eq. (22).
Using the same bosonization and vertex operator construction \r\Gep, it follows
that this is the representation $7$ of $(G_2)_1$. It is left as an exercise
to verify that these fields indeed give the correct bosonization of the 
$7$ representation of $(G_2)_1$, as described in ref. \r\Gep.

The entire foregoing construction of the new quotient theories can
be substantially
generalized. Consider the theory,
$$U\approx G_{k_1} \times G_{k_2} \times \ldots \times G_{k_r},\e$$
where $G$ is any Lie algebra, and the $k_i$ are the levels which are
taken to be arbitrary
rational numbers. For $k$ which is not a positive integer, $G_k$ stands
for the realizations of the affine theories at a fractional level, see
section (5).  The primary fields in the theory are
$$\Phi^{\lambda_1,\lambda_2,\ldots,\lambda_r}(z)
=\prod_i G^{\lambda_i}(z),\e$$
where $\lambda_i$ are highest integrable weights of the affine Lie algebra
$\hat G$ at the level $q_i$ where $k_i=q_i/p_i$ and $\gcd(p_i,q_i)=1$,
Now consider the primary field
$$J_\Lambda(z)=\prod_i G^{q_i \Lambda}(z),\e$$
where $\Lambda$ is a cominimal fundamental weight of $G$, i.e.,
$\Lambda\theta=1$, where $\theta$ is the highest root. The field $J_\Lambda$
is a product of primary fields in each of the current algebras.
According to the fusion rules, eq. (4.12), the field $J_\Lambda$ acts on the
primary fields of the theory as the external automorphism,
$$J_1\times \Phi^{\lambda_1,\lambda_2,\ldots,\lambda_r}=
\Phi^{\sigma(\lambda_1),\sigma(\lambda_2),\ldots, \sigma(\lambda_r)},\e$$
where $\sigma$ is the external automorphism of $\hat G$  associated
to the cominimal weight $\Lambda$. The dimension of the field $J_\Lambda$ is
computed from eq. (4.22),
$$\Delta_\Lambda={\Lambda^2\over2} \sum k_i,\e$$
and is an integer provided that $\sum k_i=d$, where $d$ is any integer
such that $d M^* \i M$.  We shall henceforth assume that this is the case.
Then, $J_\Lambda(z)$, for all the cominimal weights $\Lambda$, are taken
to be an extended currents of the theory.
The fusion
rules of $G$ imply that the currents $J_\Lambda$, where $\Lambda$
is any of the cominimal weights, generate a closed
operator product algebra, where the fusion rules form a group algebra,
with a group which is
isomorphic to the center of $G$. The operator products of the currents
$J_\Lambda$ with the rest of the fields
follow from the fusion rules eq. (4.12), and are
$$J_s(z_1) \Phi^{\lambda_1,\lambda_2,\ldots,\lambda_r}(z)=
\sum_d (z_1-z)^{\delta+d} O_d(z),\e$$
where $d$ is any integer, $O_d(z)$ is some field in the conformal block
$\Phi^{\sigma_s(\lambda_1),\sigma_s(\lambda_2),\ldots,\sigma_s(\lambda_r)}$,
and the dimension $\delta$ is
$$\delta=\sum_i \Delta(\lambda_i)-\Delta(\sigma(\lambda_i)),\e$$
as computed from eq. (4.22). It follows that the currents $J_\Lambda$ have
an abelian ($U(1)$) monodromy with respect to all the fields in the theory
and a well defined mutual semi--locality exponent which is equal to
$\delta$. In particular, we may consistently eliminate from the
theory all the fields  for which $\delta$ is not an integer,
keeping only fields which are local with respect to all the
currents $J_\Lambda$. The theory, so obtained, has an extended current algebra,
$\cal O$, which is generated by the affine currents of $G\times G\times\ldots
\times G$,
along with the extended currents, $J_\Lambda$. The blocks of the theory,
organized with respect to this extended current algebra, correspond to the
classes of admissible (i.e., integral $\delta$, eq. (33), for all the
cominimal weights) $r$-plets
$(\lambda_1,\lambda_2,\ldots,\lambda_r)$, modulo the action of
$\wig\sigma$. The characters of the theory are
$$\chi^{\lambda_1,\lambda_2,\ldots,\lambda_r}=
\sum_{\sigma\in C}\prod_i \chi^{\sigma(\lambda_i)}_i(\tau),\e$$
where $\chi^\lambda_i(\tau)$ stands for the character
of the representation $\lambda_i$ of the theory $G_{k_i}$, in
the appropriate realization. The modular invariant toroidal partition
function of the theory is
$$Z=\sum_{\lambda_i,\bar\lambda_i} \prod_i N_{\lambda_i,\bar\lambda_i}^{(i)}
\chi^{\lambda_1,\lambda_2,\ldots,\lambda_r}(\bar\tau)
\chi^{\bar\lambda_1,\bar\lambda_2,\ldots, \bar\lambda_r}(\tau)^*,\e$$
where the sum is over the admissible $r$-plets $(\lambda_1,
\lambda_2,\ldots,\lambda_r)$, (i.e., $\delta$ is an integer),
taking exactly one representative from each of the cosets
modulo the action of $\wig\sigma$. $N_{\lambda,\bar\lambda}^{(i)}$
are arbitrary modular invariants for each of the sub-theories $G_i$, such that
$c(\lambda_i)=c(\bar\lambda_i)$, where $c(\lambda)$ is the value of
$\lambda$ under the center of $G$, or $c(\lambda)=\lambda \mod M_l$,
where
$M_l$ is the long root lattice. The diagonal modular invariant
corresponds to
$N_{\lambda_i,\bar\lambda_i}^{(i)}=\delta_{\lambda_i,\bar\lambda_i}$,
when inserted into eq. (35).

We can still generalize this construction in a number of ways.

1) We may let $C$ stand for any subgroup of the center, taking only
   a subalgebra of the currents for the extended algebra.

2) We may choose different groups instead of $G$ in eq. (27),
and taking $C$
to be any subgroup of a center of the product group $\otimes_i G_i$,
which is $\oplus_i C_i$, where $C_i$ is the center of $G_i$. In short,
   $C\subset \oplus C_i$ can be any subgroup. For the extended currents
we take $J=\prod_i J_{\Lambda_i}$, where $\Lambda_i$ is a cominimal weight
of $G_i$, and the $r$-plets $(\Lambda_1,\Lambda_2,\ldots,\Lambda_r)$
go over all elements of $C$.

3) Note that there are no limitations on the levels $k_i$, eq. (27),
which may be any rational numbers.
In particular, the construction carries through, even, when
there are fixed points in the action of $\tilde\sigma$.

The construction above applies to the cases (1--3), mutatis mutandis,
in a straightforward manner. It is left as an exercise for the reader
to verify this. Albeit, there is one difference when two or more
of the sub--theories are non--bosonic. This is due to the fact that
there are more than one conformal blocks in each of the sub--theories,
and eq. (23) needs to be modified to,
\def\calF{{\cal F}}
$$\calF^\Phi_p(z_1,z_2,\ldots,z_r)=\sum_{p_i} d^p_{p_1,p_2,\ldots,p_r}
\prod_i
\calF^i_{p_i}(z_1,z_2,\ldots,z_r),\e$$
where $\calF^i_p(z_1,z_2,\ldots,z_r)$ is the $p$th conformal block of the
correlation function $\langle \prod_s G_i^{\lambda_s}(z_s)\rangle$
in the $G_i$ theory, and $d^p_{p_1,p_2,\ldots,p_r}$ are some coefficients.

We may actually take any of the $G_i$ to be a free boson theory, or a vector
boson $\vec\phi$ propagating on the lattice $M$ at the level $k$.
The primary fields, $\lambda$, are given by the elements of $M^*/kM$.
The
external automorphisms of the bosonic theory are shifts on the lattice,
$\sigma_a(\lambda)=\lambda+a$, where $\lambda$ is any primary field, and
$a$ is any element of $M^*/kM$. Clearly, the external automorphisms form
the group $M^*/kM$, which may be thought of as the center. For the product
theory,
$\otimes_i G^i_{k_i}\times B$, where $B$ stands for the bosonic theory,
we find the center group $C=\oplus_i C_i\oplus {M\over k M^*}$. We may take
any subgroup, $H$, of this center, $H\i C$. The current associated to the
element $h=(h_1,h_2,\ldots,h_r,\vec a)\in H$, where $h_i$ is a cominimal weight
of $G_i$, and $\vec a$ is some shift in $M^*/kM$,
is $J_h=\prod_i J_{h_i} \exp(i\vec a\vec\phi
/\sqrt k)$, where $J_{h_i}$ are the primary fields $G_i^{q_ih_i}$ associated
to the the cominimal weight $h_i$. The fact that $H$ is a subgroup, implies
that the currents $J_h$, $h\in H$, form a closed operator product algebra.
The rest of the discussion follows as in the non--abelian case, with the
appropriate obvious modification of the mutual semi--locality
exponent $\delta$, eq. (32). Note that the conformal field theories described
in sect. (2) are precisely $SU(m)_k\times U(1)_{m(m+k)}$, where the $U(1)$
piece  corresponds to one free boson propagating on the
root lattice of $SU(2)$ at the level $m(k+m)$. The automorphism $\wig\sigma$
is taken, in this case, to be $\wig\sigma=(\sigma_1,\sigma_{k+m})$,
where $\sigma_1$ is the fundamental weight automorphism of $SU(m)$, and
$\sigma_{k+m}$ is the shift by $k+m$ on the lattice. Note, that $\wig\sigma$
generates a $Z_m$ group $H$, which is a subgroup of the center group, $C
\approx Z_m\times Z_{m(k+m)}$, i.e., $H\i G$. The discussion
in sect. (2) may be
considered as an example of the above general construction.

The realizations presented here describe all the known
rational conformal systems (except, perhaps, some rational bosonic orbifolds)
along with many new ones.
In particular, the usual coset models
$G/H$
\REF\GKO{P. Goddard, A. Kent and D. Olive, Phys. Lett. 152B (1986) 105;
Comm. Math. Phys. 103 (1986) 105}
\REF\KMQ{D. Kastor, E. Martinec and Z. Qiu, Phys. Lett. 200B (1988) 434}
\REF\BNY{J. Bagger, D. Nemeschansky and S. Yankielowicz, Phys. Rev. Lett.
60 (1988) 389}
\REF\Douglas{M. Douglas, Caltech preprint, CALT-68-1453 (1987)}
\r{\ZF-\Douglas,\Gep}
can be described as $G_k\times H_{-k}$ along the lines above.
Similarly, theories corresponding to the quotient theories
$G/C$ where $C$ is a subgroup of the center, are described by the
case of $r=1$.
It is intriguing to note that the form, eq. (27), is particularly suitable
for a description of the theories as Chern-Simons models, which are  based
on the groups $G_i$ and at the levels $k_i$. This is in line with
the ideas of ref. \r\Imbimbo\ which describes $SU(2)_k$, $k$ rational
from a Chern--Simons approach, and with those of
\REF\Gaw{G. Felder, K. Gawedzki and A. Kupiainen, Comm. Math. Phys. 117 (1988)
          127}
\REF\Zoo{G. Moore and N. Seiberg, Phys. Lett. 220B (1989) 422}
refs. \r{\Gaw,\Zoo}\
which attempt to elucidate cosets models in this language.
\def\calF{{\cal F}}
\def\pr#1{{#1^\prime}}
\def\bra{\langle}
\def\ket{\rangle}
\def\cm#1#2#3#4#5#6{{C_{#1\,#2} \,\left[\matrix{#3&#4\cr#5&#6\cr}\right]}}
\mysec{Fusion IRF}
In this section we will give the general formulae for the Boltzmann
weights of solvable interaction round the face (IRF) lattice models,
which will be constructed out of the data of the general rational conformal
field theory (RCFT) and in particular the braiding relations on the sphere.
(For reviews on braiding relations in RCFT, see, for example,
\REF\Frohlich{J. Frohlich, In Como 1987 proceedings, Differential
geometry methods in theoretical physics 219--259; B. Schroer, same volume,
219--259; J. Frohlich, Cargese Summer Inst. 1987, 71--100; L. Alvarez--Gaum\'e,
same volume, 1--11}
\REF\LGrev{L.A. Gaum\'e, G. Sierra and C. Gomez, In Physics and mathematics
of strings, Knizhnik memorial volume, pp. 16--184}
\REF\Mrev{G. Moore and N. Seiberg, Trieste spring school 1989, p.1}
refs. \r{\Frohlich,\LGrev,\Mrev}.)
More precisely, as we will show, at the extreme UV limit, the IRF
models Boltzmann weights give a representation of the braid group,
which coincides with the conformal braiding algebra.
This is precisely
the conformal limit of integrable massive soliton systems. In particular,
the kink operators exchange algebra reduces to the chiral vertex operators
exchange algebra.
Thus, we are able
to recover the full $S$ matrix or Boltzmann weight for any rapidity,
making use of the algebraic properties of the braiding matrices, i.e.,
the n-CB algebra defined in the sequel, eq. (9.2).

The general fusion IRF model is defined as follows. Consider a square lattice
(fig. 7.1) whose vertices are labeled by the primary fields of some rational
conformal field theory. Further, restrict the admissible configurations
by the condition that $N^a_{xb}>0$, where $x$ is some fixed fundamental
primary field, $N^a_{xb}$ is the fusion coefficient, and where $a$ and $b$
label the two vertices on the same link. The partition function of the theory
is
then defined as
\def\bw#1#2#3#4#5{{w\left(\matrix{#1&#2\cr#3&#4\cr}\bigg\vert #5\right)}}
$$Z=\sum_{\rm configurations} \prod_{\rm faces} \bw a b c d u,\e$$
where $\bw a b c d u$ is the face Boltzmann weight, which vanishes unless
the state primary fields $a,b,c$ and $d$, which are defined on the vertices,
(fig. 7.1), obey the admissibility condition.
Here, $u$ is the spectral parameter, which  labels a continuous
parameter family of IRF models (it is not to be confused with temperature
and other relevant parameters, which are implicit in $w$).
Alternatively, we shall use the notation
$$\bw a b c d u =\dw a b d c.\e$$
The key to the  solvability of the models is in the fact
that the transfer matrices commute for different values of the spectral
parameter $u$. The mathematical condition ensuring this is the star triangle
equation (STE) \r\Book,
$$\twoline{\sum_c \bw g c a b u \bw c  e b  d  {u+v} \bw g f c e v =}{
\sum_c\bw a c b d v
\bw g f a c  {u+v} \bw f e c d  u,}$$
which is depicted in fig. (7.2). There are three types of solutions
to the  STE relation, eq. (3), elliptic, trigonometric and constant.
We shall concentrate here mainly on the trigonometric case which is the
one relevant for soliton systems. The different types of solutions are
specializations of each other, expressing different critical limits,
but the converse (i.e., how to construct an
elliptic solution from a trigonometric one and a trigonometric
solution from a constant one) is
not generally obvious. Actually, for subsequent use, it is convenient to
introduce a slightly more general notion of a fusion IRF model, where
for the horizontal  links, we allow
one primary field $h$ (fixed) and for the vertical links another fixed
primary field, $v$.  The admissibility condition then becomes,
$$N_{hb}^a N_{vd}^b N_{hd}^c N_{vc}^a > 0, \e$$
for the face defined in eq. (2).  Every pair of primary fields, $v$
and $h$, will denote a different IRF model
based on the same RCFT. The STE relation stays essentially the same,
having changed
only the admissibility condition\foot{See eq. (10.7) for the explicit
form of the modified STE.}. In the sequel, we shall exhibit solutions for the
STE for all possible pairs, $h$ and $v$.

It is more convenient to rewrite the STE in an index free notation,
using face transfer matrices, fig. (7.3). The states on the diagonal,
fig. (7.3), are denoted by $\vert a_1,a_2,\ldots,a_n \ket$, where the $a_i$
are the heights, i.e., different primary fields. The face transfer matrix, $X_i
(u)$, is defined through its matrix elements
$$\bra a_1,a_2,\ldots,a_i,\ldots,a_n\vert X_i(u) \vert
a_1,a_2,\ldots,\pr a_i,\ldots,a_n\ket=\bw {a_{i-1}} {\pr a_i} {a_i} {a_{i+1}}
u.\e$$
The STE is equivalent to the fact that the
face transfer matrices, $X_i(u)$, obey the Young Baxter (matrix) equation
(YBE),
$$ X_i(u) X_{i+1}(u+v) X_i(v)= X_{i+1}(v) X_i(u+v) X_{i+1}(u).\e$$
$$X_i(u) X_j(v)=X_j(v) X_i(u),\qquad{\rm for \ \vert i-j\vert\geq2}.\e$$
Now, we may consider as a special limit of eqs. (6,7) the case where
all  the spectral parameters go to infinity (henceforth, termed the
ultra--violet (UV) limit),
$u,v\rarrow i\infty$. In this case, the YBE assumes the form,
$$X_i X_{i+1} X_i=X_{i+1} X_i X_{i+1},\e$$
where $X_i$ is the limit,
$$X_i=\lim_{u\rarrow i\infty} f(u) X_i(u),\e$$
and where $f(u)$ is some irrelevant function ensuring a finite limit.
Another limit which gives the braid
relation is the infra-red (IR) limit, $u=v=0$. We
will assume that $X_i(0)\propto 1_i$, the trivial solution, the regularity
condition. As will
be discussed in section (10), this condition
expresses the fact that the
corresponding soliton system has no scattering at zero rapidity.
The relations eq. (8), along with eq. (9), form the so called braid group.
Denote by $\sigma_i$ the braiding of the $i$ and $i+1$ strands, fig.
(7.4).
It is well known that the braiding generators $\sigma_i$ obey the
relations
$$\eqalign{\sigma_i
\sigma_{i+1}\sigma_i&=\sigma_{i+1}\sigma_i\sigma_{i+1},\cr
\sigma_i\sigma_j&=\sigma_j\sigma_i,\qquad{\rm for\ } \vert
i-j\vert\geq2,\cr}\e$$
and that there are no other relations (i.e., these are the defining relations
for the group). We conclude that the $X_i (u)$ form a representation of the
braid group at the extreme UV and IR limits, eqs. (8,9). On the sphere ($S^2$)
the braid group enjoys the additional relation
$$\sigma_1\sigma_2\ldots\sigma_{n-1}\sigma_{n}^2\sigma_{n-1}\ldots\sigma_2
\sigma_1=1,\e$$
which is depicted in fig. (7.5), and is obvious by undoing the braid around
the north pole. As we shall see, the $X_i$ form a projective representation
of the braid group on the sphere.

Let us turn now to rational conformal field theory. Let $\psi_i(z,\bar z)$
denote some primary fields. Consider the correlation function,
$$\langle\prod_i \psi_i(z_i,\bar z_i)\rangle=\sum_{p,q} D_{p,q}
\calF_p(z_1,z_2,\ldots,z_n)^* \calF_q(z_1,z_2,\ldots,z_n),\e$$
where $p$ and $q$ label the different conformal blocks, and $D$ is some
matrix of coefficients, which counts the different blocks appearing in the
correlation functions. $\calF_q(z_1,z_2,\ldots,z_n)$ are holomorphic
function (more precisely, holomorphic sections of the
conformal blocks vector bundle) in each of the $z_i$, and
can be depicted pictorially by tree diagrams where the external legs are
the $\psi_i$, and different blocks correspond to different intermediate fields.
Different presentations of the blocks must give the same physical correlation
function, which implies that the different presentations are the same up to
a linear transformation on the blocks. In particular, the four point blocks,
$\calF_p(z_1,z_2,z_3,z_4)$, are labeled by an intermediate primary field $p$
such that $N^i_{pk}=N^j_{pl}=1$, where $N$ is the fusion coefficient,
fig. (7.6).
(In case that $N=N^p_{lj}N_{pki}\geq2$, the block
is labeled by the pair $p,a$ where $1\leq a\leq N$.)

It follows that
$$\calF_p(z_1,z_2,z_3,z_4)=\sum_{\pr p} C_{p\pr p}\left[\matrix{j&k\cr i&l\cr}
\right] \calF_{\pr p} (z_1,z_3,z_2,z_4).\e$$
where $i,j,k$ and $l$ label the four primary fields in the correlation
function,
and $C$ is the conformal braiding matrix, fig. (7.7). The braiding matrix
$C$ obeys the hexagon identity,
\REF\MS{G. Moore and N. Seiberg, Phys. Lett. 212B (1988) 451}
$$\twoline{\sum_c \cm i c a j b p\cm p d a k c l \cm c {\pr c} j k b d=}{
\sum_{\pr p} \cm p {\pr p} j k i l \cm i {\pr i} a k b {\pr p} \cm
{\pr p} d a j {\pr c} l.}$$
An index free notation for the braiding operation is obtained with the
aid of chiral vertex operators \r\MS. Define the operator $\Phi^j_{ik}(z)$,
where $j$, $i$ and $k$ label the different primary fields, to be the map of
Hilbert spaces $\Phi^j_{ik}: \calH_k\rarrow\calH_i$ defined by
$$\Phi^j_{ik}(z)=\langle b\vert a(z)\vert c\rangle,\e$$
where $a$, $b$ and $c$ are any fields in the respective conformal blocks
$i$, $j$ and $k$, (denoted as ${\cal H}_i$, etc.). Here we assume that $N^i_{jk}
=1$. (If $N^i_{jk}>1$, we add, as usual, a label for the different
couplings, which ranges to $N^i_{jk}$.)
Graphically, the chiral vertex operator is depicted in fig. (7.8). The matrix
$C$
can now be interpreted as the braiding matrix of two chiral vertex operators,
$$\Phi_{s_1}(z)\Phi_{s_2}(w)=\sum_{s_3,s_4} R_{s_1,s_2,s_3,s_4}
\Phi_{s_3}(w) \Phi_{s_4}(z),\e$$
where $s_a$ stands for the triplet of primary fields, $s_a=\left({j_a\atop
i_a\,k_a}\right)$, and the $R$ matrix is defined by
$$R_{s_1s_2s_3s_4}=\delta_{i_1i_3}\delta_{k_2k_3}\delta_{j_2j_3}\delta_{k_1i_2}
\delta_{k_3i_4} \delta_{j_4 k_4}\cm {k_1} {k_3} {j_1} {j_2} {i_1} {k_2},\e$$
and where the delta functions ensure the proper sewing of the intermediate
Hilbert spaces.

In passing, note the remarkable resemblance of eq. (16) with the
Fadeev--Zamolodchikov kink operators exchange algebra,
$$K_a(u) K_b(0)=\sum_{ c d} S_{a,b,c,d}(u)  K_d(0) K_c(u),\e$$
where $a,b,c$ and $d$ label different kinks,
$u$ is the rapidity, and $S_{a,b,c,d}(u)$
is the scattering matrix
for the kink process $a+b\rarrow c+d$ (see section (10)). In fact, we will
establish in the sequel that the two formulations are very closely related,
and that the conformal braiding matrix can be thought of as the UV limit of the
kink
$S$ matrix, eq. (18).\foot{
Eq. (18) thus suggests that the kink operators
(which are the composite fields creating the kinks out of the vacuum)
smoothly extrapolate the chiral vertex operators, at the extreme UV
(i.e., critical) limit. This is particularly notable, since the kink
operators, and their correlation functions, is a relatively ill understood
subject, whereas in the RCFT this is very well understood. We hope that this
observation can be used to understand better this slightly abstruse
problem.}

Now, the associativity of the conformal braiding algebra, eq. (16), implies
the YBE,
$$R_{12}R_{23}R_{12}=R_{23}R_{12}R_{23},\e$$
where $R_{i,j}$ acts on the $i,j$ Hilbert spaces in the triplet $V\times V
\times V$, and is an identity in $k$. We changed the definition of $R$ into operator notation:
the matrix element of $R_{12}$ between $s_1$ and  $s_2$ to $r_1$ and  $r_2$ is
$R_{s_1s_2r_1r_2}$.
The Young Baxter equation,
eq. (19) is equivalent to the hexagon identity, eq. (14). It follows, that the
$R$ matrix forms a representation of the braid group, eq. (8), where
we identify $R_{i,i+1}$ with the operation $\sigma_{i}$ of braiding the
$i$ and $i+1$ strands.

Let us return now to the IRF lattice models. At the limit of $u\rarrow i\infty$
the face transfer matrices $X_i(u)$ form a representation of the braid group.
Similarly, the $R$ matrix form a representation of the same braid group.
It is only natural to identify the two representations. More precisely,
consider the most general IRF Boltzmann weight,
$$w_{s_1,s_2}\left(\matrix{a&b\cr c&d\cr}\bigg\vert u\right),\e$$
which is the weight attached to the face with $a$, $b$, $c$ and $d$ at the vertices,
where $a$, $b$, $c$ and $d$ are some primary fields, and $s_1$ and $s_2$ are
arbitrary primary fields which label the model, fig. (7.1).
The admissibility condition reads, $N^{a}_{s_1,b}=N^{b}_{s_2,d}=
N^{d}_{s_1,c}= N^{c}_{s_2,a}=1$.

Now, we identify  for $u\rarrow i\infty$ the Boltzmann weight $w$, eq.
(20), with the conformal braiding matrix $R$ (up to an irrelevant factor needed to ensure a finite limit),
$$\lim_{u\rarrow i\infty} w_{s_1,s_2}\left(\matrix{a&b\cr c&d\cr}\bigg\vert u
\right)=R_{t_1,t_2,t_3,t_4},\e$$
\def \vop#1#2#3{{\left({#1\atop #2\,#3}\right)}}where
$t_1=\vop {s_1} a b$, $t_2=\vop {s_2} b c$, $t_3=\vop {s_1} d a$, and
$t_4=\vop {s_2} c d$. We used all the six delta functions
in the definition of $R$, eq. (18), showing that
$w$ contains all the information
in the $R$ matrix, and is an alternative presentation for it.
Recall that the $R$ matrix vanishes unless $N^{i_a}_{j_a k_a}>0$.
The crux of
the identification, eq. (21), is
that, very importantly, and lo and behold, the Boltzmann weight
satisfies the precise
fusion admissibility conditions, eq. (4). Further, since the $R$ matrix
satisfies the YBE eq. (18), it follows that $w$ satisfies, for any
$s_1$ and $s_2$ the star-triangle relation eq. (3). It follows that $w$,
as defined, corresponds to a family of integrable
lattice models, which are based
on the fusion rules of any RCFT. We shall label the model as
${\rm IRF}({\cal O},h,v)$, where $\cal O$ is any RCFT, and $h$ and $v$ are
arbitrary primary fields which give the horizontal and vertical admissibility
conditions.

Next, we wish to exhibit trigonometric solutions of the fusion STE
relations eq. (3), which at the limit $u\rarrow i\infty$ coincide with the
Boltzmann weights described above. To do so, we shall first make the
simplifying
assumption that the link primary fields $h$ and $v$ are `fundamental',
i.e., the operator product of $h$ with $v$ contains exactly two
primary fields\foot{The one block case correspond to a
trivial $U(1)$
monodromy (which is a property of free boson theories),
in which case $X_i$
is one dimensional, and thus proportional to the trivial solution
$X_i=1_i$.}.
In terms of the fusion ring,
$$hv=\psi_1+\psi_2,\e$$
where $\psi_i$ are two primary fields. The case of more than two
primary fields is discussed in section (9).  In most of the following, for
notational simplicity, we shall assume that $h=v=x$, a fixed primary
field for both the horizontal and the vertical couplings. The arguments
below do not depend, however, on this assumption.
Denote by $\Delta_x$, $\Delta_1$
and $\Delta_2$ the dimensions of these fields. For example, in the  case
of $SU(N)_k$, the field  $x$ can stand for the fundamental ($N$)
representation.
The fact that $x^2=\psi_1+\psi_2$ implies that the four point function
$\langle i\, x\, x\, j \rangle$ has only two conformal blocks labeled
by the $\psi_i$, $i=1,2$. It follows that the $R$ matrix is
two dimensional, using the $N$ move, fig. (7.9).
{}From fig. (7.9), it is clear that the eigenvalues of $R$ are always given
by
$$\lambda_i=\epsilon_i e^{i\pi (\Delta_h+\Delta_v-\Delta_i)},\e$$
where $\epsilon_i=\pm1$ signify whether the coupling is symmetric
or anti--symmetric, and the equation holds for arbitrary $h$ and $v$
\REF\AG{L. Alvarez--Gaum\'e, C. Gomez and G. Sierra, Nucl.Phys. B330
(1990) 347}\r\AG.
It follows that the $R$ matrix obeys the fixed polynomial equation,
$$\prod_{p=1}^n (R_i-\lambda_p)=0,\e$$
where $\lambda_i$ are given by eq. (23).
In the two block case, eq. (21), $R_i$ obeys a quadratic equation,
eq. (24), which by choosing normalizations,
$$\sigma_i=R_i e^{-i\pi(\Delta_h + \Delta_v)} e^{i\pi (\Delta_1+\Delta_2)/2},$$
is set in the more standard form
$$\sigma_i^2=1 + (q^{-1} - q) \sigma_i,\e$$
The braid group relations, eq. (10) along with the quadratic relation,
eq. (25), are the defining relations for the Hecke algebra of type $A_N$.
The parameter $q$ depends on the RCFT and the particular fields $h$ and
$v$.  Introduce also the notation,
$$q=e^{2i \hat\lambda},\qquad {\rm and\ } l={\pi\over\hat\lambda},\e$$
where $\hat\lambda$ is termed the crossing parameter and $l$ which is
a rational number is termed the `level'.  Then, from eq. (23), (assuming
$\epsilon_1 \epsilon_2=- 1$, which is always the case),
$$\hat\lambda = {\pi (\Delta_1 - \Delta_2)\over 2}.\e$$

For example, in the $SU(N)_k$ we find $l=k+N$ in the $N$--$\bar N$
channel, and
for the $(r,s)$ minimal model
\REF\BPZ{A.A. Belavin, A.M. Polyakov and A.B. Zamolodchikov, Nucl. Phys.
B241 (1984) 333}\r\BPZ\
with the $(2,1)$ generator, we find
$l=s/r$, and with the $(1,2)$ generator $l=r/s$.

Following ref. \r\AG\ we can actually express $\Delta_x$ in
terms of the conformal
dimensions $\Delta_i$.  Using the braid relation on the sphere, eq.
(11), taking a trace, and using eq. (23), we find,
$$e^{i\pi(8\Delta_x-3\Delta_1-3\Delta_2)}=1,\e$$
where we used also the fact that the relation, eq. (11), has  the value
$e^{-4\pi i \Delta_x}$, i.e., $R$ forms a projective representation of
the sphere braid group,
which follows from the behavior of the four point function at infinity.

It is convenient to define the generators $U_i$,
$$U_i=q^{-\half}-\sigma_i,\e$$
which obey the alternative presentation of the
Hecke algebra,
$$\manyeq{U_i^2&=\beta U_i,\cr
          U_iU_j&=U_jU_i,\qquad{\rm if\ } \vert i-j\vert\geq 2,\cr
          U_iU_{i+1}U_i-U_i&=U_{i+1}U_iU_{i+1}-U_{i+1},}$$
where, $\beta=2\cos(\hat\lambda)$.

Define now the following face transfer matrix,
$$X_i(u)=\sin(\hat\lambda-u)\cdot 1_i+\sin u \cdot U_i,\e$$
where $1_i$ is the unit face transfer matrix.
Clearly $X_i(u)$ still represents a fusion IRF face transfer matrix, with
the correct admissibility conditions. Further, a direct calculation using
the Hecke algebra, eq. (30), readily shows that for all $u$, the face
transfer matrices obey
the Young--Baxter equation,
$$X_i(u)X_{i+1}(u+v)X_i(v)=X_{i+1}(v)X_i(u+v)X_{i+1}(u),\e$$
and
$$X_i(u)X_j(v)=X_j(v)X_i(u),\qquad{\rm for \ }\vert i-j\vert\geq2.\e$$
Thus the $X_i(u)$ form a trigonometric solution of the Young--Baxter
equation. Further, it is clear that $\sigma_i =
X_i=\lim_{u\rarrow i \infty} e^{iu} X_i(u)$,
and thus $X_i(u)$ indeed extrapolates the conformal braiding matrix. We find
also $X_i(0)\propto 1_i$, which is another standard requirement of IRF models.
(Connected, as mentioned, to the fact that the corresponding kink system should
have
no scattering at zero rapidity, see section (10).)
The resulting Boltzmann weights $\bw a b c d u$ obey the STE relation,
implying that the transfer matrices commute for different values of  $u$,
and thus the lattice theory is solvable.
This completes the construction of the trigonometric ${\rm IRF}({\cal O},x,x)$
models for any RCFT, $\cal O$, and for a fundamental field $x$.
Evidently, the entire discussion applies, as is, for the case of
different horizontal and vertical couplings, $h\neq v$, as long as the
number of fields in the product $hv$ is equal to two.
For more than two fields,
one can construct the trigonometric solutions in a number of ways.
The simplest, perhaps, is using the so called fusion procedure of IRF
models ref.
\REF\FIRF{P.P. Kulish, N.Yu Reshetikhin and E.K. Sklyanin, Lett. Math. Phys.
5 (1981) 393}\r\FIRF, (for applications of the fusion procedure, see e.g.,
\REF\Date{E. Date, M. Jimbo, T. Miwa and M. Okado, Lett. Math. Phys. 12
(1986) 209; Phys. rev. B35 (1986) 2907; Nucl. Phys. B290 [FS20] (1987) 231}
\r\Date),
which is equivalent
to the bootstraping of soliton
systems (for a review see, for example,
\REF\Klas{T.R. Klassen and E. Melzer, Nucl. Phys. B338 (1990) 485}\r\Klas),
or the fusion of $R$ matrices in RCFT. Alternatively, one
may proceed directly, by choosing the relevant ansatz. This is described
in section (9). The results of either of the methods are the same, and,
in any event,
the fundamental Boltzmann weights contain all the information regarding
the others, via the fusion procedure.

Another question is generalizing the trigonometric solutions to
full off critical, i.e., elliptic, solutions, which leads to a
solvable thermalization of the lattice models.
For the two block case,
we again expect this to be straightforward, essentially replacing
each of the sines with the appropriate theta function,
or
replacing $\sin u$ with $\Theta_1(u;p)$ in the notation of eq. (8.27),
where $p$ is a parameter which measures the distance from
criticality\foot
{The complication stems from the fact that some of the `$1$'s have to be
replaced by an appropriate $\Theta$ function, $\Theta(u;p)$, such that
$\Theta(u;0)=1$. I thank T. Miwa for a discussion on this.},
(see more in section (8)).
\def\pr#1{{\twidle#1}}
\def\bw#1#2#3#4#5{{w\left(\matrix{#1&#2\cr#3&#4\cr}\bigg\vert #5\right)}}
\mysec{Graph State IRF and RCFT}

Consider a general IRF model defined on a square lattice, fig. (7.1). Assume,
for simplicity, that the horizontal and vertical admissibility conditions
are the same, and that the states of the model are labeled by the elements
$a\in S$, where $S$ is some finite set. The admissibility condition can
be represented as a set of pairs $(a,b)$ such that $a$ is adjacent to $b$.
Alternatively, define the set $K\subset S\times S$ by
$$K=\left\{ (a,b)\vert b\sim a\right\},\e$$
where we used the notation $b\sim a$ to denote that $(b,a)$ is an admissible
pair.
Note that the order is important, and $b\sim a$ does not necessarily implies
that $a\sim b$. 
Now, the pair $(S,K)$ defines an oriented graph, with the possibility of
one\foot{Actually, in general the number of lines may be more then one.
For fusion IRF these denote the number of couplings $N_{na}^b$ which can
be any positive integer. Fig. (8.1,d) is such an example.}\
self loop on each vertex. The points of the graph are the elements
of $S$,
and the lines go from all $a$ to $b$ such that $(a,b)\in K$. We put an arrow
on the line to indicate the direction, unless both $(a,b)$ and $(b,a)$ are in
$K$, in which case we omit the arrow. We thus get a graphical representation
of the model.

Assume that the IRF is a fusion theory associated to some conformal
field theory, i.e., IRF$({\cal O},x,x)$, where $\cal O$ denote the RCFT,
and $x$ the primary field used to construct the model, as in section
(7). Recall, that the admissibility condition is, $a\sim b$ iff
the fusion coefficient obeys $N^b_{x,a}=1$. Then, according to the rules
above we find the oriented graph
$K=\left\{(a,b)\big\vert N^b_{x,a}=1\right\}$,
which will be termed the admissibility graph for the model IRF$({\cal O},x)$.
In fig. (8.1) examples of fusion admissibility graphs are depicted.
%
%
The graphs are convenient visualizations for the model.

Now, recall the defining relations of the Hecke algebra eq. (7.30),
$$\manyeq{H_i H_{i+1} H_i-H_i&=H_{i+1} H_i H_{i+1}-H_{i+1},\cr
          H_i H_j&=H_j H_i\qquad{\rm if\ } \vert i-j \vert\geq2,\cr
          H_i^2&=\beta H_i.}$$
Given any representation of the Hecke algebra, we can build an IRF model
as explained in the previous section.
The face transfer matrices $X_i(u)$ are
$$X_i(u)=\sin(\lambda-u)\cdot 1_i+\sin u \cdot H_i,\e$$
where the crossing parameter $\lambda$ obeys $\beta=2\cos\lambda$.
The $X_i(u)$ so defined satisfy the YBE, eqs. (7.6,7.7). We can define
a representation of the Hecke algebra, for any graph, by choosing
the ansatz,
$$\langle a_{i-1},a_i, a_{i+1}\vert H_i\vert a_{i-1},a_i^\prime, a_{i+1}\rangle
=\delta_{a_{i-1},a_{i+1}} \left[{\psi(a_i)\psi(a_i^\prime) \over \psi(a_{i-1})
\psi(a_{i+1})}\right]^{1\over2},\e$$
where the states on the right and left are admissible. Now, the first two
relations of the Hecke algebra, eq. (2), are easily verified for all
choices of $\psi(a)$. Further, it can be checked that
$$H_iH_{i\pm1}H_i-H_i=0,\e$$
which is a stronger form of the braiding relation. The algebra satisfied by
the $H_i$ is the Hecke algebra, with a quotient by the ideal
generated by eq. (5), and is
termed the Temperly--Lieb algebra.
The third relation, $H_i^2=\beta H_i$, is equivalent
to
$$\sum_{b\atop (b,a)\in K} \psi(b)=\beta\psi(a),\e$$
which is an eigenvalue problem for the connectivity matrix, where the
eigenvectors are $\psi(a)$, and the eigenvalue is $\beta$. Further,
we should require, due to eq. (6), that $\psi(a)\neq0$ for all $a$, which
limits the possible allowed eigenvectors. The IRF model, so obtained, will be
referred to as a graph--state IRF. Such models have been described
before, in particular, in association with the A-D-E Dynkin graphs
\REF\Pas{V. Pasquier, J. Phys 20 (1986) L217, L221}
\r{\Pas,\Zuber,\rev}.
Now assume that the model is
the fusion IRF$({\cal O},x)$. Then eq. (6) becomes,
$$\sum_b N^b_{x,a} \psi(a)=\beta \psi(b),\e$$
i.e., the $\psi$'s diagonalize the fusion rules with respect to $x$.
The solution to this eigenvalue problem is well known ref. \r{\DGCMP,
\Verlinde} being given in terms of the points of the fusion variety,
$$\psi_\gamma(a)={S_{a,\gamma}\over S_{0,\gamma}},\quad {\rm and} \quad
\beta={S_{x,\gamma}\over S_{0,\gamma}},\e$$
where $\gamma$ is any primary field, labeling the different solutions,
and $S$ is the conformal $S$ matrix.
The eigenvalue problem is then equivalent to the
well known relationship between the fusion rules and the $S$ matrix,
$$\sum_b N^b_{x,a}{S_{b,\gamma}\over S_{0,\gamma}}=
         {S_{x,\gamma}\over S_{0,\gamma}} {S_{a,\gamma}\over S_{0,\gamma}},\e$$
which simply means that the value of the primary fields on the
points of the fusion variety $[a](x_\gamma)=\psi_\gamma(a)$ represent the
fusion
algebra. To summarize, the IRF$({\cal O},x)$ graph--state model Boltzmann weights
are
$$w_\gamma\left(\matrix{a&b\cr c&d\cr}\bigg\vert u\right)=
  \delta_{b,c}\sin(\lambda-u)+\delta_{a,d} \sin u\left[{S_{b,\gamma}
S_{c,\gamma}
\over S_{a,\gamma}S_{d,\gamma}}\right]^{1\over2},\e$$
where $\gamma$ is any primary field, which labels the different graph--state
IRF
models associated to the fusion rules of ${\cal O}$, and the crossing parameter
$\lambda$ is
$$\beta=2\cos\lambda={S_{x,\gamma}\over S_{0,\gamma}}.\e$$
Note that, for the sake of consistency, we must choose a
primary field $\gamma$, such that $S_{a,\gamma}\neq0$, for all primary fields
$a$. Otherwise, there are no restrictions on $\gamma$\foot{Of course,
requiring a sensible probabilistic interpretation of the Boltzmann
weights implies that $\omega$, eq. (10), must be real and positive for
all $a,b,c$ and $d$.  This puts severe limitations on the allowed graphs and
choices of $\gamma$.  However, there is no reason to think that the
general, non-probabilistic theory, does not represent a sensible field
theory for any values of $\omega$.  A similar situation is encountered
in discretized models with topological terms, e.g., $CP^n$ sigma model
in two dimensions
with a topological term, $\theta \not= 0$.}.

The graph--state Boltzmann weights obey a number of important properties,
which are evident from their definition eq. (10),

\noindent Regularity:
$$\bw a b c d 0=\sqrt{\rho(0)}\delta_{b,c},\e$$
Time reversal:
$$\bw a b d c u=\bw c b d a u,\e$$
Crossing symmetry:
$$\bw a b d c {\lambda-u}=\left[{S_{b,\gamma} S_{d,\gamma}\over
S_{a,\gamma} S_{c,\gamma}}\right]^{1\over2} \bw b c a d u,\e$$
Unitarity:
$$\sum_e \bw a e d c u \bw a b e c {-u}=\rho(u)\delta_{b,d},\e$$
where $\rho(u)=F(u)F(-u)$ is some function (termed the unitarizing factor).
Using the face transfer matrices $X_i(u)$, the unitarity property
may be written as
$$X_i(u)X_i(-u)=F(u) \cdot 1_i.\e$$
The r.h.s. of eq. (16) becomes
$$X_i(u)X_i(-u)=
[\sin(\lambda-u)1_i+\sin u H_i] [\sin(\lambda+u) 1_i-\sin u H_i ],\e$$
where $H_i$ obeys the Hecke algebra, eq. (2). Using the Hecke relation,
$H_i^2=\beta H_i$ and $\beta=2\cos\lambda$, it follows immediately
that eq. (16) holds, with
$$\rho(u)=\sin(\lambda+u)\sin(\lambda-u),\e$$
and thus the IRF models based on any Hecke algebra, via
the ansatz eq. (7.31), are always unitary. In particular, all
the graph state fusion IRF models described in this section are unitary.

The graph--state construction gives us an easy way to build an IRF model
from any given rational conformal field theory. In general, however, it is not
a true fusion IRF, since the configurations appearing in $\bw a b c d u$
are not all the ones that are allowed by the admissibility condition,
but rather, only the ones for which $a=d$ or $b=c$. For special graphs,
i.e., trees with no self loops except at the end points,
it is clear that the admissibility
conditions indeed imply that $a=d$ or $b=c$. Thus also, for such theories,
the state--graph model
is the correct fusion IRF. Further, it can be shown, that up to irrelevant
factors, the general Hecke ansatz always reduces to a state graph model for
trees. This gives us an immediate way to express the conformal braiding
matrix in terms of the torus $S$ matrix in such theories, since both
give rise to same IRF model. Comparison of eqs. (7.30) and eq. (10) shows that,
up to an irrelevant overall factor,
$$R_{t_1,t_2,t_3,t_4}=q^{-1/2}\delta_{b,c}-
\left[ {S_{b,\gamma} S_{c,\gamma}\over
S_{a,\gamma} S_{d,\gamma}}\right]\delta_{a,d},\e$$
for some $\gamma$, up to irrelevant normalizations of the primary
states\foot{In section (9) we will derive a more general form of this
relation, eq. (9.20), and its special case, eq. (9.22),
based on the connection between
the crossing properties of $w$ and
the RCFT torus modular transformations, eq. (9.20).  Eq. (19) is the two
block case of
eq. (9.22), which holds for any RCFT.}.

At this point, the reader might wonder about the parameter $\gamma$,
which labels apparently different solutions of the STE for the
same fusion rules. From the perspective of the correspondence with the RCFT,
this might seem puzzling, since a given RCFT is supposed to give rise
to a unique IRF model. The resolution of this is that, indeed, different
choices of $\gamma$ correspond to different conformal field theories!
Recall from section (5) that for a given set of fusion rules, there are
a number of possibilities for consistent $S$ matrices, which give rise
to these fusion rules. The different $S$ matrices are related by
some permutation of the primary fields, $p$, such that $\hat S_{ab}=
N_b S_{a,p(b)}$, where $N_b$ are some non--vanishing normalizations.
The allowed permutations are limited by $S_{ab}=S_{ba}$ and
$S^2=C$, which, however, are fulfilled for various permutations. Now,
if we take $b=1$ (the unit primary field), we find that,
$\hat S_{a,1}=N_1 S_{a,\gamma}$, where $\gamma=p(1)$. Since
$\hat S_{a,1}\neq0$,
this must be so, also, for $S_{a,b}$. Assume that a permutation $p$
exists such that $\gamma=p(1)$ for all $\gamma$ such that $S_{a,\gamma}\neq0$
(and for all $a$), which obeys the RCFT properties listed above. Then
we may assume, without any loss of generality, that $\gamma=1$, the unit
primary field, by replacing $S\rarrow \hat S$ and $\gamma\rarrow 1$.
Then, also, we find a unique correspondence between the Boltzmann
weights and the RCFT. We shall, thus, in the sequel, assume without any
loss of generality, that $\gamma=1$.

Example (a) :
Consider the case of IRF$(SU(2)_k, [1])$, where $[1]$ stands for the
fundamental representation, fig. (8.1,a). As was discussed in section (5)
we may replace $SU(2)_k$ with $SU(2)_{k/p}$, for any integer $p$ such that
$\gcd(p,2(k+2))=1$. The $S$ matrix for $SU(2)_{k/p}$ is, using eq. (5.1)
$$S^p_{m,n}={2\over\sqrt{k+2}}\sin\left[{\pi p(m+1)(n+1)\over
k+2}\right],\e$$
where $m,n=0,1,2,\ldots,k$, are the (twice) isospins of the different
primary fields, and $S^1_{mn}$ is the usual $S$ matrix for $SU(2)_k$.
All the different $p$'s give the same fusion rules (c.f., section (5)).
We find,
$$S_{m,n}^p={\hat S}_{m,n},\e$$
where $\hat S$ is a modular matrix for another theory, with the same fusion rules.
The condition $\gcd(p,2(k+2))=1$ is precisely the one needed to ensure that
$S^p_{m,n}\neq0$. Eq. (21) then shows that, indeed, we can assume without any loss of
generality that $p=1$, by replacing $S^p$ with $\hat S$.

The state graph IRF$(SU(2)_{k/p},[1])$ Boltzmann weights are obtained by
substituting the $S$ matrix, eq. (20), into eq. (10),
$$\twoline{\bw a b c d u =\dw a b d c=}{
\sin(\lambda-u)\delta_{bc}+\delta_{ad}\sin u
{\sqrt{\sin[\pi p(b+1)/(k+2)]\sin[\pi p(c+1)/(k+2)]} \over
       \sin[\pi p(a+1)/(k+2)]},}$$
where $\lambda={\pi p\over k+2}$, and $a,b,c$ and $d$ obey the admissibility
conditions.
Explicitly the Boltzmann weights, for $p=1$, are,
$$\bw m {m\pm1} {m\pm1} {m\pm2} v=\dw {m} {m\pm1} {m\pm2} {m\pm1}=
{[1-v]\over [1]},\e$$
$$\bw m {m\pm1} {m\pm1} m v=\dw {m} {m\pm1} {m} {m\pm1}={[\pm v+m+1]
\over [m+1]},\e$$
$$\bw m {m\mp1} {m\pm1} m v=\dw {m} {{m\mp1}} {m} {m\pm1}=
\pm {[v]\over [1]} {\sqrt{[m+2][m]}\over [m+1] },\e$$
where we used the variable $v=(k+2)u/\pi$, the symbol
$[v]=\sin[\pi v/(k+2)]$,  the states are $j=0,1,2\ldots,k$,
Boltzmann weights where $j$ is outside of this range vanish,
and where we have divided all the Boltzmann
weights by $\sin[{\pi \over k+2}]$. The model for $p\neq1$ is
obtained by replacing
$k+2$ with $(k+2)/p$ in the definitions of $v$ and $[v]$, and no other
changes.
The Boltzmann weights eqs. (23-25) are identical to the
ones originally given in ref.
\REF\ABF{G.E. Andrews, R.J. Baxter and P.J. Forrester, J. Stat. Phys. 35
(1984) 193}\r\ABF\
for $p=1$, and
form the so called restricted height model. We conclude that the restricted
height model is a graph state fusion IRF associated to $SU(2)_k$. Similarly,
other values of $p$ give rise to generalizations of the model. Note,
however that for $p>1$ some of the Boltzmann weights become complex, and
thus no longer allow for a probability interpretation. The partition function
remains,
however, real and positive, and thus the models appear to represent
bona--fide field theories. 

Example (b): Consider the conformal field theory $SU(2)_{(2k+1)/p}/
SU(2)_{p/(2k+1)}$
described in sects. (4,5,6). Again, take  the field $x]$ to be the
fundamental
representation. The graph of the theory is depicted in fig. (8.1,b).
Substituting the $S$ matrix, eq. (4.6), into the general graph--state
representation, eq. (10), we find
$$S_{m,n}={2\over\sqrt{2k+1}} \sin\left[{\pi p(m+1)(n+1)\over 2k+1}\right],\e$$
where $m,n=0,2,\ldots,2k$. The Boltzmann weights are identical to those
of $SU(2)_{2k+1}$, with the identification $n\equiv 2k+1-n$, and no other
changes.

Example (c): The hard hexagon model on a square lattice (ref.
\REF\Baxter{R.J. Baxter, J. Phys. A 13 (1980) L61} \r\Baxter).
The states of the classical hard hexagon model are labeled by the
variables $0$ and $1$. The Boltzmann weights are expressed in terms of
the elliptic function $\theta_1(u;p)$,
$$\theta_1(u;p)=2p^{1/4}\sin u\prod_{n=1}^\infty [1-2p^{2n}\cos(2u)
+p^{4n}](1-p^{2n}),
\e$$
where $p$ is a parameter which measures the distance from criticality.
The Boltzmann weights of the integrable hard hexagon model are (ref.
\r\Baxter)
$$\manyeq{
w_1&=\dw 0 0 0 0={\theta_1(3\lambda-u)\over\theta_1(3\lambda)},\qquad
w_2=\dw 1 0 0 0=\dw 0 0 1 0 ={\theta_1(\lambda-u)\over\theta_1(\lambda)},\cr
w_3&=\dw 0 0 0 1=\dw 0 1 0 0 ={\theta_1(u)\over\sqrt{\theta_1(\lambda)
\theta_1(2\lambda)}},\cr
w_4&=\dw 1 0 1 0 ={\theta_1(4\lambda-u)
\over \theta_1(4\lambda)},\qquad w_5=\dw 0 1 0 1={\theta_1(2\lambda-u)\over
\theta_1(2\lambda)},}$$
where we used the short hand notation, $\theta_1(u;p)\equiv\theta_1(u)$,
and where $\lambda={\pi\over5}$ is the crossing parameter.
The model becomes critical for $p=0$, in which case the five amplitudes become
the trigonometric solutions (using $\lim_{p\rarrow0}\theta_1(u;p)/(2p^{1/4})=
\sin u$),
$$\twoline{
w_1={[3\lambda-u]\over[3\lambda]},\qquad w_2={[\lambda-u]\over [\lambda]},
\qquad w_3={[u]\over \sqrt{[\lambda][2\lambda]}},}{
w_4={[4\lambda-u] \over [4\lambda]},\qquad
w_5={[2\lambda-u]\over [2\lambda]},}$$
where we defined $[u]=\sin u$. It is straight forward to verify that these
Boltzmann weights are identical to the state--graph IRF of $(G_2)_1$,
given in example (b), with $k=p=1$, up to multiplying by an overall factor
of $\sin\lambda$, and identifying the states by $0=[2]$ and $1=[0]$.

Example (d): The Ising model. This well known model can be presented
as a graph--state fusion IRF theory,  IRF(SU$(2)_2),
[1])$, example (a), with $k=2$. In this case the state variables are
$j=0,1,2$. Since a square lattice is a bipartite graph, and so is
the admissibility graph, the configurations on the lattice break into
two disjoint sets, where all even (odd) sites on the lattice are occupied
by $j=1$. Choosing the latter possibility, we may decimate the odd
lattice sites where $j=1$, to get a square lattice where the state
variables are $S=j-1=\pm1$. The four Boltzmann weights of the fusion
IRF are related to the Ising horizontal and vertical couplings,
$J_1$ and $J_2$, by (e.g., for a review, see \r\rev),
$$w_1(u)=y_1e^{-J_1}=\dw 1 2 1 0=\dw 1 0 1 2={\theta_1(u+\lambda;p)\over
         \theta_1(\lambda;p)},\e$$
$$w_2(u)=y_1e^{J_1}=\dw 1 0 1 0=\dw 1 2 1 2 ={
         \theta_1(\lambda-u;p)\over \theta_1(\lambda;p)},\e$$
$$w_3(u)=y_2e^{-J_2}=\dw 0 1 2 1=\dw 2 1 0 1=\epsilon {\theta_1(u;p)\over
         \theta_1(2\lambda;p)},\e$$
$$w_4(u)=y_2e^{J_2}=\dw 0 1 0 1=\dw 2 1 2 1={\theta_1(2\lambda-u;p)\over
         \theta_1(2\lambda;p)},\e$$
where $\lambda=\pi/4$, $\epsilon=\pm1$,
and $y_1$,  $y_2$ are some constants. The relationship
of the temperature parameter $p$ and $u$ with the Ising couplings $J_1$ and
$J_2$ is then seen to be,
$$e^{2J_1}={\theta_1(\lambda-u;p)\over \theta_1(\lambda+u;p)},\qquad
e^{2J_2}={\theta_1(2\lambda-u;p)\over\theta_1(u;p)}.\e$$
The model becomes critical for $p=0$, giving the two Ising couplings
(horizontal and vertical ones) at the fixed point. Now, it is
a straightforward exercise, which is left to the reader, to verify that
at the critical limit, $p\rarrow0$,
the Boltzmann weights $w_1,w_2,w_3$ and $w_4$ are identical to the fusion IRF
ones
for $SU(2)_{2}$, where we set in example (a), $k=2$
and $p=\epsilon$.

Example (e):
Let us return now to the previous section, where the relationship
between RCFT and their corresponding IRF lattice models was discussed.
As a particular example, consider the case of $SU(2)$ where the braiding
matrix, $R$, have been computed, ref.
\REF\TK{A. Tsushiya and Y. Kanie, Lett. Math. Phys. 13 (1987) 303}
\REF\AGlett{L. Alvarez--Gaum\'e, C. Gomez and G. Sierra, Phys. Lett.
220B (1989) 142}
\r{\TK,\AGlett}. From the IRF Boltzmann weights
described in section (7), we may compute the $u\rarrow  \infty$ limit,
which should correspond to the conformal braiding matrix, as explained
in section (7). We find,
$$R\left(\dw j{j\pm1}j{j\pm1}\right)=
    \pmatrix{-q^{-(j/2+3/4)}[1]/[j+1] &
    \sqrt{q^{-1/2} [j][j+2]}/[j+1] \cr
    \sqrt{q^{-1/2} [j][j+2]}/[j+1] &
    q^{(j/2+1/4)}[1]/[j+1] \cr}   .\e $$
The brackets $[x]$ are defined by eq. (36) with $N=2$.
The $R$ matrix so computed is, indeed, also, identical to the
$R$ matrix obtained from the  $SL(2)_q$ universal $R$ matrix \r{\AGlett}, and
is identical to the conformal braiding matrix of $SU(2)_k$,
up to an irrelevant similarity transformation (which depends on the
normalizations of the primary blocks). We conclude that the model
IRF($SU(2) ,[1]$) gives an alternative derivation for the restricted
height model, example (a).

Example (f):
The braiding matrix of $SU(N)_k$ in the $N$-$N$ channel
are computed in ref.
\REF\GFnext{D. Gepner and J. Fuchs, to appear}
\r\GFnext, where the general calculation for the braiding matrices
of any two block RCFT are found as solutions of the corresponding
Riemann monodromy problem. The result for $SU(n)_k$ can be expressed
in terms of the quantity $q=
e^{2\pi i/(k + N)}$ and the function
$$[x] : = {q^{x/2} - q^{-x/2}\over q^{1/2} - q^{-1/2}} = {\sin
({\pi x\over k + N})\over \sin ({\pi \over k + N})} \,\, . \e$$
The braiding matrix then assumes the form,
$$B\left(\dw\lambda{\lambda+e_m}{\lambda+e_m+e_l} {\lambda+e_l} \right)
={q^{-1/2N}\over [a_{lm}]} \pmatrix{-q^{-a_{lm}/2}&
\sqrt{[a_{lm} + 1] [a_{lm} - 1]}\cr
\sqrt{[a_{lm} + 1] [a_{lm} - 1]} & q^{a_{lm}/2},\cr},\e$$
where $\lambda$ is any weight at the level $k$, $f_u$ its signature sequence
of the Young tableau
(see sect. (2) for the definition), where $u=1,2,\ldots,N$,
$a_{lm}=f_l-f_m+m-l$, and
$e_l=\lambda_l-\lambda_{l-1}$, where $\lambda_l$ is the fundamental weight,
and we defined $\lambda_0=\lambda_N=0$.

Note, that for the special case of $SU(2)$, $N=2$, the Braiding matrix
agrees precisely with the one described earlier, eq. (35). As in the
case of $SU(2)$, we can now proceed to construct the lattice model
IRF(SU$(N)_k,N,N)$ using the formalism of section (7).  The crossing
parameter is given by, eq. (7.25),
$\lambda=\pi(\Delta_1-\Delta_2)/2={\pi N/2\over k+N}$, where $\Delta_1$ and
$\Delta_2$ are the dimensions of the two fields exchanged in this channel.
The generator of the Hecke algebra is, as usual,
$$H=q^{-1/2}-e^{-i\pi(\Delta_1+\Delta_2-2\Delta_N)}B,\e$$
and it obeys the usual Hecke relation,
$$H^2=\beta H,\qquad {\rm where\ } \beta=2\cos\lambda.\e$$
Explicitly, the Hecke algebra elements are given by
$$H_i=\dw \lambda{\lambda+e_k}{\lambda+e_j+e_l}{\lambda+e_j}=
(1-\delta_{jl}) {\left[s_{jl}(\lambda+e_j) s_{jl}(\lambda+e_k)
\right]^{1\over2}\over s_{jl}(\lambda)},\e$$
where $s_{jl}(\lambda)=\sin[({\pi\over k+N}) (e_j-e_l)\cdot \lambda]$.
This expression agrees precisely with the representations of the Hecke algebra
described by Wenzl
\REF\Wenzl{H. Wenzl, Representations of Hecke algebras and subfactors, Univ. of
Pennsylvania Thesis (1985), Invent. Math. 92 (1988) 349}\r\Wenzl.
Substituting this representation into eq. (7.31),
we find the Boltzmann weights of the
trigonometric lattice IRF model, which are,
$$\eqalign
{\dw \lambda{\lambda+e_\mu}{\lambda+2e_\mu}{\lambda+e_\mu}&=[1-u],\cr
\dw \lambda{\lambda+e_\mu}{\lambda+e_\mu+e_\nu}{\lambda+e_\mu}
&={[a_{\mu\nu}+u]\over [a_{\mu\nu}]},\cr
\dw \lambda{\lambda+e_\mu}{\lambda+e_\mu+e_\nu}{\lambda+e_\nu}&=
p[u] {\sqrt{[a_{\mu\nu}+1][a_{\mu\nu}-1]}
\over [a_{\mu\nu}] },\cr}\e$$
where $p=\pm1$ corresponds to two different solutions (given by $B$ or its
complex conjugate matrix, which is the braiding matrix of $SU(N)_{-1}$).

Amazingly, these are precisely the Boltzmann weights of the $SU(N)$
models described in ref. \r\Jimbo,
at the trigonometric limit.
This fully illustrates
the universal connection described in this
paper between IRF models and RCFT, giving, in this particular
instance, the Boltzmann weights of the solvable lattice models
IRF(SU$(N)_k,N,N)$.

One can easily extend the results described here to other
modular invariants of $SU(N)$, and their extended algebras. It is known
that the same blocks appear in all modular invariants, and that the
problem of writing the braiding matrices is a simple sesquilinear
re--juxtapositioning of the conformal blocks
(see, for example,
\REF\Kato{A. Kato and Y. Kitazawa, Nucl. Phys. B319 (1989) 474}
\REF\Sotkov{G. Sotkov and M. Stanishkov, Phys. Lett. 215B (1988) 674}
\REF\Green{B.R. Greene, C.A. Lutken and G.G. Ross, Nucl. Phys. B325 (1989) 101}
\REF\Trivedi{M.R.  Douglas and S.P. Trivedi, Nucl. Phys. B320 (1989) 461}
\REF\Jur{J. Fuchs, A. Klemm, Ann. Phys. 194 (1989) 303; J. Fuchs, A. Klemm
  and C. Scheich, Z. Physik 46 (1990) 71;
J. Fuchs, Phys. Rev. Lett 62 (1989) 1705}
\r{\Kato-\Jur}).
Substituting the so obtained braiding matrices into eq. (7.31),
we would find  new solvable IRF models, and
an explicit solution for their Boltzmann weights. In fact, the Boltzmann
weights
of such models are known only in the case of $SU(2)$ (the ADE cases)
\REF\Pasquier{V. Pasquier, J. Phys. A20, (1987) L217}, and some
quotient (orbifold) cases \REF\Fenthree{P. Fendley,
J. Phys. A22 (1989) 4633} \r{\Fendley,\Fenthree,\Zuber}.
For the general modular invariants of $SU(N)$
the problem of finding such Boltzmann weights have been posed in ref.
\r\Zuber\ but none have been found so far. Our approach provides a
straightforward
solution for these Boltzmann weights. Clearly, the
method is applicable to all rational conformal field theories, and
the calculations described in the considered 
examples and in ref. \r\GFnext\ can readily be carried out
for any theory.
\mysec{General Fusion IRF Models}
\par
Let us now proceed to describe the general fusion IRF model based on
any rational conformal field theory, $\cal O$, and any pair of primary
fields, $h$ and $v$, denoted as ${\rm IRF}({\cal O},h,v)$. In
section (7) we have described the two block case, $hv=\psi_1+\psi_2$,
using Hecke algebra properties. Our purpose in this section is to describe the
general case,
$$h \cdot v=\sum_{r=0}^{n-1} \psi_r,\e$$
where $n$ is the number of primary fields, and $\psi_r$ are the intermediate
primary fields in the block. We shall begin by generalizing the Hecke
algebra to a more general one, that will be termed the $n$-conformal
braiding ($n$-CB) algebra. This algebra is a quotient algebra of
the braid group
algebra, and was already described in sect. (7). Recall the defining
relations eqs. (7.19, 7.24),
$$\manyeq{R_iR_{i+1}R_i&=R_{i+1}R_iR_{i+1},\cr
          R_iR_j&=R_jR_i\qquad{\rm for\ } |i-j|\geq2,\cr
          0 &= \prod_{p=0}^{n-1} (R_i-\lambda_p),\cr}
$$
where the parameters $\lambda_p$ are fixed, specify the algebra, and
are arbitrary. Recall also from sect. (7) that the braiding matrix
of any RCFT obeys the $n$-braiding algebra relations, eq. (2), with
the eigenvalues which are,
$$\lambda_p=\epsilon_p e^{i\pi (\Delta_h+\Delta_v-\Delta_p)},\e$$
where $\epsilon_p=\pm1$, in accordance to whether $\psi_p$ appears
symmetrically ($+1$) or anti-symmetrically ($-1$) in the product,
and where $\Delta_h$, $\Delta_v$ and $\Delta_p$ denote the dimensions of the
corresponding fields. We shall assume that the fields are so arranged
that
$$\epsilon_p=(-1)^{p-1},\e$$
Clearly,
the case of $n=2$ corresponds to the Hecke algebra described
earlier, eq. (7.30) with
$$\beta={\lambda_0+\lambda_1 \over \sqrt{\lambda_0\lambda_1}}.\e$$
Special cases of the $n$-CB algebra appeared in some IRF hierarchies, ref.
\r\rev\ and references therein,
but it has not
been previously defined, in general (arbitrary parameters).

Consider now the braiding matrix of the primary fields $h$ and $v$,
$R^{h,v}_i$,
which obeys the $n$-CB algebra eq. (2). We can now define the projection
operators,
$$P^a_i=\prod_{p\neq a} \left[ {R_i-\lambda_p\over \lambda_a-\lambda_p}
\right],\e$$
where $a=0,1,\ldots,n-1$ labels the corresponding eigenvalue.
{}From the $n$-CB algebra, the projection operators obey the relations,
$$\eqalign{\sum_{a=0}^{n-1} P^a_i&=1,\cr
           P_i^a P_i^b&=\delta_{ab} P_i^a,\cr
           \sum_{a=0}^{n-1} \lambda_a P_i^a&=R_i.\cr}\e$$

Our purpose is to introduce trigonometric solutions of the YBE,
$X_i(u)$,
$$X_i(u)X_{i+1}(u+v) X_i (v)=X_{i+1}(v) X_i(u+v) X_{i+1} (u).\e$$
Further, we shall require that at the extreme UV limits these
solutions would coincide with the conformal braiding matrix,
$$\lim_{u\rarrow i\infty} e^{i(n-1) u} X_i(u)=R_i,\e$$
and that the correct admissibility conditions would be satisfied for
all values of the spectral parameter $u$, i.e., the
face transfer matrix $\langle a, b, d\vert X_i(u) \vert a,c,d\rangle$
vanishes unless $N_{a h}^b N_{b v}^c N_{c h}^d N_{d v}^a>0$,
where $N^i_{jk}$ are the fusion coefficients.
The above requirements essentially imply the following ansatz for the
face transfer matrix,
$$X_i(u)=\sum_{a=0}^{n-1}  f_a(u) P_i^a,\e$$
where $f_a(u)$ are the eigenvalues of $X_i(u)$, and are scalar functions of the
spectral parameter $u$. On physical grounds we shall also demand the
regularity relation,
$$X_i(0)\propto 1_i.\e$$
The set of constraints (10,11) are fulfilled provided that the functions
$f_a(u)$ are such that
$$\eqalign{f_a(0)&=1,\cr
           \lim_{a\rarrow i\infty} f_a(u)&=K e^{-i\pi\Delta_a},\cr}
  \e$$
  where $K$ is some constant.
Then the face transfer matrices $X_i(u)$ obey the correct UV and IR limits,
as can be verified using the relations obeyed by
the projections operators, eq. (7).

It can be shown, via considerable algebraic computations, that all
the relations specified so far, and in particular the YBE, eq. (8),
hold provided that
$$f_a(u)=\left[ \prod_{r=0}^{a-1} \sin(\zeta_r-u)\right]
         \left[ \prod_{r=a}^{n-2} \sin(\zeta_r+u)\right],\e$$
where the parameters $\zeta_i$,
$$\zeta_i=\pi(\Delta_{i+1}-\Delta_i)/2,\e$$
are the `crossing parameters'. It can be checked that the functions,
$f_a(u)$, indeed, obey the relations, eq. (12),
$$f_a(0)=\prod_{a=0}^{n-2} \sin \zeta_i,\e$$
and
$$\lim_{u\rarrow i\infty} X_i(u) e^{(n-1)iu}=e^{-i\pi(\Delta_h+\Delta_v)}
    R_i.\e$$
We conclude that the face transfer matrices $X_i(u)$, eq. (10), indeed
give rise to the Boltzmann weights of the solvable lattice model ${\rm IRF}
({\cal O},h,v)$ associated to an arbitrary
conformal field theory $\cal O$ and arbitrary primary fields $h$ and $v$.

As a side remark, note that we need not restrict ourselves to primary
fields $h$ and $v$, but can, in fact, take any algebraic combination of
primary fields, $h=\sum c_i p_i$, where $c_i$ are some constants, and
$p_i$ are the primary fields, and similarly for $v$. In other words,
$h$ and $v$ can be taken to be any elements of the fusion ring, and
the admissibility condition are taken as $a\sim b$ iff $ah=nb+\ldots$,
where $n>0$ labels the couplings, and the product is in the fusion
ring. (Alternatively $n=(a|hb)>0$, using the inner product of the
fusion ring ref. \r\DGCMP). We would then get an IRF model associated to
{\rm arbitrary} elements of the fusion ring, $h$ and $v$. The $R$
matrix, is simply the corresponding linear combination of the
separate $R$ matrices,
$$R_{h,v}=\sum_{n,m} c_n d_m R_{n,m},\e$$
where $c_n$ and $d_n$ are arbitrary complex coefficients, which
express the fields $h$ and $v$ in the basis of the primary fields.
The YBE relation is evidently satisfied, since $R_{h,v}$ is the
braiding of the fields $h$ and $v$, and thus corresponds to the
braid group on the punctured sphere.
Evidently, the general $R$ matrices satisfy a fixed polynomial
equation, with some eigenvalues $\lambda_p$, 
since this is so for each of the component
matrices in eq. (17).  Hence, projection operators can be defined as in
eq. (6).  The rest of the
construction of the Boltzmann weights $X_i(u)$ follows as before,
and these are given by the same formula, eq. (10), with the replacement
of the old projection operators by the new ones, and no other change.
The projection operators are now simply labeled by $P_{h,v}^p$ for each
of the primary fields $h$ and $v$ and each of the fields appearing in
their product, $p$. Thus we can still associate primary fields and dimensions
to each channel, and proceed as before. We find that the braiding matrix is
$$X_i(u)=\sum_{n,m} c_n d_m e^{i\pi(\Delta_n+\Delta_m)}
\sum_a P_{n,m}^a f^a_{n,m}(u),\e$$
where the functions $f^a_{n,m}(u)$ are defined as before, eq. (13), with
the obvious replacements, and the projection operators are the same ones
for each of the channels. Thus $X_i(u)$ is a linear combination of the
face transfer matrices in each of the channels,
$$X_i(u)=\sum_{n,m} c_n d_m e^{i\pi(\Delta_n+\Delta_m)} X_i^{n,m} (u),\e$$
where $X_i^{n,m}$ is given by eq. (10), and is the face transfer matrix for
the model IRF(${\cal O},n,m)$. Evidently, $X_i(u)$ obeys the same properties
as before, eqs. (9,11).

Returning from this digression, consider now the case where $h=\bar v$ (or more
generally as in the
discussion above, where $hv=1+\ldots$, and where $h$ and $v$ are arbitrary
elements of the fusion ring). Then it follows that the unit operator
is one of the fields $\psi_i$ exchanged in the channel, and we can assume
without any loss of generality that this field is $\psi_0=1$. Now,
the parameter $\lambda=\zeta_0=\pi (\Delta_1-\Delta_0)/2$ is the
crossing parameter of the model. In other words, the following
highly non--trivial relation holds,
$$\bw a b c d {\lambda-u}=\left[{S_{b,0} S_{c,0}\over S_{a,0}
S_{d,0} }\right]^{1/2} \bw b d a c  u,\e$$
where $S$ is the torus modular matrix, and
$\left[ {S_{b,0} S_{c,0}\over S_{a,0} S_{d,0}}\right]^{1/2}$
is the universal crossing multiplier, and where we have used the
standard face transfer matrix notation,
$$\bw a b c d u=\langle a,b,d |X_i(u)|a,c,d\rangle.\e$$

A particular consequence of the crossing relation, is the expression
for the projection operator $P_0$ which reads
$$P_0=\left[ \prod_i {\sin\zeta_i\over
\sin (\zeta_i+\lambda)}\right]
\left[ {S_{b,0} S_{c,0}\over S_{a,0} S_{d,0}} \right]^{1/2} \delta_{a,d},\e$$
which follows by applying the crossing relation to $X_i(0)$, and by noting
that while $X_i(0)\propto 1_i$, $X_i(\lambda)\propto P_0$.
Similarly, relations among all projection operators can be derived,
using the various crossing points.  Note that eq. (8.19) is a special
case, for trees, of the general eq. (22).

From the expression for $X_i(u)$, eqs. (10,13), it is straightforward to
compute,
$$X_i(u) X_i(-u)=1_i \sum_a f_a(u)f_a(-u)=n\rho(u) 1_i,\e$$
where
$$\rho(u)=\prod_{r=0}^{n-2} \sin(\zeta_r-u) \sin(\zeta_r+u).\e$$
It follows that the IRF theories are always unitary, i.e., they obey
the property eq. (8.15), with a unitarity factor which is $\rho(u)$.
Evidently also eq. (11) (the regularity) holds. Consequently the
IRF models described in this section obey all the standard properties
of IRF theories, eqs. (8.12--8.18). These properties will be of importance
in the next section where soliton scattering theories are constructed,
based on the fusion lattice models.

Example: Consider the case of IRF(SU$(2)_k,[2],[2])$ where $[2]$
corresponds to the adjoint representation of $SU(2)_k$ WZNW RCFT.
The projection operators can be calculated from the known structure constants,
\REF\ZF{A.B. Zamolodchikov and V.A. Fateev, Yad. Fiz. 43 (1986)
1031} ref. \r\ZF, of the RCFT.
Alternatively, the projection operators are
related to the corresponding $SU(2)_q$ quantum group $6-j$ symbols
\r{\AGlett}. Explicitly, these are as follows
\REF\Etim{V. Pasquier, Comm. Math. Phys. 118 (1988) 355}\r\Etim.
Denote the projection operators for a given face in the notation,
$$\langle a,b,d | P^i | a,c,d \rangle=P^i\left(\dw a b d c \right).\e$$
Denote also $P^i\left(\dw a b d c\right)$
by the matrix $(P^i)_{bc}=\sigma_{bc}$, and let $a=j$.
We then find
\def\mox#1#2#3{\line{ \hfil$\displaystyle{#1}$\hfil$\displaystyle{#2}$
\hfil$\displaystyle{#3}$\hfil}}

$$\mox{P^0}{d=a\pm1} {\sigma=0,}$$
$$\eqalign{\mox{P^0}{d=a}{\pmatrix{\sigma_{j-1,j-1}&\sigma_{j-1,j}&
               \sigma_{j-1,j+1}\cr
               \sigma_{j,j-1}&\sigma_{j,j}&\sigma_{j,j+1}\cr
               \sigma_{j+1,j-1}&\sigma_{j+1,j}&\sigma_{j+1,j+1}\cr}
=}\cr
\qquad  \mox{}{}{
         {1\over[3] [2j+1]}\pmatrix{[2j-1]&*&*\cr
                         \sqrt{[2j+1][2j-1]}&[2j+1]&*\cr
                         \sqrt{[2 j+3] [2 j-1]} & \sqrt{[2 j+3] [2 j+1]} & [2 j+3]\cr},}\cr}$$
$$\eqalign{
\mox{P^1}{d=a+1}{\pmatrix{\sigma_{j,j}&\sigma_{j,j-1}\cr
                                     \sigma_{j+1,j}&\sigma_{j-1,j+1}}=}\cr
\mox{}{}{{[2]\over[4][2j+4]}\pmatrix{[2j]&\sqrt{[2j][2j+4]}\cr
                           \sqrt{[2j][2j+4]}&[2j+4]\cr},}\cr}$$
$$\eqalign{
\mox{P^1}{d=a-1}{
\pmatrix{\sigma_{j,j}&\sigma_{j,j-1}\cr
                     \sigma_{j-1,j}&\sigma_{j-1,j-1}\cr}=}\cr
\mox{}{}{{[2]\over[4][2j]}\pmatrix{[2j+2]&-\sqrt{[2j+2][2j-2]}\cr
                                   -\sqrt{[2j+2][2j-2]}&[2j-2]\cr},}\cr}$$
$$\eqalign{
\mox{P^1}{d=j}{
\pmatrix{\sigma_{j-1,j-1}&\sigma_{j-1,j}&
               \sigma_{j-1,j+1}\cr
               \sigma_{j,j-1}&\sigma_{j,j}&\sigma_{j,j+1}\cr
               \sigma_{j+1,j-1}&\sigma_{j+1,j}&\sigma_{j+1,j+1}\cr}
=}\cr\mox{}{}{
              {[2]\over[4]}\pmatrix{
              1-{[2]\over[2j][2j+1]}&*&*\cr
              -\sqrt{[2j-1]\over[2j+1]} {q^{2j+1}+q^{-2j-1}\over[2j]}&
             {2+q^{4j+2}+q^{-4j-2}\over[2j][2j+2]}&*\cr
             -{\sqrt{[2j-1][2j+3]}\over[2j+1]}& \sqrt{[2j+3]\over[2j+1]}
             {q^{2j+1}+q^{-2j-1}\over[2j+2]}&
             \left(1-{[2]\over[2j+1][2j+2]}\right)\cr}
,}\cr}$$
where the matrices are symmetric, $q=e^{i\pi\over k+2}$,
$$[x]={q^x-q^{-x}\over q-q^{-1}},\e$$
and $P^2=1-P^0-P^1$.
Here $a$,$d$ and $j$ stands for the isospin of the primary fields
and are in the range $0,\half,1,\ldots,{k\over2}$.

{}From these projection operators we find,
$$\lambda=\zeta_0=\pi(\Delta_1-\Delta_0)/2={\pi \over k+2},\e$$
and
$$\zeta_1=\pi(\Delta_2-\Delta_1)/2={\pi\over2}{4\cdot6-2\cdot4\over
     4(k+2)}={2\pi\over k+2}=2\zeta_0.\e$$
Substituting into eq. (10) we find the face transfer matrix,
$$X_i(u)=P^0 \sin(\zeta_0+u)\sin(\zeta_1+u)+
       P^1  \sin(\zeta_0-u)\sin(\zeta_1+u)+
       P^2 \sin(\zeta_0-u)\sin(\zeta_1-u).\e$$

It can be explicitly checked, involving numerous relations among the
matrix elements,
that the crossing property, eq. (20), which is a powerful constraint,
is indeed obeyed, where the $S$ matrix is the usual one for $SU(2)$,
eq. (8.20), with $p=1$. Indeed crossing symmetry, eq. (20) puts severe constraints on the
Boltzmann weights, eq. (25). In fact, $P^0$ is determined entirely by eq. (20), in accordance with
eq. (22).
Similarly $P^2$ and $P^1$, are almost fully (but not completely) determined by
eq. (20).

The face transfer matrix $X_i(u)$, eq. (27), derived here by specializing
the universal eq. (10),
is identical to the trigonometric solutions of the YBE equation
previously found in ref. \r\Jimbo, for this particular case
\r\Etim. Similarly, all the solutions of the YBE based on various
quantum groups are special cases of eq. (10), as applied to the
corresponding WZNW model. This includes a host of models studied in
literature (for a review, see, e.g., \r\rev).
\def\Re{\mathop{\rm Re\,}\nolimits}
\def\Im{\mathop{\rm Im\,}\nolimits}
\mysec{Fusion Soliton Theories}
Consider a soliton or kink system based on some fusion IRF model.
The solitons are labeled by some primary fields in the theory, $h$, $v$,
$p$,
etc. The vacua of the theory are also labeled by the primary fields, where
each primary field, $a$, corresponds to a unique vacuum. The $h$ solitons
interpolate between any two vacua $a$ and $b$ in the theory, such that
the fusion coefficient $N_{ha}^b$ does not vanish.
The two particle scattering amplitude for the solitons $h$ and $v$ is
described by the process $(a|h|b)+(b|v|d)\rarrow (a|v|c)+(c|h|d)$,
where $(a|h|b)$ stands for the $h$ soliton interpolating between the
$a$ and $b$ vacua, and so forth. We denote this scattering amplitude by
\def\S#1#2#3#4#5#6#7{S_{#1,#2}
\left(\matrix{#3&#4\cr#6&#5}\bigg\vert #7\right)}
$$\S h v a b d c \theta,\e$$
where $i\pi \theta$ is the relative rapidity of the two incoming
particles, fig. (10.1). Here, $a,b,c$ and $d$ label the different admissible
pairs, which obey, $N_{ha}^bN_{hc}^dN_{va}^cN_{vb}^d>0$.

The $S$ matrix of any such theory (regardless of whether it
corresponds to a fusion of RCFT) needs to have the following properties.
\par\noindent Regularity, which expresses the absence of scattering at
zero relative rapidity:
$$\S h v a b d c 0=\delta_{b,c}.\e$$
Unitarity:
$$\sum_e \S h v b c e a \theta \S {\bar v} {\bar h} e c d a {-\theta}=
\delta_{b,d}.\e$$
Crossing symmetry:
$$\S h v a b d c \theta=\S v {\bar h} b d c a {1-\theta}=
  \S {\bar v} h c a b d {1-\theta}.\e$$
Charge conjugation:
$$\S h v a b c d \theta={\S {\bar h} {\bar v} d c a b \theta}^*.\e$$
Time reversal:
$$\S h v a b d c \theta=\S v h a c d b \theta.\e$$
In addition, for an integrable scattering theory the $S$ matrix must
obey the factorization equation (YBE) \r{\ZamZam,\rev},
$$\twoline{
\sum_g \S h v a b g f \theta \S p v f g d e {\theta+\phi}
\S p h g b c d \phi = }{
        \sum_g \S p h f a g e \phi \S p v a b c g {\theta+\phi}
        \S h v g c d e \theta,}$$
which is depicted in fig. (10.2). Note
that the factorization equation (7) is identical to the STE relation for
the corresponding IRF model eq. (7.3), and that fig. (10.2) is dual to the
STE fig. (7.2), by replacing each of the vacua domains by a vertex,
and connecting the vacua if and only if a soliton interpolates between
them.  Thus an $S$ matrix which satisfies the factorization equation
is a Boltzmann
weight for a solvable lattice model. Our purpose is to build the $S$ matrix
from the IRF Boltzmann weights described in sections (7,8,9). Thus, we relist
here the properties of the Boltzmann weights, denoted as before by
\def\w#1#2#3#4#5#6#7{w_{#1,#2}
\left(\matrix{#3&#4\cr#6&#5\cr}\bigg\vert #7\right)}
$\w h v a b d c u$.

Regularity:
$$\w h v a b d c 0=\sqrt{\rho(0)}\delta_{b,d},\e$$
Unitarity:
$$\sum_e \w h v  b c e a u \w {\bar v} {\bar h} e c d a {-u}=
\rho_{h,v}(u)
        \delta_{b,d}.\e$$
Crossing symmetry:
$$\w h v a b d c u=\left({\psi(a)\psi(d)\over \psi(b)\psi(c)}
\right)^{1\over2}
\w v {\bar h} b d c a {\lambda-u}.\e$$
Charge conjugation and time reversal:
$$\w h v a b c d u={\w {\bar h} {\bar v} d c a b u}^*=\w v h a c d b
u.\e$$
The parameter $\rho_{h,v}(u)$ is called the unitarity factor. $\lambda$
is the crossing parameter, and $\psi(a)$ is the crossing multiplier.
For a graph--state IRF model we found that $\psi(a)$ is an eigenvalue
of the connectivity matrix and $\beta = 2\cos(\lambda)$ is the eigenvalue.
For a fusion IRF model IRF$({\cal O}, h,\bar h)$ we have the universal
crossing multiplier $\psi(a)=S_{a,1}/S_{h,1}$, where $S$ denotes the toroidal
modular function, eq. (9.20).
For an IRF model based on a two block product $hv=\psi_1+\psi_2$ we
have from sect. (7),
$$\w h v a b d c u=\delta_{b,c}\sin(\lambda-u)+\sin u \cdot R_{h,v}
\pmatrix{a&b\cr c&d\cr},\e$$
where $R_{h,v}$ is the Hecke algebra element, eq. (7.29), and
where $\lambda$ is the crossing parameter,
$\lambda=\pi(\Delta_2-\Delta_1)/2$,
where $\Delta_i$ are the dimension of the fields in $hv$ product,
and which depend on the choice of fields $h$ and $v$.
The generalization of the Boltzmann weight eq. (12) to more than two blocks
is described by eqs. (9.10, 9.13). The unitarity factor is given by eq.
(9.25)
We take the following general ansatz for the $S$ matrix,
$$\S h v a b d c \theta=F_{h,v}(\theta) \left[
{\psi(b)\psi(c)\over\psi(a)\psi(d)}\right]^{\theta/2}
\w h v a b d c {\eta_{h,v}\lambda_{h,v} \theta},\e$$
where $\lambda_{h,v}$ is the crossing parameter in this channel,
$\eta$ is some parameter, to be determined, and $F_{h,v}(\theta)$
is an overall function.

Evidently, $S$ obeys all the properties listed, except possibly, for
crossing and unitarity. In particular, the factorization equation, eq.
(7),
which is a consequence of the STE is obeyed by the Boltzmann weights (as long
as
we keep $\eta_{h,v}\lambda_{h,v}$ fixed for all $h$ and $v$, which will be
assumed), which by itself is a consequence of the Hexagon identity obeyed
by the conformal braiding matrix, eq. (7.13). Unitarity and crossing, will
be achieved by properly setting the parameters $\eta_{h,v}$ and the
functions $F_{h,v}(\theta)$.

Let us first assume that the field $h=v=x$ is a real field, $x=\bar x$,
and that this is a two block case, $x^2=\psi_1+\psi_2$.
This applies to all the graph--state models, as well as all RCFT with
a real fundamental field. Setting in this case, $\eta=1$ and $\lambda=
\lambda_{x,x}=\pi p/q$, where $p$ and $q$ are some
strange integers. We find
that unitarity and crossing hold provided that the function $F(\theta)=
F_{x,x}(\theta)$ obeys
$$\eqalign{F(1-\theta)&=F(\theta),\cr
           F(\theta)F(-\theta)&={1\over\sin[\lambda(1-\theta)]
           \sin[\lambda(1+\theta)]}.\cr}\e$$
The system of equations (14) can be solved for any rational $\lambda/\pi$
giving a universal minimal\foot{The minimality here denotes least number
of zero and poles in the strip $0\leq \Re \theta < 2$.} solution $F_\lambda
(\theta)$. Note that for an RCFT, indeed
$\lambda/\pi={\Delta_1-\Delta_2\over 2}$ and is thus a rational number.
To present a closed form solution for $F(\theta)$ we need to distinguish
the cases, where $q$ is odd, or $q$ is even. For odd $q$ we find the minimal
solution,
$$F(\theta)^{-1}=\sin[\lambda(1+\theta)]\prod_{m=2}^{(q-1)/2}
{ \sin[\lambda(m-(-1)^m\theta)]\over \sin[\lambda(m+(-1)^m\theta)] }.\e$$
To write the solution for an even $q$ in a closed form, we define the
special function $H(z)$ by
$$H(z)=\prod_{m=1}^\infty \left[ {
\big(1-{z\over2m}\big)^{2m}
\big(1+{z\over2m-1}\big)^{2m-1}  \over
\big(1+{z\over2m}\big)^{2m}
\big(1-{z\over2m-1}\big)^{2m-1} }\right].\e$$
It can be checked that the terms appearing in this product are of the
form $1+\alpha(z)/m^2+O(m^{-3})$,
and thus the product converges absolutely
for any $z$ which is not an integer. Thus, the special function $H(z)$ is
a holomorphic function, whose only poles and zeros in the complex plane
are for an integer $z$. The order of the zero at $z=N$ is $N$ for an
even $N$, and $-N$ for an odd one, with the convention that a zero of
order $-k$ is a pole of order $k$. It can be checked that $H(z)$ obeys
the functional equations,
$$\eqalign{H(z)H(-z)&=1,\cr
           {H(1-z)\over H(z)}&=\tan\left({\pi z\over2}\right).\cr}\e$$
Now, the function $F(\theta)$ for $\lambda=2\pi p/q$, even $q$, becomes
$$F(\theta)=H(\pi\theta)^{-2p/q}
\prod_{m=1}^{q-1} (\sin[\lambda(\theta-m)])^{(-1)^m
p_m p/q},\e$$
where the integers $p_m$ are defined by,
$$p_m=qs_m-2m,\qquad s_m=\cases{0&\quad $m=0$\cr
                                1&\quad $1\leq m<q-1$\cr
                                2&\quad $m=q-1$\cr}.\e$$
Note that $F(\theta)$ is a holomorphic function with no branch
lines (or `cuts')
with poles and zeros only for $z=\pi m$, where $m$ is some integer.

Note now that we may multiply the minimal solution, eq. (15,18), by any
solution of the `homogeneous'  equation
$$\eqalign{
&Q(\theta) Q(-\theta)=1,\cr
&Q(\theta)=Q(\lambda-\theta),\cr}\e$$
and, still, eq. (14) holds.  The general solution to eq. (20) is
$$Q(\theta) = {h(\theta)\over h(-\theta)} ~ {h(\lambda-\theta)\over
h(\lambda+\theta)},\e$$
where $h$ is any holomorphic function.  Such a solution $Q(\theta)$ is termed a
`CDD' ambiguity, and it needs to be determined, for the
full $S$ matrix to be found.

The only way to determine, with any confidence, the CDD factor, eq. (21),
is by some additional physical
input. (In the literature, more often than not, this is
done by trial and error on conjectured $S$ matrices).
In particular, we may use the soliton bound state content, along
with the known masses of the solitons, eq. (3.11), to find the poles
of $F(\theta)$ in the physical sheet $0 \leq \theta <1$.  Such
poles correspond to bound states of the scattered particles,
whose mass is
$$M_B^2 = M_h^2 + M_v^2 - 2M_h M_v \cos \left({\pi
\theta}\right),\e$$
where $M_u$, $M_v$ and $M_B$ are the masses of the $h$ and $v$
solitons, along with their bound states, denoted by $B$.  The equation
(22), along with the known solitonic spectrum, determines all the
physical poles in the regime $0\leq \Re \theta <1$.  Further, for
the fundamental solitons, we expect no higher order poles, as these
correspond to higher order scattering processes.  Thus, we form a
complete knowledge of the poles in the physical sheet  $0 \leq \Re
\theta<1$.  These are allowed only for $\Im \theta = 0$, and then only
for the values given by eq. (22). For a unitary theory, the residues at
the poles must be positive. For non--fundamental solitons, the residues
can be calculated using diagrams involving the constituent solitons.

Equipped with this knowledge, we may proceed to determine the suitable
function $h(\theta)$ up to such $Q(\theta)$'s which have no zero
or poles in the
physical domain $0 \leq \Re \theta < 1$.  Such $Q$'s  are called
Z-factors, and will always be  assumed to be equal to one, unless
otherwise specified.

We may multiply, the solution $F_\lambda(\theta)$ by the suitable
$Q(\theta)$, in such a way as to remove all the poles from the physical
sheet. This resulting solution is termed the strictly minimal
solution.  For $\lambda = {\pi\over q}$, the solutions eq. (15) and (18)
are strictly minimal.  For $\lambda = {2\pi p\over q}$, where $p>1$, we
need to use a non-trivial $h$, which is easily found to be of the form,
$$h(\theta)= \prod_i \sin (\lambda \mu_i-\theta),\e$$
where $\mu_i$ are parameters specifying for the location of the poles of
$F_\lambda (\theta)$ on the physical sheet, which come in pairs
at $\mu_i$ and  $1-\mu_i$, as follows from the crossing symmetry, eq.
(4).  Thus, at last, determining the complete $S$ matrix.

For, example, in all soliton theories based on state-graph IRF models,
we expect no bound states for the fundamental solitons.  Hence, the full $S$
matrix is given by the strictly minimal solution, eq. (13).

Now, note that we have chosen the value of $\eta=1$. However other values
may be valid as well. Albeit, these correspond to choices of different RCFT,
which have the same fusion rules. As explained in sections (4,5), each set of
fusion rules lead to a number of {\rm different RCFT}. The parameter $\eta$
can be set equal to one, {\it always}, provided that we build the Boltzmann
weights around {\rm all} the possible RCFT's with the given set of fusion
rules.
For example, $SU(2)_3$ can be replaced by $SU(2)_{3/4}$, giving two different
$S$ matrices, with the same exact solitonic content. Which of all these
choices correspond to the actual $N=2$ supersymmetric field theory we are
interested in? Usually, one may use charge conjugation, eq. (5), to eliminate
all choices of the RCFT (for a real particle), by demanding that $S$ is real.
However, there are some very rare instances where an ambiguity remains,
and cannot be dispelled without
additional physical input. In fact, the Hard hexagon
model, which is based on the RCFT
$SU(2)_3/ SU(2)_{1/3}$, as discussed in section (8), example (c), has
such an ambiguity, since the RCFT $SU(2)_{3/4}/SU(2)_{4/3}$ equally well
leads to
real Boltzmann weights along with
the same fusion rules. The first $S$ matrix
was proposed in ref.
\REF\Tri{A.B. Zamolodchikov, Princeton preprint PUPT 1195 (1990)}\r\Tri,
while the other was suggested in ref.
\REF\Smir{F.A. Smirnov, Int. J. Mod. Phys A6 (1991) 1407}\r\Smir. Both
are correct $S$ matrices, but they correspond to different physical systems.
In general, these ambiguities can be resolved only with an additional physical
input\foot{
In this particular example, both $S$ matrices were suggested to describe
the $\phi_{2,1}$ perturbation of the tri--critical Ising model. It is still not
resolved which is the correct one for this specific theory, though some support
for the $S$ matrix of ref. \r\Smir\ was reported in ref.
\REF\KMtwo{T. Klassen and E. Melzer, Cornell and Stony Brook preprint,
CLNS-92-1130, ITP-SB-92-01, (1992)} \r\KMtwo.}.
\def\wig{\wiggle}
\def\sh{\mathop{\rm sh}\nolimits}

Let us turn, now, to soliton scattering amplitudes in non--real fusion
theories.
Consider the fundamental soliton of the theory, labeled by the primary field
$p$, and assume for simplicity that this is the two block case,
$$\manyeq{\psi_p^2&=\psi_1+\psi_2,\cr
          \psi_p \psi_{\bar p}&=1+\psi_3,}$$
where $\psi_i$ denotes the respective primary fields, and the product is
in the fusion ring sense. Now, there are two different crossing parameters
we can define, in accordance with the theory of section (7). These are
$$\lambda={\pi \Delta_3\over 2},\qquad \wig\lambda={\pi(\Delta_2-\Delta_1)
\over 2},\e$$
where $\Delta_i$ denotes, as before, the dimensions of the primary fields.
Now, we have four different soliton scattering amplitudes, $S_{p,p}$,
$S_{p,\bar p}$, $S_{\bar p,p}$ and $S_{\bar p,\bar p}$ to consider.
These are related via the crossing and charge conjugation, eq. (4,5),
to the basic amplitude $S_{p,p}$ but need to obey four different factorization
equations, eq. (7). Note that for amplitudes involving $S_{p,\bar p}$
and $S_{\bar p,p}$ the YBE no longer holds and it is modified to a
quadratic relation.
It follows that the general Hecke ansatz holds only for the $S_{p,p}=
S_{\bar p,\bar p}^*$ amplitudes. For $S_{p,p}$ we adopt the same ansatz
as before, eq. (13),
\def\wigl{{\wiggle\lambda}}
$$\S p p a b d c \theta=
   F(\theta) \left({S_{b,0} S_{c,0}\over S_{a,0} S_{d,0} }
\right)^{\theta/2}
 \w p p a b  d c {\eta\theta} ,\e$$
where $S_{a,b}$ is the usual toroidal modular matrix, $F(\theta)$ is the
unitarizing factor that needs to be determined, and the Boltzmann weight
$w$ is given by the usual fusion IRF Hecke ansatz, eq. (7.31),
$$\w p p a b d c \theta=\sin[\wigl -\lambda\theta)]\delta_{b,c}+
    \sin(\lambda\theta) H\pmatrix{a & b\cr c &d\cr},\e$$
and we absorbed the crossing multiplier into the definition of the
rapidity. Again, we may assume $\eta=1$\foot{The reason for it, as before,
that different values of
$\eta$ correspond to different choices of RCFT's which have the same
fusion rules, and thus with no loss of generality we may take $\eta=1$,
along with choosing a different RCFT.}.
$H$ is the Hecke algebra element associated to the braiding matrix
$B_{p,p}$, and is defined by equation (7.29).
{}From the discussion of section (7), it follows that $w$, and consequently
also
$S$, obey the YBE, eq. (7.32). The other amplitudes can now be determined
by the crossing symmetry, eq. (4), where the only  new amplitude
is $S_{p,\bar p}$, due to charge conjugation. Using crossing, we find,
$$\S p {\bar p} a b d c \theta= \S p p c a b d {1-\theta}.\e$$
It remains to verify that the modified YBE, eq. (7) holds with this ansatz,
and that indeed the modular matrix, again corresponds to the crossing
multiplier. In other words, we need the highly non--trivial modified YBE,
$$\twoline{
\sum_g \S p {\bar p} a b g f \theta \S p {\bar p}  f g d e {\theta+\phi}
\S p p  g b c d \phi = }{
        \sum_g \S p p  f a g e \phi \S p {\bar p} a b c g {\theta+\phi}
        \S p {\bar p} g c d e \theta.}$$
We conjecture that  indeed, eq. (29) holds with the ansatz eq. (26). 

The other modified YBE relations
follow immediately from eq. (29)
and charge conjugation. Note that $S_{p,\bar p}$
is {\it not} of the general Hecke ansatz, as it is not a solution of the
YBE, but the modified relation eq. (29).
However, for any number of
blocks, it remains of the form eq. (9.10), expressed in terms of the same
projector operators,
albeit with a different set of eigenvalue
functions $f_a(\theta)$, which can
be deduced from the crossing property eq. (9.20).
Now, the only remaining non--trivial property that needs to be
checked is the unitarity, eq. (3), which is implemented by a judicious
choice of the function $F(\theta)$. Unitarity holds provided that
$F(\theta)$ obeys
$$\manyeq{  F(1-\theta)F(1+\theta)&=\left[ \sin\lambda(1- \theta)
    \sin\lambda(1+\theta) \right]^{-1}\cr
            F(\theta) F(-\theta)&=\sin[\wigl-\lambda\theta]^{-1} \sin [
     \wigl+\lambda\theta]^{-1}.\cr}$$
This set of equations can be solved for any $\lambda$ and $\wig\lambda$.
It is convenient to parametrize the crossing parameters via the
variables $n$ and $k$,
$$\lambda={\pi n \over 2(n+k)},\qquad \wigl={\pi \over n+k},\e$$
which are arbitrary real numbers, which will be termed the ``rank'' and the
``level'', respectively.
(In RCFT the rank and the level
are, of course, always rational).
The minimal solution to the set of equations (30), can be found in a closed
form, as in the real case eqs. (15,18),
via pole counting. For brevity sake, we omit the detail.
Alternatively, $F(\theta)$ may be written as
$$F(\theta)=a_n(\theta) f_{n,k}(\theta),\e$$
where $f_{n,k}(\theta)$ has no poles in the physical sheet, $0 \leq \Re
\theta<1$, and is given by the expression
$$\twoline{
f_{n,k}(u)=\exp\bigg( 2\int_0^\infty {\d x\over x} \,
{\sh(nux/2) \over \sh[(k+n)x] \sh(nx) }
\big\{ \ch(kx) \ch (nxu/2)-
}{
\ch[(n+k-2) x] \ch [nx(u/2-1)] \big\} \bigg),
}$$
whereas $a_n(u)$ is the Koberle--Swieca scattering matrix \r\KS, of
the $\hat A_n$ type,
$$a_n(u)={\sin \pi\left({u\over2}+{1\over n}\right) \over
\sin \pi\left({u\over2}-{1\over n}\right)}.\e$$
All the poles in the physical sheet are contained in $a_n(u)$, while
$f_{n,k}(u)$ has no poles or zeros in the physical sheet. Thus in the
$p$-$p$ channel there is a unique bound state, whose mass is given by,
using eq. (3.11),
$$m_2=2m_1\cos({\pi\over n})=2m_1\cos({2\pi\lambda\over\wigl}).\e$$
The rest of the amplitudes are found by the fusion procedure of IRF,
or the bootstrap of the soliton amplitudes as in refs. \r{\FIRF,\Date}.

Let us exemplify the foregoing discussion with the $SU(n)_k$ fusion
soliton theory. The Hecke algebra elements
were given in section (8), example (f).
The two crossing parameters are readily calculated
using the dimension formula, eq. (4.22),
$$\lambda={\pi\over n+k},\qquad \wigl={\pi n\over 2(n+k)}.\e$$
Again  $n$ is identified with the rank an $k$ with the level, as 
in eq. (31). Further,
the modular matrix element $S_{\lambda,0}/S_{0,0}$ is the
quantum dimension of the representation $\lambda$, and
may be expressed by a quantum
Weyl formula (this holds for all the groups; see, e.g, \r\AG),
$${S_{\lambda,0}\over S_{0,0}}=\prod_{\alpha>0} {[(\lambda+\rho,\alpha)] \over
     [(\rho,\alpha)]},\e$$
where we defined $[x]=\sin({\pi x\over n+k})$, the product ranges
over all the positive roots $\alpha$, and $\rho$ is the usual half the sum
of the positive roots. At $n+k\rarrow \infty$ the quantum dimension becomes
the usual dimension of the representation $\lambda$, while eq. (37)
becomes the usual Weyl's dimension formula. The bound states of the theory are
the solitons corresponding to the level one (i.e., fully anti--symmetric)
representations, $p_t=\lambda_t$, $t=1,2,\ldots,n-1$, and
where $\lambda_t$ are the fundamental
weights. The masses of the solitons, are given by the usual Koberle--Swieca
relation,
$$M_t=m \sin\left({\pi t\over n}\right ),\e$$
where $m$ is the mass scale of the theory.
The relevant  part of the $S$ matrix,
i.e., the part that has poles in the physical sheet,
which are the analogues of the $a_n(u)$ piece for all the amplitudes,
is given by the usual $\hat A_n$ Toda $S$ matrices,
see for example ref. \r\Klas. Note now that this mass spectrum agrees
precisely with the one derived from the LG potential, eq. (3.11), for this
particular case, where the potential $V$ is given by \r\DGCMP
$$V=\sum_{i=1}^n q_i^{n+k},\e$$
written in terms of the anti--symmetric functions
$$x_t=\sum_{i_1<i_2<\ldots<i_n}  q_{i_1} q_{i_2}\ldots q_{i_n},\e$$
which correspond
to the fully anti--symmetric representations, along with
the constraint $x_n=1$. Substituting into eq. (3.11),
we recover the mass formula
eq. (38), which was previously conjectured in ref. \r\GS\foot{
Curiously, this mass spectrum is identical to the one found for the
class of potentials, eq. (2.1), which was discussed in
section (3), $M_{rs}$ given after eq. (3.11). The reason for this
coincidence is that the spectrum does not depend on the level $k$, and for
$k=1$ the two potentials give the same RCFT field theory, and hence also,
the same soliton scattering theory
(see the example at
the end of section (3), and the footnote there).}.

The scattering matrices described, in the above $SU(n)_k$ example,
agree precisely with those previously derived in ref.
\REF\Vega{H.J. de Vega and V.A. Fateev, Int. Journal Mod. Phys. A6 (1991) 3221}
\r\Vega, using the lattice IRF models of ref. \r\Jimbo,
where it was suggested to describe the perturbations
of the $W$ invariant coset model ${G_k\times G_1\over G_{k+1}}$, as
perturbed by the operator $\Phi^{0,0}_{\rm ad}$, where $\lambda_1$, $\lambda_2$, $\lambda_3$
for 
$\Phi^{\lambda_1,\lambda_2}_{\lambda_3}$ denote the highest
weights of $G_k$, $G_1$ and
$G_{k+1}$, respectively, using the usual coset notation. Alternatively
the theory can be described as the $W$ algebra symmetric theory, $W(A_n)$,
ref.
\REF\Luk{V.A. Fateev and S.L. Lukyanov, ``Kiev Lectures'', ITP preprint (1988)}
\r\Luk. For $k=1$ all the factors drop out, and the $S$ matrix is given
just by $a_n(u)$, and thus coincides precisely with the $\hat A_n$
matrices. For any $k$, the spectrum relevant part of the amplitude is
given by the $k=1$ amplitude, $a_n(u)$. The same holds for the potentials
eq. (2.1), which are described by the very same $\hat A_n$ scattering
matrices, for any $k$, times an `irrelevant' (i.e., no bound states) IRF
piece, which depends on $k$.

Note that we may build a more general $S$ matrix by choosing the RCFT
$SU(N)_{k/p}$ instead (c.f. section (5)). This amounts to replacing $(n+k)$
with $(n+k)/p$ everywhere, and no other changes. However, for a general $p$
the resulting amplitudes violate charge conjugation symmetry, as they are
complex.
The only allowable values are $p=\pm 1$. Albeit, we may use the duality
relation $SU(n)_k\approx SU(k)_{-n}$, to assume that $p=1$ always. (Where
the choice $p=-1$ gives the dual amplitudes of $SU(k)_n$; see the footnote
after eq. (2.40).)

Thus, we, indeed,
see a one to one correspondence between fusion LG theories and their
associated scattering
matrices, pinning down a unique RCFT with the given set of fusion rules,
for each such LG theory. The actual $S$ matrices of the $N=2$ supersymmetric
theory are a tensor product with a trivial factor, which accounts
for the doubling
of the solitons, and is independent of $k$. 
Examples of such factors, and $S$ matrices, were already
conjectured in refs.
\REF\Denis{D. Bernard and A. LeClair, Phys. Lett. 247B (1990) 309}
\REF\Fint{P. Fendley and K. Intriligator, Nucl. Phys. Nucl. Phys. B372 (1992)
533}
\r{\Denis,\Fint,\GS}.

The arguments above can be generalized straightforwardly to the case of more
than two blocks. The Boltzmann weights are built as described in section (9).
The only essential modification is in the unitarity equation, which assumes
the form, (for simplicity, we take the primary field $p$ to be real;
the variation for
the non--real case is analogous, and is omitted for brevity),
$$\manyeq{
F(v)F(-v)&=\prod_i \left[ \sin(\lambda_i-\lambda v) \sin(\lambda_i+
\lambda v)\right]^{-1},\cr
F(v)&=F(1-v),\cr}$$
where $i$ ranges over the primary fields in the product, and
where $\lambda_i$ and $\lambda$ are the relevant crossing parameters,
(given in section (9), eqs. (9.14), where $\lambda_i=\zeta_i$, and $\lambda=
\zeta_0$ is the crossing parameter),
and where we
have used also eq. (9.24) (unitarity).
The solution to this set of equations is
$$F(v)=\prod_i f_i(v),\e$$
where the functions $f_i(v)$ obey
$$\manyeq{
f_i(v)f_i(-v)&=\left[\sin(\lambda_i-\lambda v)\sin(\lambda_i+
\lambda v) \right ]^{-1},\cr
f_i(v)&=f_i(1-v),}$$
and can be solved in a closed form, in analogy with eqs. (15,18), for rational
crossing parameters.

As an example, one finds in this
way the scattering matrices associated to IRF$(SU(2),[2],[2])$ which are
given by substituting the Boltzmann weights eq. (9.27). The unitarity
equation, eq. (41), becomes,
$$\manyeq{F(v)F(-v)&=[\sin\lambda(1+v) \sin\lambda(1-v) \sin\lambda(2-v)
    \sin\lambda(2+v) ]^{-1},\cr
           F(v)&=F(1-v).\cr}$$
Assuming an odd level, $\lambda=\pi p/q$, where $q$ is the
level of $SU(2)$, and $\gcd (p,2q)=1$, we find that
$$F(v)=f_1(v) f_2(v),\e$$
where
$$f_1(v)^{-1}=\sin[\lambda(1+v)] \prod_{m=2}^{(q-1)/2}
 {\sin[\lambda(m-(-1)^mv )]\over \sin[\lambda(m+(-1)^mv)]}\e$$
(from eq. (15)), along with
$$f_2(v)^{-1}=\sin[\lambda(2+v)] \prod_{m=2}^{(q-3)/2}
   {\sin[\lambda(m+1-(-1)^m v)] \over
     \sin[\lambda(m+1+(-1)^m v)]}.\e$$
The reader is urged to complete the case of even $q$, as an exercise.
We expect no bound states for this soliton theory. Charge conjugation, for
$SU(2)$ implies that $p=\pm 1$. The solution eqs. (45-47),
is then strictly minimal,
indicating that there are, indeed, no bound states.

This concludes the description of the general soliton scattering theory, and
their scattering amplitudes. In particular, we find the $S$ matrices of all
the integrable $N=2$ supersymmetric field theories, which are constructed
from their corresponding rational conformal field theories.
It is noteworthy that all the
known examples of soliton scattering theories studied in the literature
are of the form eq. (26), and conform to this
universal expression. Thus, the discussion in this paper
gives a unified
framework for all integrable soliton systems, mapping them uniquely to
rational conformal field theories.
This completes the circle of categorical isomorphisms
put forward in the introduction, section (1),
for all the four systems described there.
\mysec{Discussion}
The theme threaded throughout this paper is the interrelationship
between rational conformal field theories,  integrable  $N=2$
scalar field theories (`Landau Ginzburg'),
Interaction round the face lattice models and factorizable solitons scattering theories.
The described correspondence found here between these four  very different physical
problems drives us to the conclusion that in effect these are equivalent
systems and leading, also, to a cross fertilization of ideas in these
much investigated subjects.

The first step, in this correspondence, is the realization that fusion rings
of rational conformal field theories give rise to integrable $N=2$
scalar field theory models, whose super--potential is the fusion potential
of the RCFT. This
correspondence need not be limited to LG theories, but one may simply
postulate an $N=2$ supersymmetric field theory, whose chiral algebra  is
identical to the fusion ring of any given RCFT. The question of whether this
fusion ring is described by a potential, and thus maps to an $N=2$
superconformal scalar field theory,
then becomes only a technical issue. It follows that we can cover
all supersymmetric integrable models and all RCFT's
by this picture.
The ground state vacua of the resulting LG theory are given by the
extrema points of the fusion potentials, which are labeled by the primary
fields. The solitons of the theory interpolate between these different primary
field vacua. The exact solitonic content may be deduced through a calculation
of the metric in moduli space. We find that the solitons are labeled by
some fundamental primary fields, and that the admissibility condition of having
two vacua connected by a given soliton is whether or not their respective
fusion coefficients vanish. Thus we are naturally led  to considering
integrable soliton systems described by the fusion rules of any
RCFT. The
$S$ matrices of the soliton systems are closely related to the Boltzmann
weights of fusion interaction round the face lattice models. We build
the Boltzmann weights from the conformal braiding matrix of the RCFT, thus
getting a direct link between RCFT and fusion IRF.

The soliton systems $S$ matrices are of the form
$$\sw a b c d u= F(u) \left({S_{b,0}S_{c,0}\over S_{a,0} S_{d,0}}
\right)^{u/2}\bw a b  c d {\eta u},\e$$
where $a$,$b$,$c$ and $d$ are primary fields labeling the vacua, $S_{a,b}$ is
the conformal modular matrix and the Boltzmann weights are built in terms of
the conformal braiding matrix $R$ as explained in sections (7,8,9). The factor
$F(u)$ is needed for unitarity, and is determined by the unitarizability
property of the Boltzmann weights, eq. (10.9).
The power part is needed to ensure crossing symmetry
of the scattering amplitudes. Together, crossing and unitarity, uniquely
determine the $S$ matrix, including the parameter $\eta$, up to
CDD ambiguities, as explained in section (10).
The actual $N=2$ amplitude  includes also a tensor with
a trivial $S$ matrix which comes from supersymmetry. Without this factor,
the above minimal $S$ matrix corresponds to different physical problems.

Now, one might ponder the direct relation between the RCFT and its
corresponding IRF model. What is, for example, the critical structure of
the IRF lattice models, and in particular, do the fixed points give back
the same RCFT we started from? There is a strong indication that this is
in fact the case, though more study of these new IRF models is required.
If so, then we have actually solved the reconstruction `fantasy'
problem of RCFT \r{\MS,\AGlett}.
Starting only from the conformal data on the sphere, we build the IRF model
based on the conformal braiding matrix. The fixed points of the lattice model
then give the full conformal field theory realizations of the original
data we started from. Thus, going through the fusion IRF models,
reconstruction of RCFT is no longer a fantasy, but a well defined technical
procedure. 
The solitons systems described here would then apply to off critical
limits of the RCFT, in the appropriate regime in parameter space.

The fusion IRF construction gives us a novel and powerful way to build
integrable lattice models satisfying the star triangle equation. It turns
out that all the known integrable IRF models with second order phase transition
point are, in fact, fusion IRF theories.
We can thus directly apply conformal field
theory ideas in the solutions of lattice models away from criticality.
There is a complete parallelism of language and tools here. The bootstrap
of soliton systems is identical to the fusion of IRF models, which itself
is equivalent to the fusion of fields in conformal field theory. The
soliton creation operators $K_a(u)$ reduce in the critical limit to
the chiral vertex operators, and so on.

A number of standing calculations will be reported elsewhere, to further
support this picture. One is the thermodynamic Bethe ansatz calculation
of the soliton systems, which will verify that these are indeed the correct
soliton systems. Another, is the general metric calculation for fusion
superconformal field theories, which is another method for deriving
the solitonic content.  We hope to return to these issues in the future.

Finally, we find here a myriad of distinct connections between what has
previously been thought of as distinct systems. Though there are
missing pieces of the puzzle, we hope that a
unified picture of rational two dimensional quantum field theory starts to
emerge on the canvas.
\endpage

\ack
I am indebted to L. Alvarez--Gaum\'e, V. Bazhanov, C. Gomez, A. Schwimmer and
A.B. Zamolodchikov for interesting discussions and to J. Schwarz for remarks
on the manuscript.
I thank the theory division of CERN and the physics department of
Rockefeller University
where part of this work was done. 
I am deeply indebted to Ida Deichaite for the
idea and insistence to publish this paper, without it this publication would not have happened. 
I am
very grateful to Vladimir Belavin for his tremendous and valuable help in preparing this manuscript.
I am  thankful to Lance Dixon for his help with the figures.

\quote
Tiger, Tiger, burning bright
In the forest of the night
What immortal hand or eye
Could frame thy fearful symmetry?
\fro{William Blake}
\endquote
\endpage
\refout
\endpage
\def\pr#1{#1^\prime}
\centerline{{\bf Figure Captions}}

\item{(2.1)} The Young tableux for the field $[r_i] = [1,2,4,5]
\equiv (s_i) = (1,2,4)$.

\item{(7.1)} A lattice interaction around the face model.  The
state variables are $a,b,c,\ldots$, and they obey the admissibility condition
$a
\sim b$, when $a$ and $b$ are on the ends of the same link.  The Boltzmann
weights are associated to the faces, $\omega \pmatrix{a & b\cr c &
d\cr}$.

\item{(7.2)} The pictorial presentation for the star triangle
equation (STE), eq. (7.3).

\item{(7.3)} Diagonal to diagonal transfer matrix, $T$, and its
associated face transfer matrix, $T = \prod_i X_i (u)$.  The face
transfer matrix element is the Boltzmann weight, $\langle
a_{i-1},\pr a_i,a_{i+1}
|X_i (u)| a_{i - 1},a_i,a_{i+1}\rangle
=w\pmatrix{a_{i-1} & \pr a_i\cr a_i
& a_{i+1}\cr}$.

\item{(7.4)} The generators of the braid group, $\sigma_i$, braid
the $i$'th and $(i + 1)$'th strands.  They obey eq. (7.9) which is one
of the Reidemeister moves, see e.g.,
\REF\Reid{K. Reidemeister, Knotentheorie (Chelsea, New-York, 1948)}
\r\Reid.
A typical braid, along with the action of
$\sigma$, are drawn.  The braids can be thought of as words in braid
group operating on the trivial braid.

\item{(7.5)} The additional braiding relation on the sphere, eq.
(11).  On the sphere this braid is equivalent to the trivial one, around
the north pole.

\item{(7.6)} The four point blocks $F_p (z_1, z_2, z_3, z_4)$ in
the $s$-channel.  The external primary fields $i,j,k$ and $\ell$ obey
$N_{p\rho}^i > 0$ and $N_{p\rho}^i > 0$.

\item{(7.7)} The braiding move on the $s$-channel blocks,
implemented by the matrix $C$, eq. (7.13).

\item{(7.8)} The chiral vertex operator $\Phi_s (z)$, where $s =
\left({a\atop bc}\right)$ is a triplet of primary fields.

\item{(7.9)} The $N$-move takes an $s$-channel block to a
$t$-channel one, and is implemented by the matrix $N_{pq} \left[{ij\atop
k\ell}\right]$.

\item{(8.1)} Examples of fusion graphs for some IRF models.

\item{(10.1)} The soliton scattering process $(a|n|b) + (b|v|d)
\rightarrow (a|v|c) + (c|n|d)$.

\item{(10.2)} The pictorial description of the factorization
equation.

\epsfsize=29cm \hskip 25 true pt

\hbox{\hfill \epsfbox{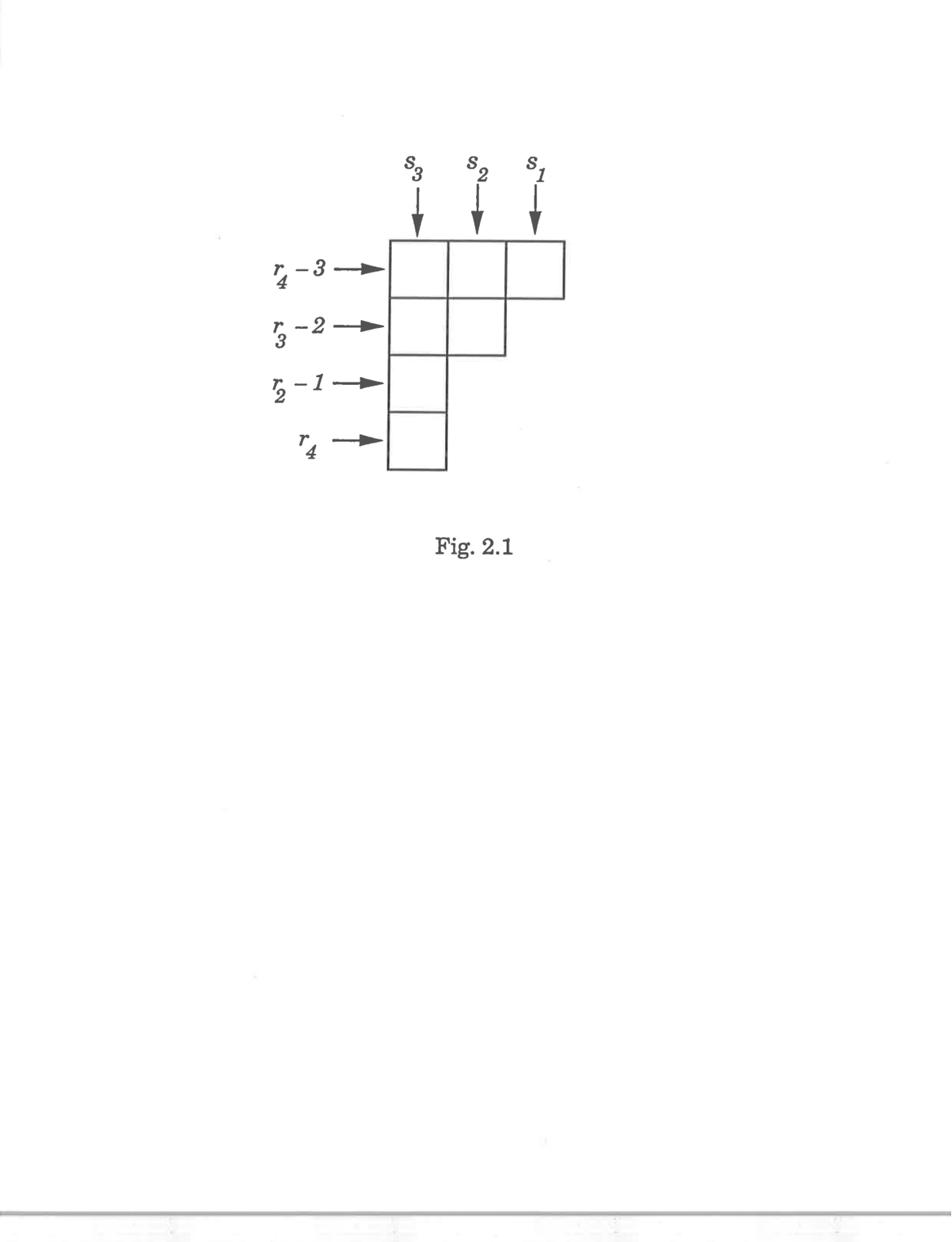} \hfill}
 \epsfsize=29cm
\hbox{\hfill \epsfbox{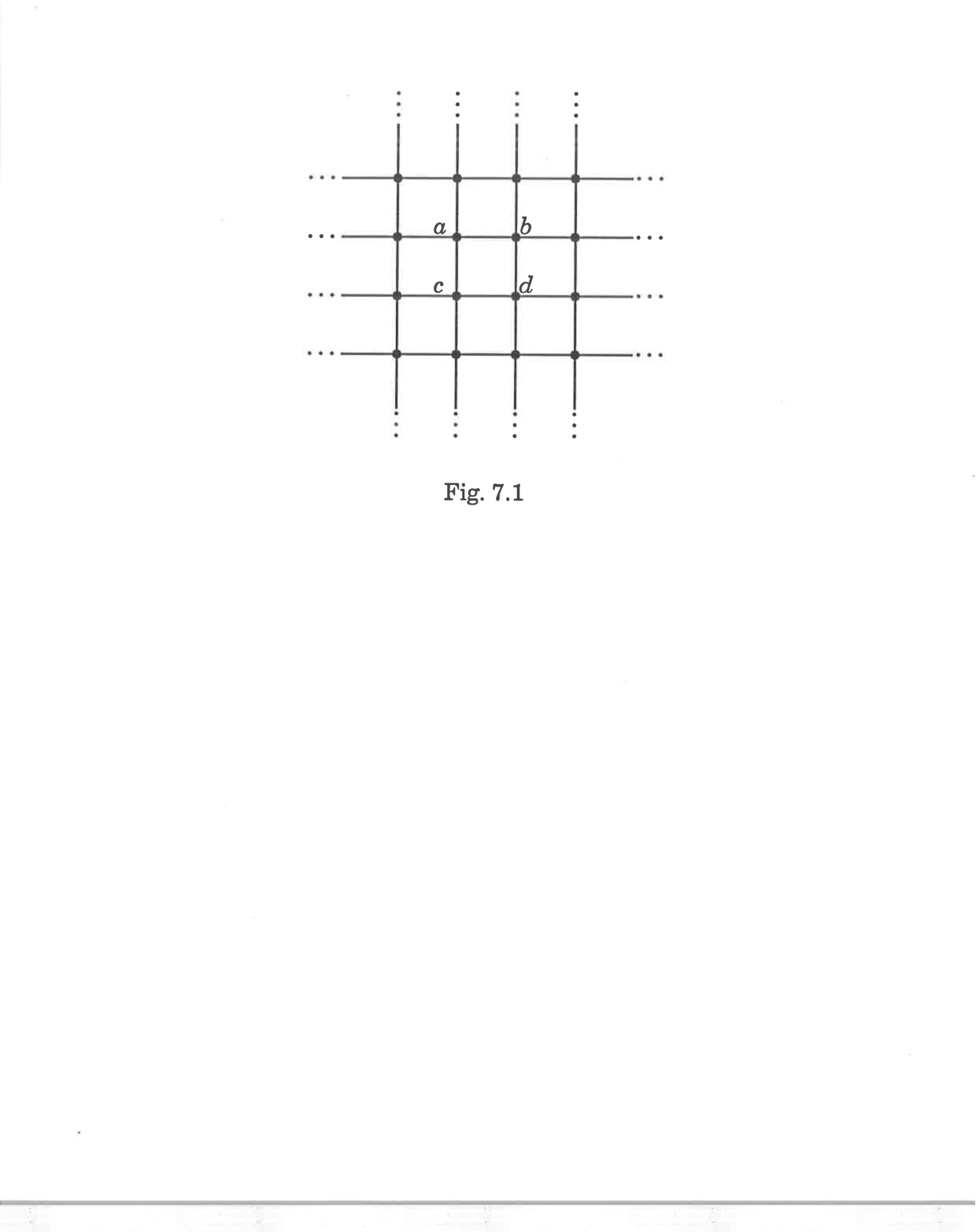}\hfill} 
\epsfsize=29cm
 
\hbox{\hfill \epsfbox{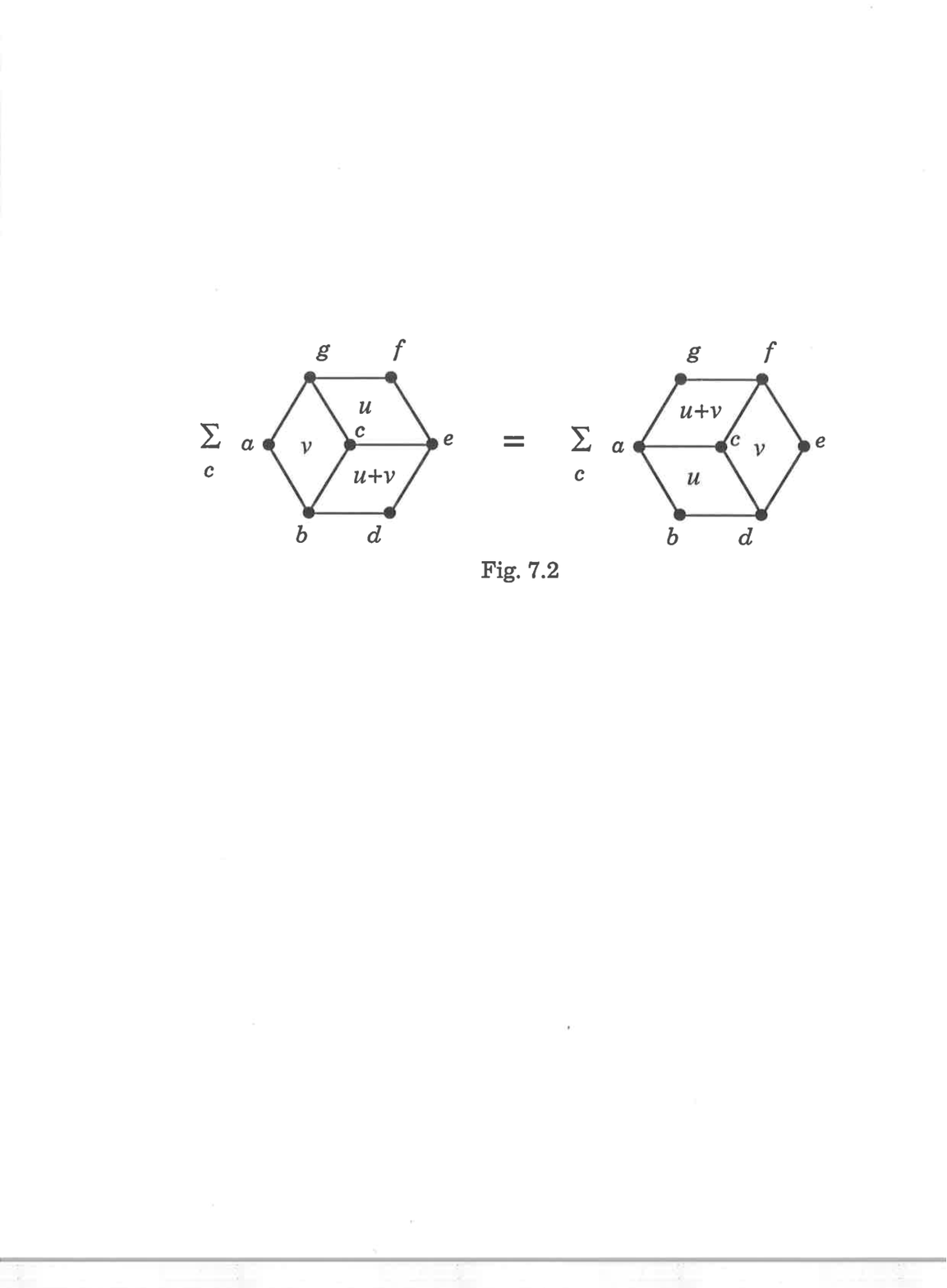}\hfill}
  \epsfsize=29cm
\hbox{\hfill \epsfbox{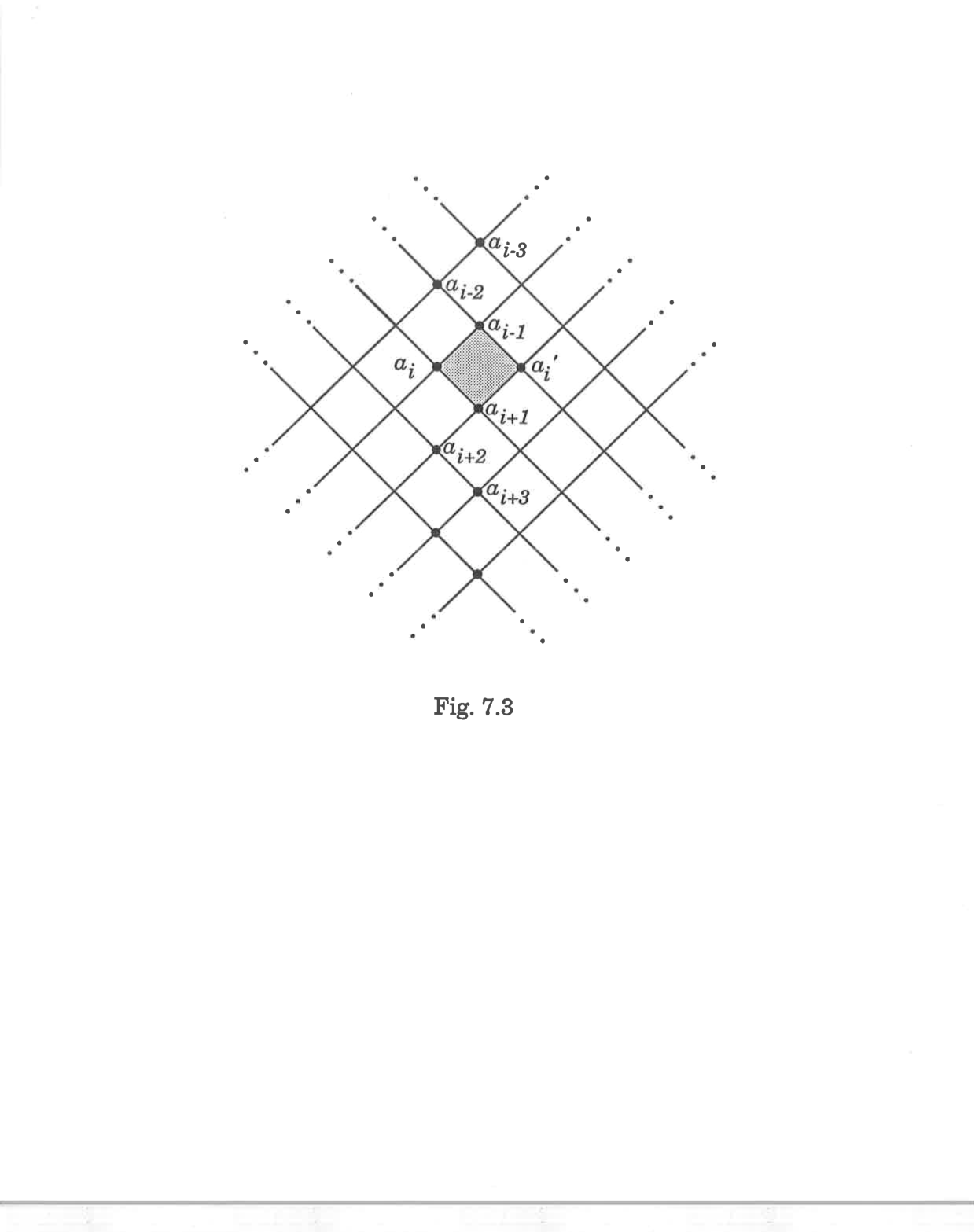} \hfill}
 \epsfsize=29cm  
   \hbox{\hfill \epsfbox{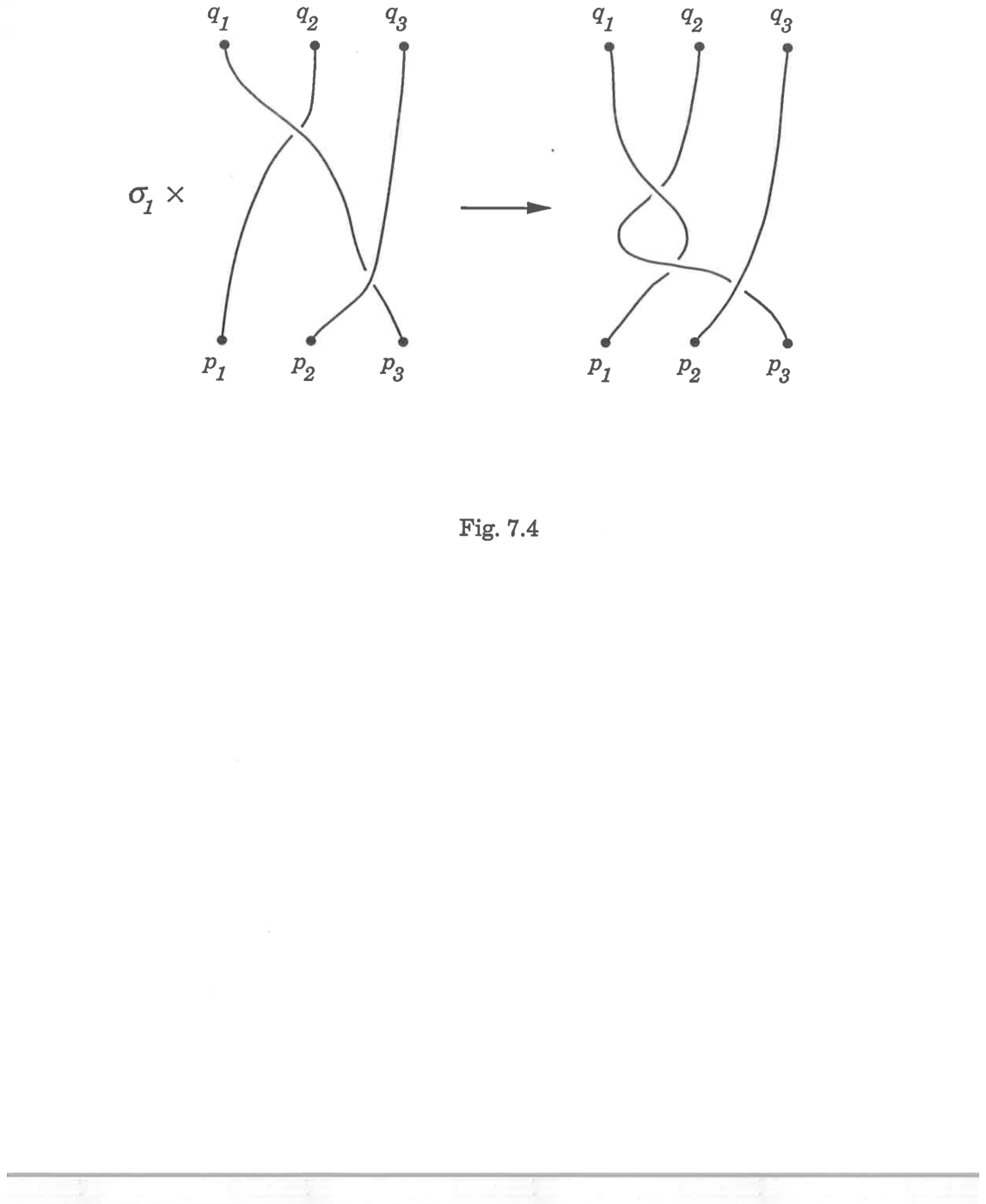} \hfill}
\epsfsize=29cm
    \hbox{\hfill \epsfbox{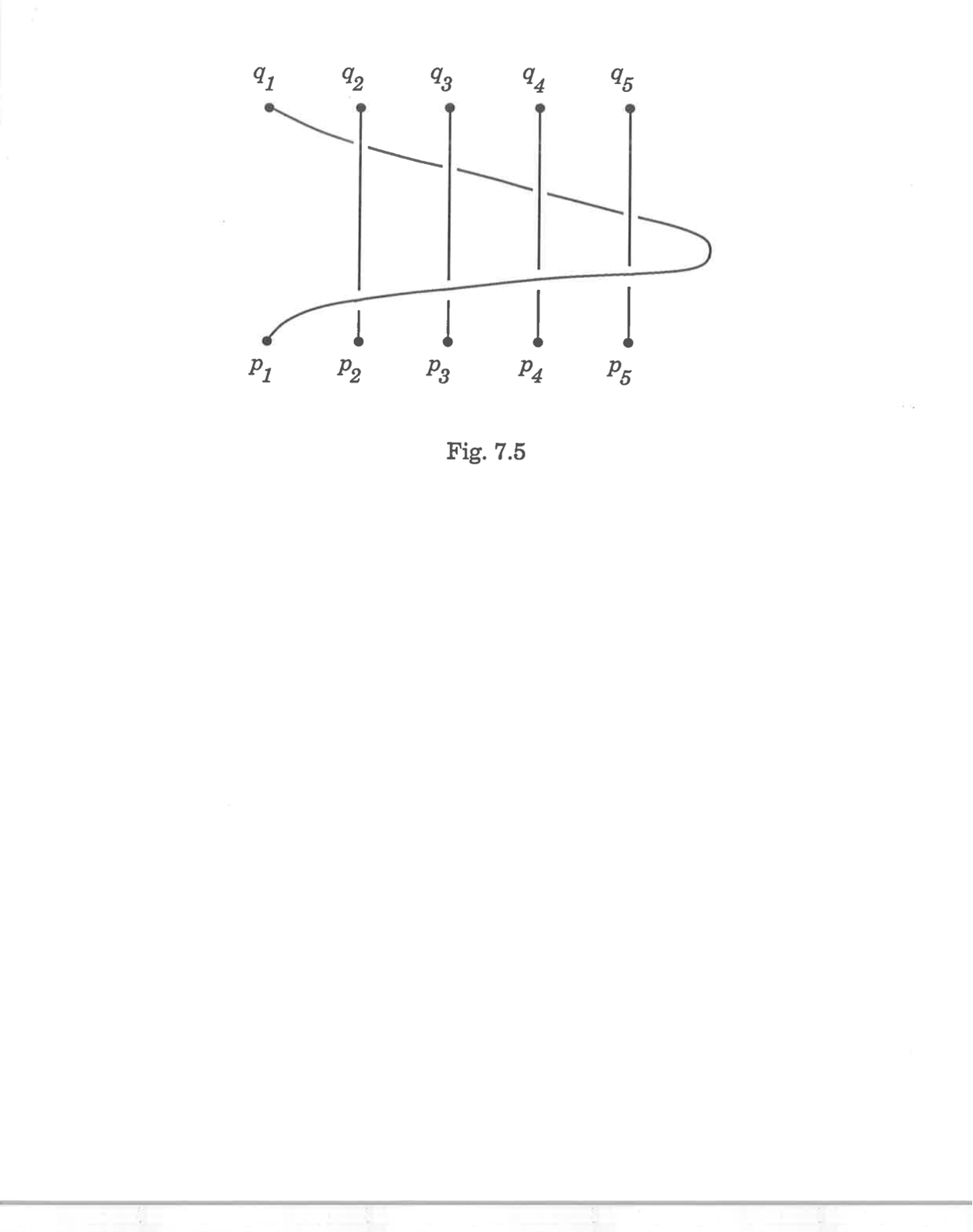} \hfill}
\epsfsize=29cm
     \hbox{\hfill \epsfbox{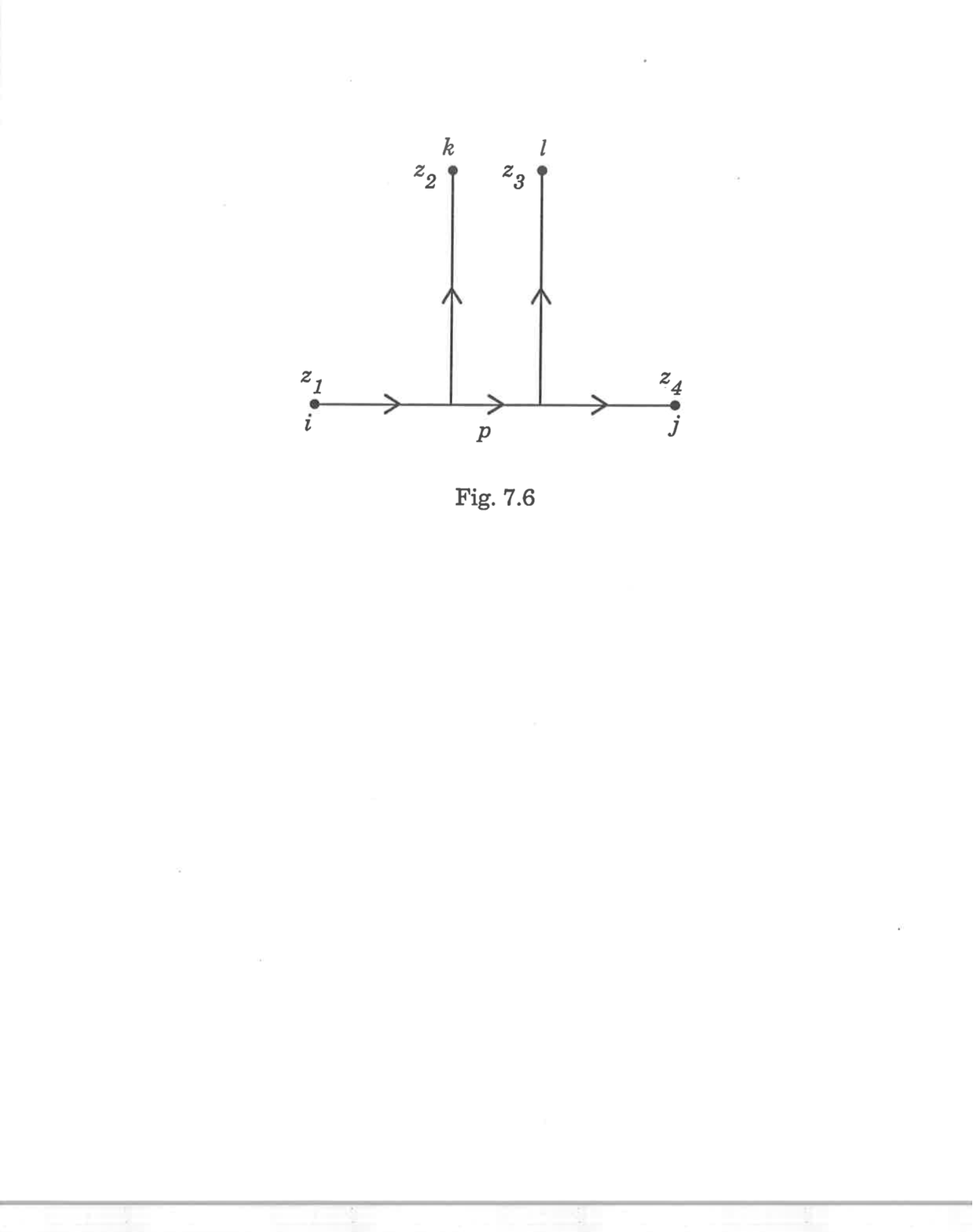} \hfill}
\epsfsize=29cm
      \hbox{\hfill \epsfbox{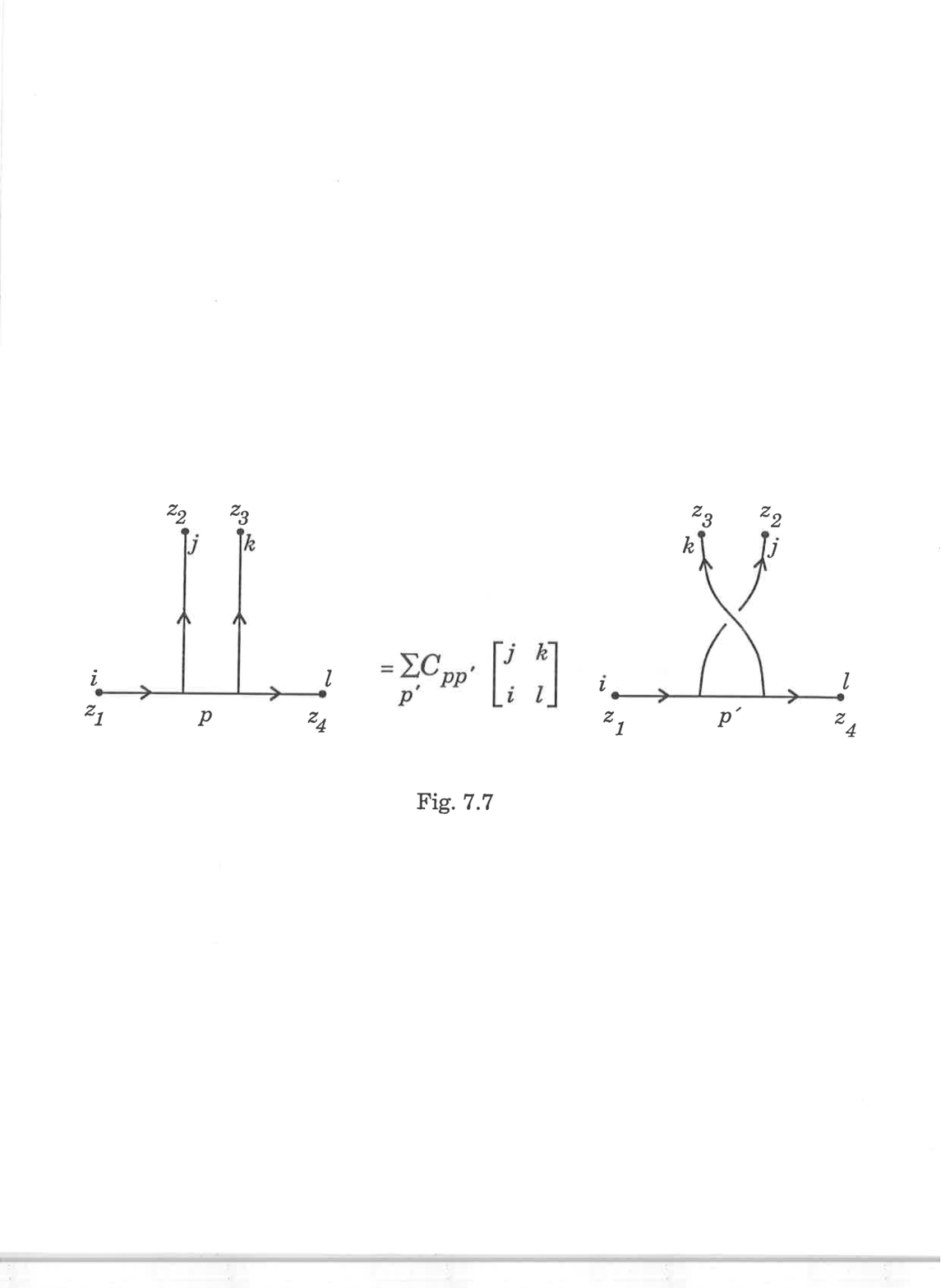} \hfill}
\epsfsize=29cm
       \hbox{\hfill \epsfbox{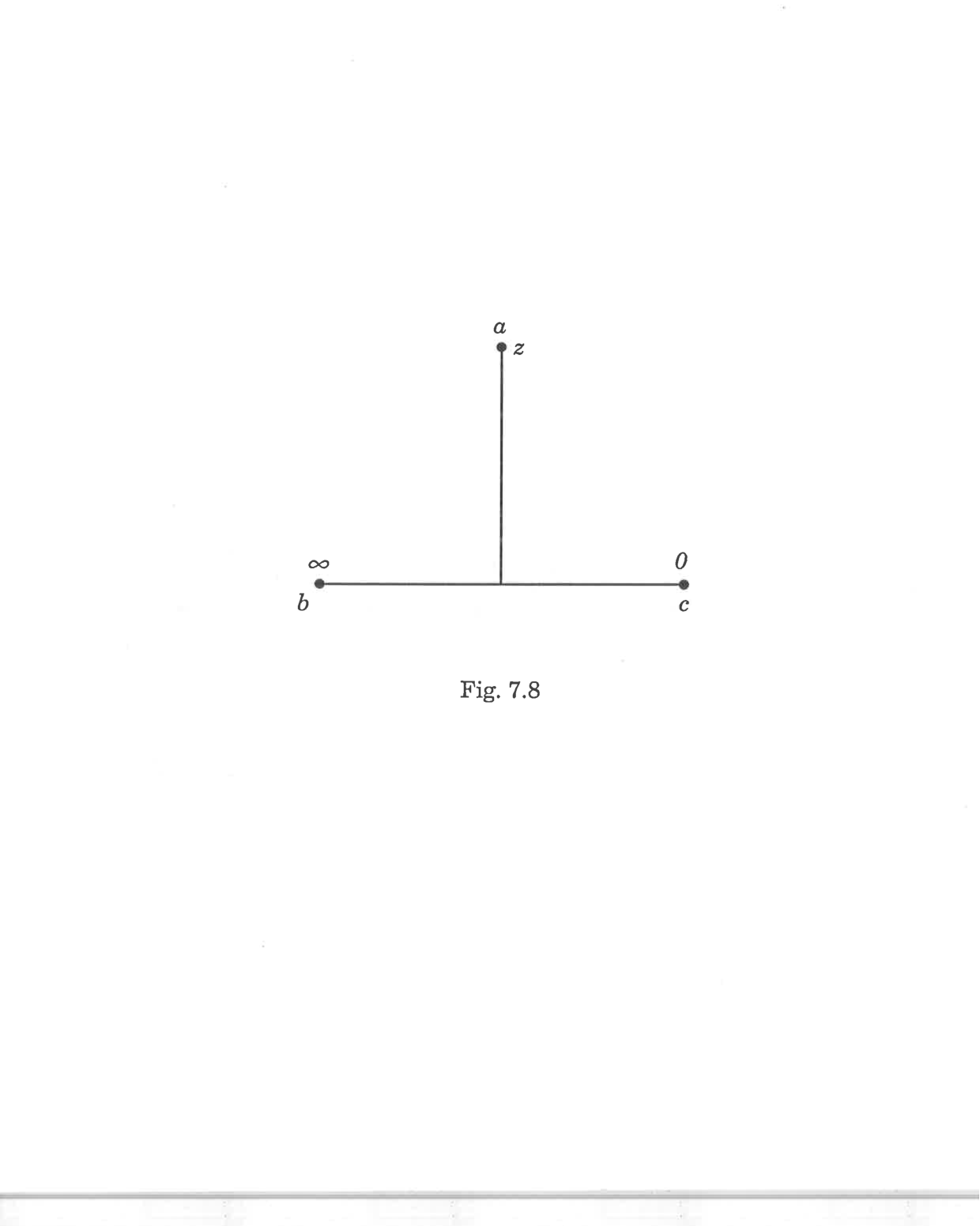} \hfill}
\epsfsize=29cm
        \hbox{\hfill \epsfbox{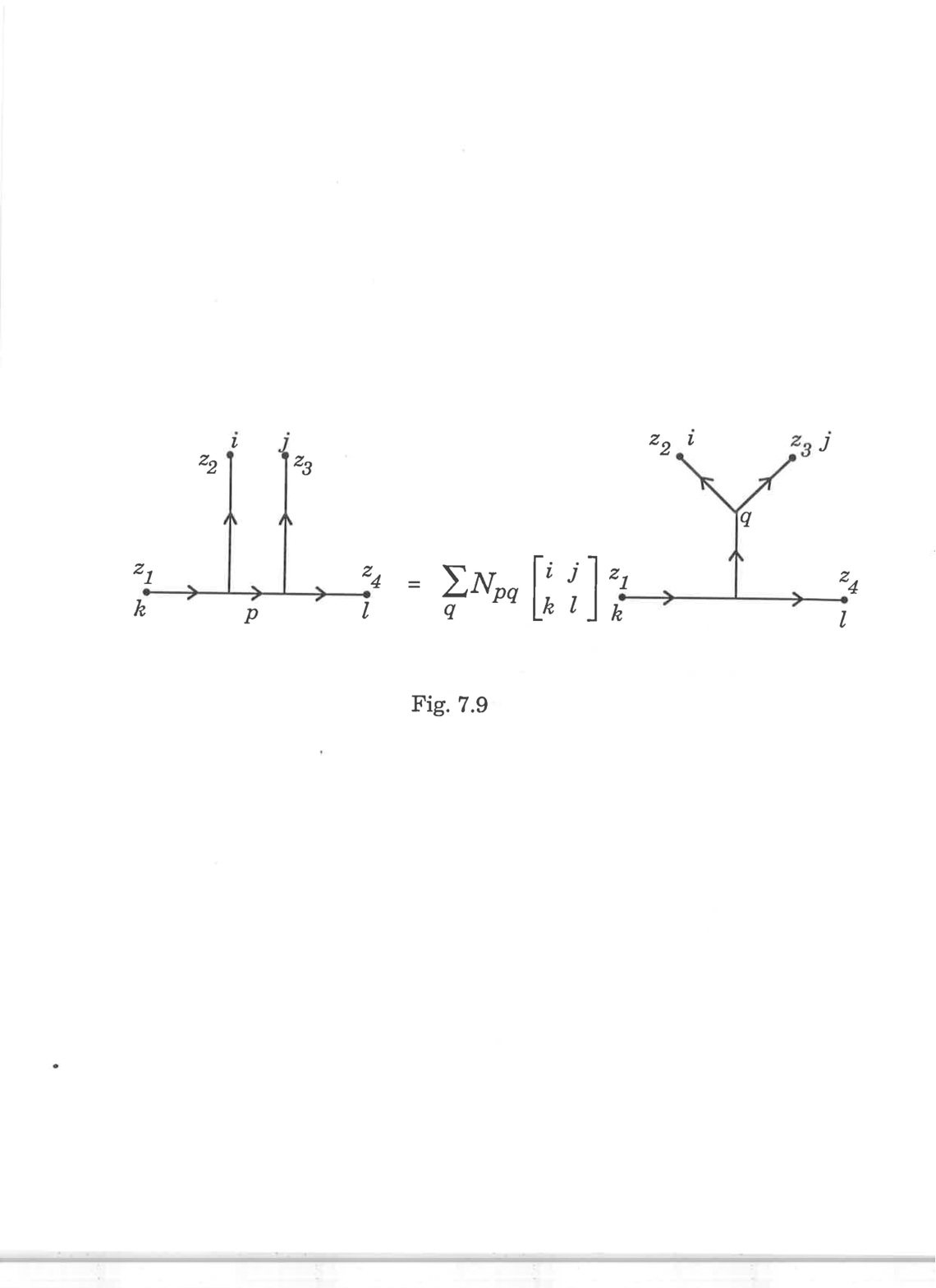} \hfill}
\epsfsize=29cm
         \hbox{\hfill \epsfbox{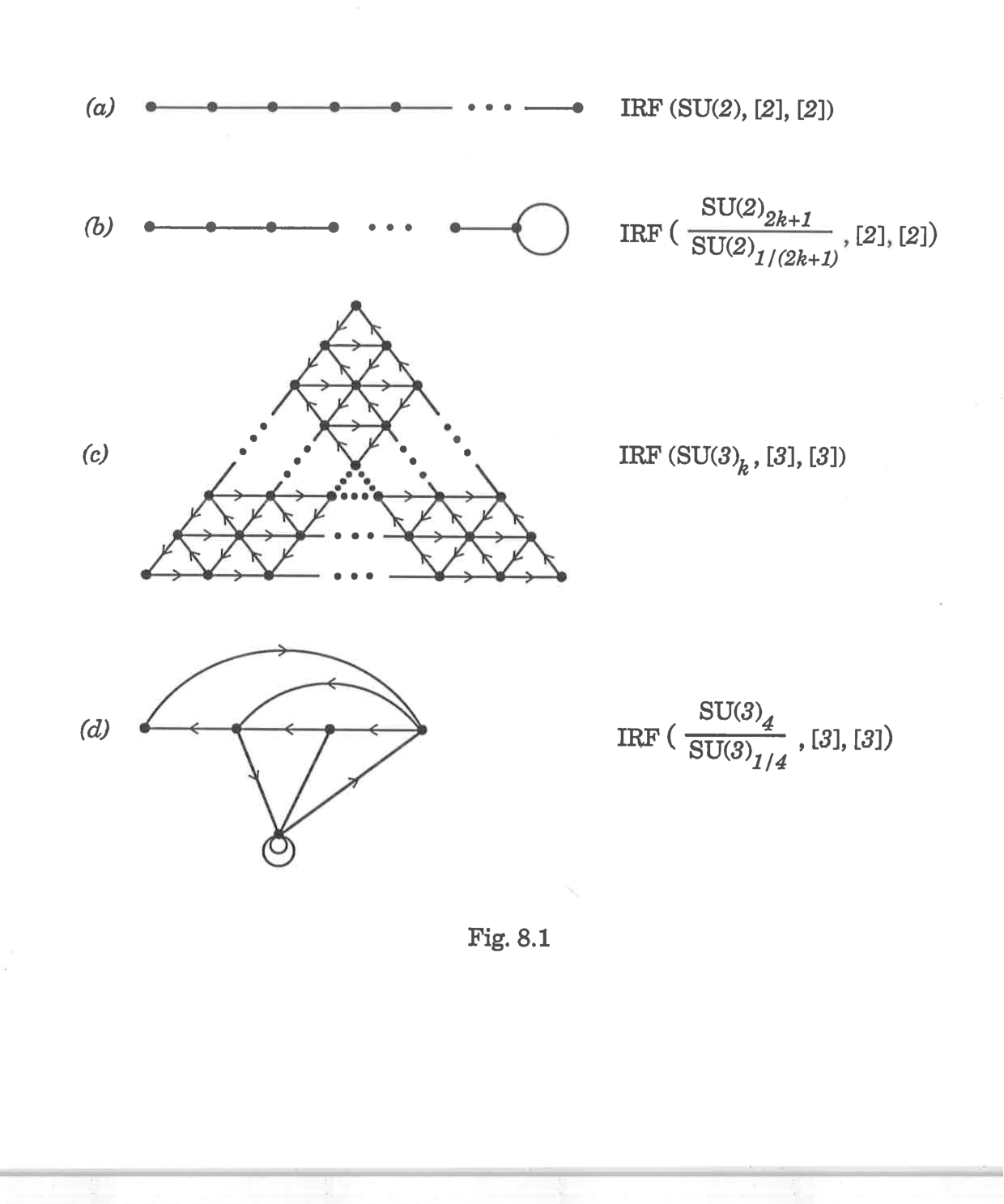} \hfill}
\epsfsize=29cm
          \hbox{\hfill \epsfbox{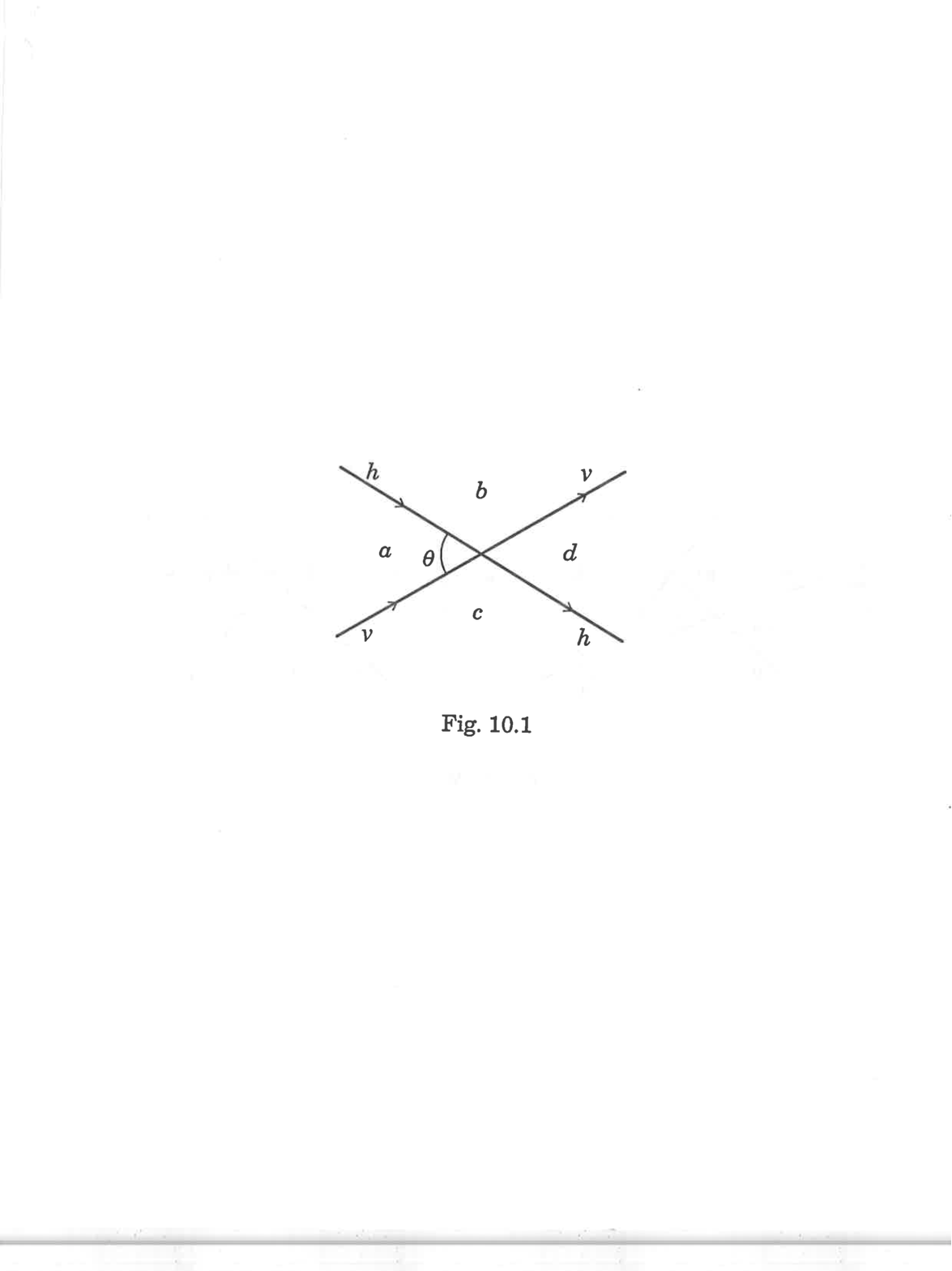} \hfill}
\epsfsize=29cm
 \hbox{\hfill \epsfbox{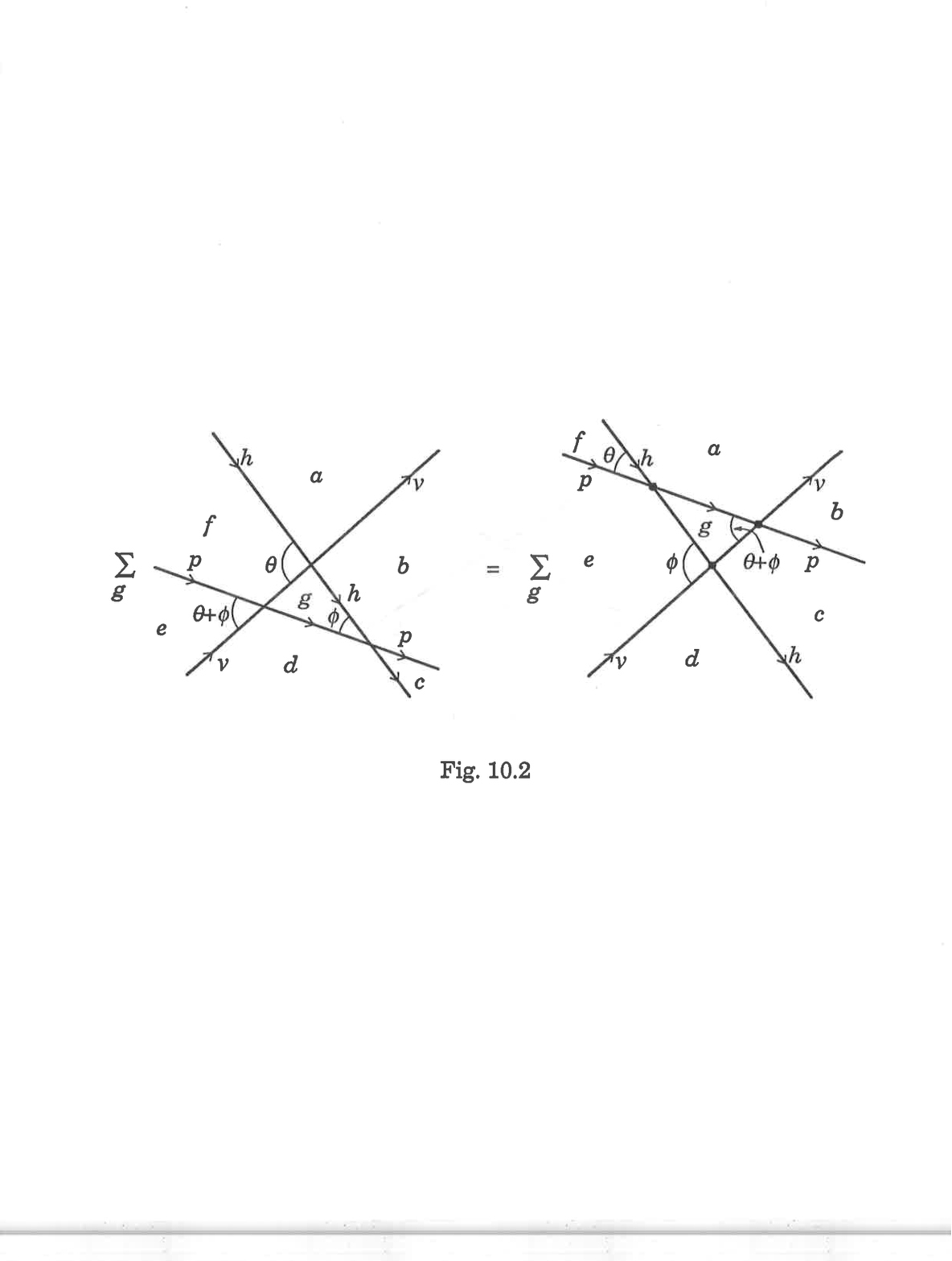} \hfill}

\bye